\newcommand{\arch}{\texttt{CyberSentinel} }

\documentclass[journal]{IEEEtran}
\IEEEoverridecommandlockouts

\def\BibTeX{{\rm B\kern-.05em{\sc i\kern-.025em b}\kern-.08em
    T\kern-.1667em\lower.7ex\hbox{E}\kern-.125emX}}
\usepackage{balance}
\usepackage{listings}
\usepackage{cite}
\let\labelindent\relax 
\usepackage{enumitem}
\usepackage{amssymb}
\newlist{todolist}{itemize}{2}
\setlist[todolist]{label=$\square$}
\usepackage{amsmath,amssymb,amsfonts}
\usepackage{graphicx}
\usepackage{textcomp}
\usepackage{xcolor}
\usepackage{url}
\usepackage{tikz}
\usepackage{amsmath}
\usepackage{titlesec}
\usepackage{multicol}
\usepackage{filecontents}
\usepackage{epsfig}
\usepackage{threeparttable}
\usepackage{graphicx}
\usepackage{amsmath}
\usepackage{bbm}
\usepackage{amssymb}
\usepackage{float}
\usepackage{latexsym}
\usepackage{booktabs}
\usepackage{wrapfig}
\usepackage[noend]{algpseudocode}
\usepackage{algorithm}
\usepackage{pgfplots}
\pgfplotsset{compat=1.8}
\usetikzlibrary{patterns}
\usepackage{lipsum}
\usepackage{algpseudocode}
\usepackage{makecell}
\usepackage{tabularx,ragged2e,booktabs,caption, subcaption}
\usepackage{notes2bib}
\usepackage[font=small,skip=0pt]{caption, subcaption}
\usepackage{amssymb}
\usepackage{pifont}
\usepackage{multirow}
\usepackage{bbm}
\usepackage{amsthm}
\usetikzlibrary{positioning, decorations.pathreplacing, calc, intersections, pgfplots.groupplots}

\newcommand{\cmark}{\ding{51}}%
\newcommand{\xmark}{\ding{55}}%
\usepackage{tikz}
\usepackage{xcolor,colortbl}

\newcommand{\rc}[1]{{\color{red!70!black}{#1}}}

\newcommand{\dgc}[1]{{\color{green!45!black}{#1}}}
\newcommand{\myparab}[1]{\vspace{0.025in}\noindent\textbf{#1}}

\newcommand{\eat}[1]{}

\newcommand{\secref}[1]{{\S\ref{#1}}}
\newcommand{\fref}[1]{\mbox{Fig.~\ref{#1}}}
\newcommand{\tref}[1]{\mbox{Table~\ref{#1}}}

\newcommand*\circled[1]{\tikz[baseline=(char.base)]{
            \node[shape=circle,fill,inner sep=2pt] (char) {\textcolor{white}{#1}};}}

\usepackage{adjustbox}
\usepackage{array}

\usepackage{hyperref}
\hypersetup{%
  colorlinks=false,
}

\pgfplotsset{
        compat=1.7,
        my ybar legend/.style={
            legend image code/.code={
                \draw [##1] (0cm,-0.6ex) rectangle +(2em,1.5ex);
            },
        },
    }

\newcolumntype{R}[2]{%
    >{\adjustbox{angle=#1,lap=\width-(#2)}\bgroup}%
    l%
    <{\egroup}%
}
\AtBeginDocument{%
  \providecommand\BibTeX{{%
    \normalfont B\kern-0.5em{\scshape i\kern-0.25em b}\kern-0.8em\TeX}}}

\usepackage{xcolor}
\definecolor{p4commentcolor}{rgb}{0.12, 0.38, 0.18}
\definecolor{p4keywordcolor}{rgb}{0.04, 0.36, 0.57}
\definecolor{p4stringcolor}{rgb}{0.17, 0.17, 0.59}
\definecolor{p4identifiercolor}{rgb}{0.56, 0.35, 0.01}

\lstdefinelanguage{p4}{
  morekeywords={action, apply, table, key, entries, header, parser, control, match_kind, state, transition, extract, set_metadata, ipv4, ethernet, ingress, egress, drop, if, execute, else, bit,actions,range,default_action,MathUnit,MathOp_t,Register,RegisterAction,hdr,exact},
  morecomment=[l]{//},
  morecomment=[s]{/*}{*/},
  morestring=[b]",
  morestring=[b]',
  sensitive=true,
  keywordstyle=\color{p4keywordcolor}\bfseries,
  commentstyle=\color{p4commentcolor}\textit,
  stringstyle=\color{p4stringcolor},
  identifierstyle=\color{p4identifiercolor},
}

\begin{document}


\title{CyberSentinel: Efficient Anomaly Detection in Programmable Switch using Knowledge Distillation
}

\author{\IEEEauthorblockN{Sankalp Mittal}
\IEEEauthorblockA{
\textit{Google, India}}}

\maketitle

\begin{abstract}
\boldmath
  The increasing volume of traffic (especially from IoT devices) is posing a challenge to the current anomaly detection systems. Existing systems are forced to take the support of the control plane for a more thorough and accurate detection of malicious traffic (anomalies). This introduces latency in making decisions regarding fast incoming traffic and therefore, existing systems are unable to scale to such growing rates of traffic. In this paper, we propose \texttt{CyberSentinel}, a high throughput and accurate anomaly detection system deployed entirely in the programmable switch data plane; making it the first work to accurately detect anomalies at line speed. To detect unseen network attacks, \arch uses a novel knowledge distillation scheme that incorporates "learned" knowledge of deep unsupervised ML models (\textit{e.g.}, autoencoders) to develop an iForest model that is then installed in the data plane in the form of whitelist rules. We implement a prototype of \arch on a testbed with an Intel Tofino switch and evaluate it on various real-world use cases. \arch yields similar detection performance compared to the state-of-the-art control plane solutions but with an increase in packet-processing throughput by $66.47\%$ on a $40$ Gbps link, and a reduction in average per-packet latency by $50\%$. 
     
\end{abstract}

\section{Introduction} \label{intro}
The internet traffic is growing at a tremendous rate. To provide a context, traffic from IoT devices more than doubled the number in 2019 \cite{cit1} and is expected to reach 24.6 billion by 2025. The limited hardware in IoT devices makes it challenging to deploy complex security mechanisms. Thus, IoT devices are in general major target for attackers \cite{alrawi2019sok}. Further, IoT devices are very large in number (\textit{e.g.}, cameras, sensors). Therefore, achieving low-cost and real-time anomaly detection on such a massive scale with growing traffic is a challenge.

Traditional anomaly detection systems are typically rule-based systems with high throughput requirements \cite{fu2021realtime}. In general, packet-level rules such as port-based or signature-based firewall rules \cite{rafique2013firma,newsome2005polygraph} are used in such detection systems. Following the application of rules on incoming traffic, offline sampling analysis \cite{tang2020zerowall} takes place. However, such systems not only fail to detect unseen network attacks but are often easily bypassed \cite{fu2021realtime,mirsky2018kitsune,290987}.

To detect unseen network attacks effectively, the use of unsupervised machine learning (ML) models is gaining popularity. Actual systems that leverage unsupervised ML mostly use an ensemble of autoencoders \cite{wang2017machine,mirsky2018kitsune} to detect malicious traffic (anomalies) or unseen network attacks. However, these systems are deployed in control planes and hence they cannot operate at a sufficient throughput to reflect practical scenarios \cite{pan2021sailfish}. Moreover, abnormal/attack/malicious traffic
\textit{usually} constitutes a very small portion of the entire traffic \cite{xiong2021anomaly}.
Uploading all traffic to the control plane for detection causes
excessive communication overhead which is inefficient from the system scalability point of view.

The emergence of programmable switches (\textit{e.g.}, P4 switches \cite{bosshart2014p4}) brings a new direction to the research in anomaly detection systems. Compared to control planes, programmable switch data planes can achieve 20x higher throughput, 20x lower latency,
and 75x faster packet forwarding rate at the same infrastructure and maintenance cost \cite{pan2021sailfish}. There have been efforts towards systems leveraging programmable switch capabilities for intrusion or anomaly detection. 

Some systems that leverage programmable switches use the data plane capabilities to collect netflow \cite{estan2004building} based features while the control plane is used for attack detection \cite{sonchack2018scaling,barradas2021flowlens,zhou2023cerberus,xing2020netwarden,seufert2024marina}. However, these systems mostly leverage supervised ML \cite{barradas2021flowlens,seufert2024marina} or statistical ML methods \cite{xing2020netwarden}. Supervised or statistical methods assume that the labels are always present in the datasets or traffic traces. In huge amounts, this is one of the major hindrances for a realistic deployment of ML-based anomaly detection systems as labeling can be very costly \cite{apruzzese2022sok}. Moreover, supervised ML-based systems fail to detect unseen attacks \cite{mirsky2018kitsune,290987} (closed-world setting) which creates issues in real-time anomaly detection \cite{apruzzese2022sok,arp2022and}. Lastly, even though these works yield increased throughput by partially leveraging data planes, the control plane still becomes a bottleneck and thus overall hurts throughput \cite{zhou2023efficient}. 

Other systems deploy intrusion or anomaly detection systems entirely in the switch data planes and can achieve high throughput. However, they are threshold filtering rule-based systems \cite{liu2021jaqen,zhang2020poseidon,accturbo} and therefore, can detect very specific types of network attacks. This is an issue from a practical anomaly detection point of view \cite{zhou2023efficient,arp2022and}.

There are also several systems that deploy decision tree-based (due to programmable switch limitations (\secref{psa})) supervised ML models (they can transform into switch-based rules) in programmable switch data planes \cite{xie2022mousika,10158739,zheng2021planter,zheng2022automating,friday2022inc,friday2022learning,Akem2023FlowrestPF,zhou2023efficient,akem2024jewel,jafri2024leo} to detect network attacks (enable in-network intelligence). This achieves high throughput and line-speed traffic processing. As stated previously, supervised methods require a large-scale anomaly dataset. Such high-quality network intrusion detection anomaly datasets are difficult to obtain \cite{arp2022and}. Moreover, these systems also have to deal with the concept drift \cite{jordaney2017transcend} of anomalous samples. This issue needs to be accounted to develop a practical anomaly detection system \cite{andresini2021insomnia,apruzzese2022sok}. In contrast, using unsupervised ML methods require us to maintain only normal/benign datasets, which are also more readily available in real-world scenarios \cite{arp2022and}.

Most recent work on programmable switch-based anomaly detection \cite{290987} deploys unsupervised Isolation Forest \cite{liu2008isolation} (iForest) model in the switch data plane for preliminary attack detection. This is followed by a more thorough anomaly detection using a state-of-the-art autoencoder in the control plane. However, there are two key issues. First, this system still takes the support of the control plane for accurate anomaly detection. Therefore, the control plane becomes the bottleneck and the overall throughput achieved is only slightly more than $50\%$ (see \secref{tdc}) of the line speed (on 40 Gbps link). Second, the data plane preliminary anomaly detection is the weaker part of the detection system as it yields high false positives (FPs) on the many intrusion detection datasets. According to practitioners \cite{277250}, this is one of the biggest problem in operational security and cyberdetection in anomaly detection systems. Moreover, we found that for many other intrusion detection based datasets, an iForest approach is not sufficient to even achieve high true positives (see \secref{dpod}). \textit{Therefore, to achieve a high throughput and accurate anomaly detection in the data planes, we have to envision state-of-the-art autoencoder detection performance in the data planes.} 

In this paper, we propose \texttt{CyberSentinel}, a high throughput and accurate anomaly detection system deployed entirely in the programmable switch data plane. To the best of our knowledge, \texttt{CyberSentinel} is the first work to envision autoencoder's anomaly detection performance at line speed.  

To design \texttt{CyberSentinel}, we must overcome the following key challenges. \textbf{\textit{(i)}} It is difficult to design an unsupervised ML model that can not only be deployed on switch data plane but can also achieve comparative performance of an autoencoder (\textit{i.e.}, achieve both high true positives and low false positives); \textbf{\textit{(ii)}} it is challenging to deploy that unsupervised model on a programmable switch data plane (that has limitations, see \secref{psa}); \textbf{\textit{(iii)}} it is challenging to extract and maintain the required flow
features on the limited switch memory.

To overcome the first challenge, we propose a novel technique of knowledge distillation \cite{gou2021knowledge} from an ensemble of autoencoders into an iForest. This is done by carefully embedding reconstruction errors from autoencoders into leaves of iForest. Our knowledge distilled iForest obtains comparable performance of an autoencoder for various anomaly detection use-cases (see \secref{exp}). We are the \textit{first} to come up with an unsupervised knowledge distillation strategy in iForests (\secref{kd}). 

To overcome the second challenge, that is, actually deploying knowledge distilled iForest in the switch data plane, we propose a \textit{variation} of whitelist rules generation strategy \cite{290987} to comply with our knowledge distillation scheme. These whitelist rules can then be installed in the data plane (\secref{iwlrg}). 

To overcome the third challenge, we borrow bi-hash algorithm and double hash table
methods from \cite{290987}. These methods are proposed to match bidirectional flow with only
$O(1)$ computational complexity, based on which burst-based (flow divided into many streams of packets called bursts)
features can be obtained to distinguish abnormal behavior (\secref{bfe}).

\myparab{Contributions. }Our contributions are as follows.
\begin{itemize}
    \item We design a novel knowledge distillation strategy that transfers "learned knowledge" from an ensemble of autoencoders into an iForest (\secref{kd}). This is done by carefully embedding the reconstruction errors from autoencoders into leaves of an iForest. 
    \item We design a variant of whitelist rules generation strategy from \cite{290987}. The small set of whitelist rules generated from our distilled iForest can easily be installed on the switch data plane (see \secref{iwlrg} for details).
    \item Following in footsteps of \cite{qu2023input}, we developed a module that can efficiently map a sequence of burst vectors (for a given flow) into a fixed length flow vector (\textit{burst-flow mapping}). This was done to make our knowledge distilled iForest (trained on burst-level features) to comply with popular autoencoders \cite{290987,mirsky2018kitsune} (trained on flow features). Due to page limit, we defer this contribution to supplementary material\footnote{We highly recommend and encourage the readers to also go through
the supplementary material for a complete understanding of this work.}. 
    \item We clarify the challenges of implementing \texttt{CyberSentinel}'s data plane on Intel Tofino switch and developing a working prototype (\secref{dpi}).
    \item We designed a control plane module for offline ML model preparation and online memory management in the data plane. See \secref{cm}.
    \item Lastly, we extensively evaluate \arch on various attack datasets \cite{mirsky2018kitsune,bezerra2018providing,koroniotis2019towards,art6,290987} and real-world IoT testbed \cite{sivanathan2018classifying,290987}. Compared to \cite{mirsky2018kitsune,zhou2023efficient,jafri2024leo,xie2022mousika,akem2024jewel}, \arch achieves superior anomaly detection which is also at par with latest work HorusEye \cite{290987}. \textit{We stress that we get similar detection performance as HorusEye without taking any support of control plane for anomaly detection.} \arch also yields $66.47\%$ higher throughput compared to HorusEye \cite{290987}. Moreover, \arch is also demonstrated to be fairly robust to adversarial attacks. See \secref{exp}.
\end{itemize}


\section{Background, Motivation and Threat Model}
\subsection{Programmable switch architecture} \label{psa}
Programmable switches such as Intel Tofino \cite{art1} have two types of processors that operate in two
different planes. In the data plane, high-speed forwarding ASICs are restricted to doing only simple arithmetic and logical computations on packets. In the control plane, CPUs are used for general-computing tasks such as controlling the packet forwarding pipeline, or for exchanging data with the ASIC through DMA.
\begin{figure}[t!]
    \centering
    \includegraphics[width=8.5cm]{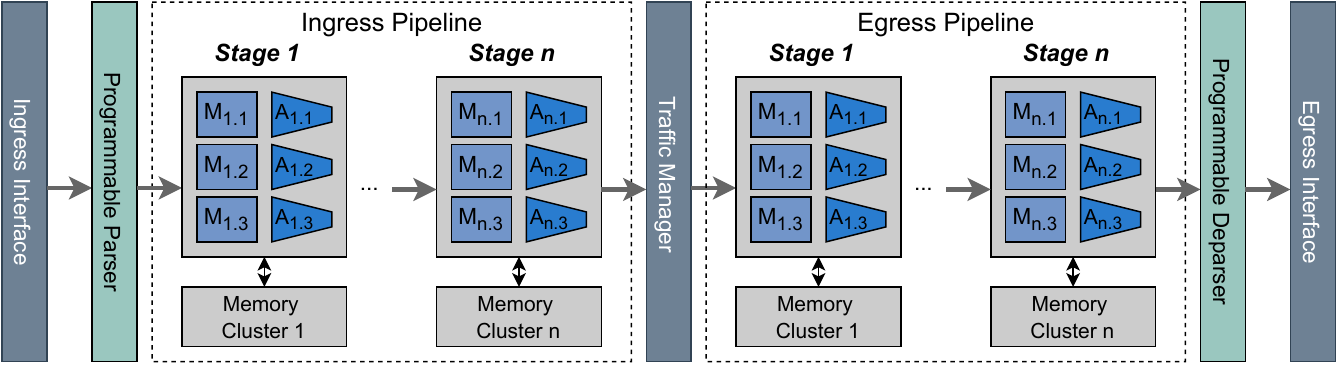}
    \caption{\textmd{Protocol Independent Switch Architecture (PISA).}}
    \vspace{-0.5cm}
    \label{fig:0}
\end{figure}
The data plane can be programmed using a protocol and hardware-independent language, such as P4 \cite{bosshart2014p4}. \fref{fig:0} illustrates Protocol
Independent Switch Architecture (PISA) \cite{art2}, where incoming packets are processed by
two logical pipelines (ingress and egress) of match+action units (MAUs) arranged
in stages. Packet headers and associated packet metadata may then
match (M) with match-action rules of one or more tables. Upon a hit, the matched rule's action is applied. 
triggering further processing by the
action (A) unit associated with the matched table’s entry. Some examples of  actions include
modifying packet header fields, or updating stateful switch memory. Tables and other objects defined in a P4 program are instantiated inside
MAUs and associated rules are populated by the control plane at run-time.

\myparab{Memory constraints. } There are restrictions on per-packet memory accesses while processing a packet in the switch pipeline. These restrictions limit the implementation of data structures defined in the P4 programs. 
More specifically, high-speed access to the switch's SRAM available in each pipeline stage (as shown in~\fref{fig:0}) enables P4 programs to persist state across packets (\textit{e.g.}, using register arrays). Unfortunately, today's programmable switches contain a small amount of stateful memory (in the order of $100$~MB SRAM~\cite{miao2017silkroad}), and only a fraction of the total available SRAM can be used to allocate register arrays. Moreover, accessing all available registers can be a complex task since the registers in one stage cannot be accessed at different stages~\cite{chole2017drmt}; this is because the SRAM is uniformly distributed amongst the different stages of the processing pipeline (see~\fref{fig:0}).

\myparab{Processing constraints. }The P4 programs installed on the switch must also use very simple instructions to process packets. 
To guarantee line-rate processing, packets spend a fixed and small amount of time in each pipeline stage (a few ns \cite{sivaraman2016packet, sharma2017evaluating}) executing simple instructions. This restricts the number and type of operations allowed within each stage. More specifically, multiplications, divisions, floating-point operations, and variable-length loops are not supported. Also, each table’s action can only perform a restricted set of simpler operations, like additions, bit shifts, and memory accesses that can quickly be performed while the packet is passing through an MAU without stalling the whole pipeline~\cite{sivaraman2017heavy}.

\subsection{Machine learning in data planes}
Deploying trained machine learning models directly on the programmable switch data planes is an active research area as it transforms the network towards line-speed traffic inference or fast malicious traffic detection. Due to the limited number of per-packet operations, various pipeline constraints, and limited switch memory, it is not possible to deploy complex ML models (\textit{e.g.}, neural networks) on the data planes. Till date, only decision tree-based models (because they are amenable to be implemented on match-action pipeline of switch) have been deployed in the data planes \cite{zhou2023efficient,Akem2023FlowrestPF,akem2024jewel,xie2022mousika,jafri2024leo} for traffic classification. Moreover, all these works are supervised learning methods. As for the unsupervised methods, while \cite{zheng2022automating} does provide a method to embed Isolation Forest in the data plane, HorusEye \cite{290987} is the only work to the best of our knowledge that \textit{actually} implements unsupervised Isolation Forest in the programmable switch. 

\subsection{Knowledge Distillation in programmable data planes}
Many works suggest using knowledge distillation \cite{hinton2015distilling,bai2022multinomial,frosst2017distilling,bai2019rectified} to transfer "learned knowledge" from complex teacher to simpler student models to obtain comparable performance of teacher models with reduced latency and increased deployment efficiencies. To envision classification performance of sophisticated ML models in data planes, it is necessary to distill their knowledge into decision trees. While many works \cite{frosst2017distilling,bai2019rectified} do follow that strategy, \cite{xie2022mousika,10158739} is the only work till date that actually deploys a distilled binary decision tree on switch data plane. However, all the prior works focus on supervised learning models. Ours is the first work to design an unsupervised learning strategy to (i) perform knowledge distillation from deep unsupervised models (\textit{e.g.}, autoencoders \cite{goodfellow2016deep}) to Isolation Forests \cite{liu2008isolation} (iForests), and (ii) methodology that is data plane friendly. See \secref{kd} for details.  

\subsection{Motivation}
Motivation for \arch is two-fold. First, there exists no prior work that can \textit{successfully} detect malicious traffic (anomalies\footnote{In this paper, terms malicious traffic and anomalies will be used interchangeably. This is because most of the traffic that we get is normal traffic.}) at line speed entirely in the data planes using unsupervised learning methods. Second, the only work \cite{290987} that tries to embed iForest to the data plane is forced to take support of complex models (autoencoders) in the control plane for accurate anomaly detection. In fact \cite{290987} claims that the data plane module is the weaker part of its traffic detection system as it yields significant false positives which are essential to be minimized \cite{277250} from an operational cybersecurity standpoint. Therefore, to envision \textit{line speed accurate anomaly detection} in data planes, it is essential to transfer the knowledge of trained deep-learning models like autoencoders to the iForest. 

\subsection{Threat model}
This paper addresses the cyber threats arising from compromised IoT devices. Due to inherent vulnerabilities of such IoT systems, attackers can gain unauthorized access and transform the devices into botnets. Once compromised, a device can be manipulated by a botmaster to execute various malicious activities, including DDoS attacks, data breaches, and the propagation of malware. Similar to prior works \cite{290987}, the paper makes specific assumptions: (i) newly manufactured IoT devices are initially secure and trustworthy, devoid of pre-installed malware or backdoors, and (ii) it assumes that all active attacks originating from IoT bots leave discernible traces within the IP layer traffic, thereby excluding considerations of attacks such as eavesdropping \cite{nakibly2012persistent} or MAC spoofing.

\section{Overview of \arch} \label{overview}
\begin{figure}[t!]
    \centering
    \includegraphics[width=9cm]{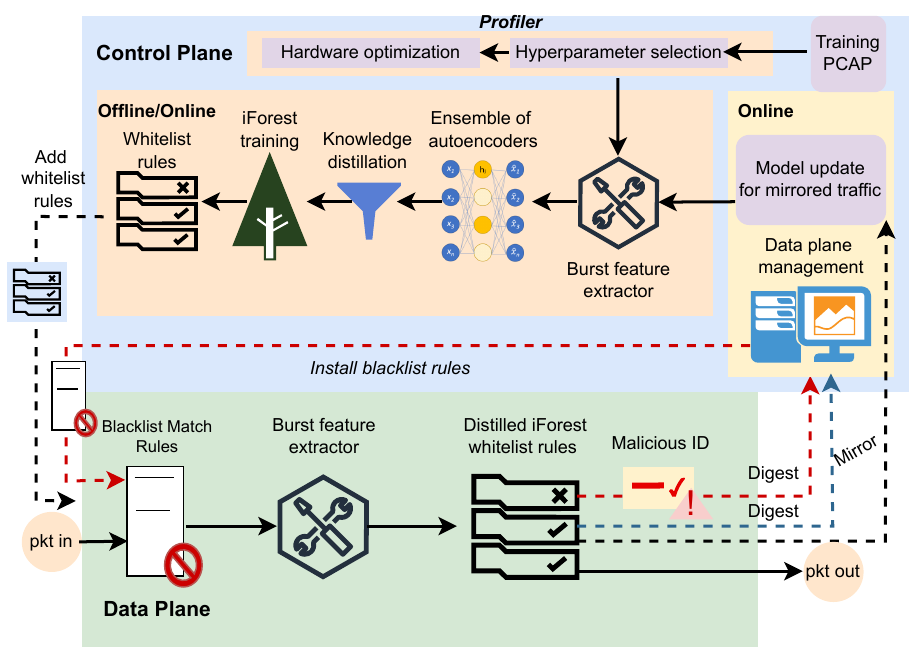}
    \caption{Overview of \arch}
    \label{fig:overview}
    \vspace{-0.5cm}
\end{figure}
\arch is a novel IoT anomaly detection system deployed in the data plane (i.e., programmable switches). The overview of \arch is shown in \fref{fig:overview}. By taking advantage of high throughput of data plane, \arch can conduct malicious traffic detection at a very high rate and reduced latency. We briefly present the two modules of \arch.  

\myparab{Data plane module.} This module first extracts burst-level features from the incoming traffic/packets using the burst feature extractor (\secref{bfe}). It detects anomalies based on the
\textit{knowledge distilled iForest model} (knowledge distillation from an ensemble of autoencoders) deployed in the form of whitelist rules. Details of knowledge distillation and whitelist rules generation are presented in \secref{kdid}. After the anomaly detection (using whitelist rules),
data plane module sends the malicious flow ID to the control
plane so that it can install appropriate black-list rules to the data plane to block incoming malicious traffic with the respective flow ID in the future.

\myparab{Control plane module. }This module consists of the following components. (a) Profiler (\secref{plr}) that performs automated hyperparameter selection for iForest model preparation offline (\textit{e.g.}, number of iTrees) to give maximum malicious traffic detection accuracy, and hardware optimization (\textit{e.g.}, number of entries in storage registers in data plane) for minimum data plane memory footprint. Once offline iForest model preparation is done, whitelist rules are generated (\secref{kdid}) and deployed in the data plane along with the P4 program. (b) The data plane management system (\secref{dpm}) that is responsible for online installation of blacklist rules to the data plane, periodically deleting old blacklist rules (as per FIFO, LRU \textit{etc.}), and updating the iForest whitelist rules in the data plane as per new incoming packets from normal traffic. 

In summary, \arch is the only work to detect anomalies (malicious traffic) using distilled iForest model \textit{entirely} in the data plane. We next describe the two modules in \secref{dm} and \secref{cm} in detail.
\section{Data Plane Module} \label{dm}
We provide the details of the following components of a \arch data plane module.
\subsection{Knowledge distilled iForest design} \label{kdid}
\begin{figure*}
    \centering
    \includegraphics[width=19cm]{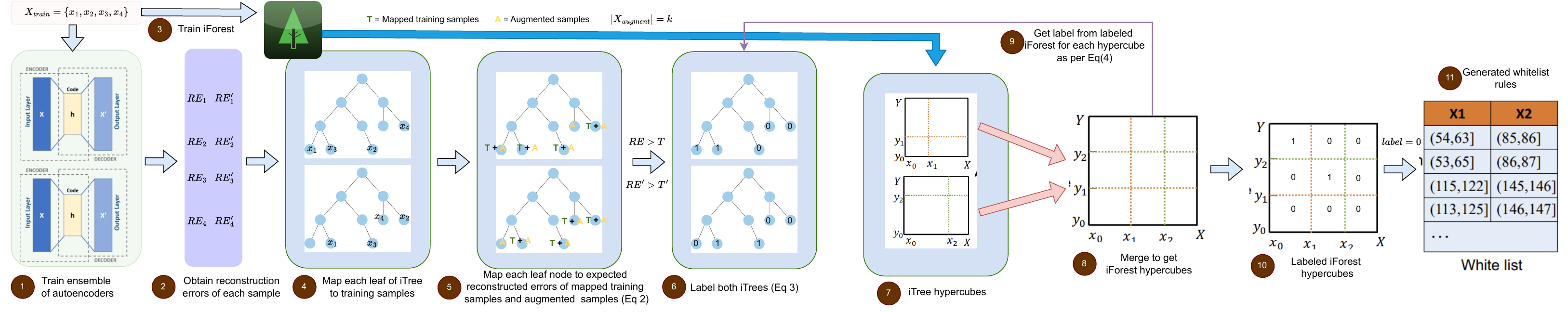}
    \caption{\textbf{Knowledge distillation of autoencoders into iForest and whitelist rules generation.} (1) We first train an ensemble of autoencoders and (2) collect reconstruction errors by feeding each training sample in the trained ensemble. (3) We then train iForest model itself. We then (4) map each leaf node of trained iForest's iTrees with the respective training samples. (5) Next, we embed each leaf node with expected reconstruction errors (over mapped training samples and augmented samples) according to Eq 2. (6) We convert combination reconstruction errors into a label (0 or 1) based on Eq 3. (7) Meanwhile we derive iTree hypercubes from trained iForest by taking the cartesian product of all possible feature boundaries for each iTree \cite{290987}. (8) We then merge iTree hypercubes into an iForest hypercubes. (9) We label each hypercube of iForest hypercubes by consulting labeled iForest as per Eq 4. (10) Once labeled, we merge adjacent hypercubes having same label. (11) Lastly we derive whitelist rules for hypercubes whose $label = 0$.}
    \vspace{-0.5cm}
    \label{fig:wl}
\end{figure*}
We first discuss the novel design of knowledge distilled iForest ML model. We perform knowledge distillation of an ensemble of autoencoders while training iForest model. Once model is prepared offline, we generate the whitelist rules. This procedure is shown below and the visual illustration is presented in \fref{fig:wl}.
\subsubsection{Knowledge distillation} \label{kd}
\myparab{Primer. }Due to the limited computation capacity and memory of
programmable switches, deploying sophisticated ML models in these
devices encounters great challenges. To this end, the idea
of learning a lightweight student model from a sophisticated
teacher model is formally popularized as knowledge distillation \cite{gou2021knowledge}. There have been many works on knowledge distillation \cite{hinton2015distilling,bai2022multinomial,frosst2017distilling,bai2019rectified,xie2022mousika,10158739} but almost all of them consider supervised ML models. Ours is the first work to distill knowledge from an ensemble of autoencoders into an iForest. We provide two distillation approaches out of which we adopt the one that is data plane friendly. 

\myparab{Training ensemble of autoencoders. }We first train an ensemble of $r$ autoencoders on $m$ features. We represent an ensemble of trained autoencoders as follows: $AE(\textbf{x}; \textbf{$\theta$}) = \{x_u' \in \mathcal{R}^m \}_{u=1}^r$. In other words, an autoencoder $AE_u$ is trained to reconstruct a benign sample $x_u \in \mathcal{R}^m$. $x_u' \in \mathcal{R}^m$ is the reconstructed sample. A trained autoencoder consists of two parts: an encoder that maps an input sample to an intermediate representation $E:x \rightarrow \mathcal{R}^d$, where $x \in \mathcal{R}^m$ is an input sample. The second part is a decoder that maps the intermediate representation of the encoder to the reconstructed sample $D:E(x) \rightarrow \mathcal{R}^m$, where $x' = D(E(x))$ and $x' \in \mathcal{R}^m$. For inferring traffic as malicious or benign, the autoencoder uses \textit{reconstruction error}. Reconstruction error for a test sample $x_{test}$ is given by,
\begin{equation}
    RE(x_{test}) = \sqrt{\frac{1}{m}\sum_{i=1}^m(x_{test,i}' - x_{test,i})^2}
\end{equation}
which is also RMSE error. A sample $x_{test}$ is malicious if $RE(x_{test}) > T$, where $T$ is RMSE threshold. Otherwise, it is benign (normal traffic). Knowledge distillation of an ensemble of autoencoders to iForest is tricky. One of the approach could be to train $r$ iForests on intermediate representations $\{E_u(X_{train})\}_{u=1}^r$ (where $E_u(X_{train}) \in \mathcal{R}^{N \times d}$) of $r$ autoencoders instead of training on benign samples $X_{train} \in \mathcal{R}^{N \times m}$ (where $N$ is the number of training samples). While this approach is sound, it is not data plane-friendly. This is because every time the data plane module extracts features, it has to feed those features to the trained encoder and $E(X)$ operation cannot be performed on data planes. Therefore, we come up with an alternate solution that carefully uses reconstruction errors $RE$s to distill knowledge from autoencoders into the iForest model. 

\myparab{iForest training and knowledge distillation. }Following is the procedure to train iForest and perform knowledge distillation. This is also demonstrated in steps 1-6 of \fref{fig:wl}.
\begin{enumerate}
    \item We train iForest model on the same set of features of training samples $X_{train}$ as the ensemble of autoencoders.
    \item Once the iForest model containing $t$ iTrees and sub-sampling size $\Psi$ is trained, we perform knowledge distillation as follows.  We traverse a sample $x \in X_{train}$ on each of the $t$ iTrees and reach $t$ respective leaf nodes. This way, we map $x$ to a leaf in every iTree. We repeat for all $x \in X_{train}$. Let the samples mapped to the leaf node be $X_{leaf} \subseteq X_{train}$.
    \item For each leaf of the iForest, we get a range of form $[a_i,b_i]_{i=1}^m$ for each of the $m$ features. Therefore, for each leaf node, we do uniform random sampling (without replacement) of $k$ samples (\textit{data augmentation factor}) from that features range. Let that be $A_{leaf}$ and $|A_{leaf}| = k$. We repeat for all the leaf nodes.
    \item We embed the reconstruction errors ($REs$) of the $r$ autoencoders  in each of the leaf node as follows.
    \begin{equation}
        RE_{leaf} = \bigl\{\frac{1}{k+|X_{leaf}|}\sum_{x \in X_{leaf} \cup A_{leaf}}RE_u(x)\bigr\}_{u=1}^r
    \end{equation}
    We repeat for all the leaf nodes. Thus, each of the leaf node contains $r$ values which are $r$ average reconstruction errors obtained using autoencoders. This is denoted by $\{RE_1, RE_2, ..., RE_r\}$.
    \item It is obvious that each leaf node corresponds to an inference rule. Thus we transform values on each leaf node as one-hot labels ($1$ for malicious and $0$ for benign) as $l = \{\mathbbm{1}\{RE_1>T_1\}, \mathbbm{1}\{RE_2>T_2\}, ..., \mathbbm{1}\{RE_r>T_r\}\}$, where $R_i$ is reconstruction error from autoencoder $AE_i$ and $T_i$ is RMSE threshold of $AE_i$. $\mathbbm{1}$ is the indicator function which activates when condition is true.
    \item We assign weights $\{w_u\}_{u=1}^r$ to each of the $r$ autoencoders, where $w_u \in [0,1]$ and $\sum_{u=1}^r w_u = 1$. For each leaf node (inference rule), label on the leaf node is obtained as,
        \begin{equation}
        label_{leaf} = \mathbbm{1}\{\sum_{i=1}^{r} w_i \times l_i > 0.5\}
    \end{equation}
   where $\mathbbm{1}$ is an indicator function. We repeat for all the leaves of iForest.
\end{enumerate}

\myparab{Distilled iForest inference. }Given a test sample $x$, we traverse each of the $t$ iTrees and end up at $t$ leaves (one in each iTree). We then retrieve labels from each leaf node and take the \textit{majority vote} over all the $t$ iTrees. That is, $label_{distilled} = $ \textit{majority\_vote(label\textsubscript{leaf} from $t$ leaves)}. Since original structure of iForest is retained, we also obtain label from iForest using anomaly score of test sample $x$ as $label_{IF} = \mathbbm{1}\{ score(x) < 0.5\}$. \textit{score(x)} is the anomaly score\footnote{anomaly score of a sample $x$ in original iForest is given by $score(x)=2^{-\frac{\mathbbm{E}(h(x))}{c(n)}}$, where $\mathbbm{E}(h(x))$ is the expected path length traversed by $x$ over all iTrees and $c(n)$ is the normalization factor based on samples in dataset $n$.} of $x$ obtained from original iForest using expected path length \cite{liu2008isolation}. Whether to prefer prediction from distilled iForest or original iForest depends on task-to-task basis. Let us say we have an arbitrary mapping function $g$, then final label of sample $x$ is given by
\begin{equation}
    label(x) = g(label_{IF}(x),\;label_{distilled}(x))
\end{equation}
In our experiments (\secref{exp}), we choose $label(x) = label_{IF}(x) \times label_{distilled}(x)$ to retain high TPR of original iForest while getting low FPR advantage from knowledge distillation.

\myparab{Rationale. }Embedding expected reconstruction errors of autoencoders into the leaves of trained iForest helps store the knowledge of autoencoders in the iForest without disturbing its structure. This helps preserve the original iForest as well. 

For more details and clarity on our knowledge distillation scheme, refer to the supplementary material. We briefly demonstrate the consistency of our knowledge distillation process while deferring the details to supplementary material. 

\myparab{Consistency of knowledge distillation. }We demonstrate the consistency and fidelity of our novel knowledge distillation algorithm using consistency metric $C$ defined as follows,
\begin{equation}
    C = \frac{1}{N} \sum_{i=1}^N \mathbbm{1}\{iForest_{distilled}(x_i) = Autoencoder(x_i)\}
\end{equation}
where $N$ is the number of samples in the validation/test set. We distill the knowledge of Magnifier \cite{290987} into the iForest. For this experiment, we do not consider anomaly score from the original iForest. In other words, the label of distilled iForest can be obtained for this experiment by simply traversing all $t$ iTrees for a sample $x$ and taking a majority vote of the labels obtained on all iTrees. 
\begin{figure}[hbt!]
    \centering
    \includegraphics[width=9cm]{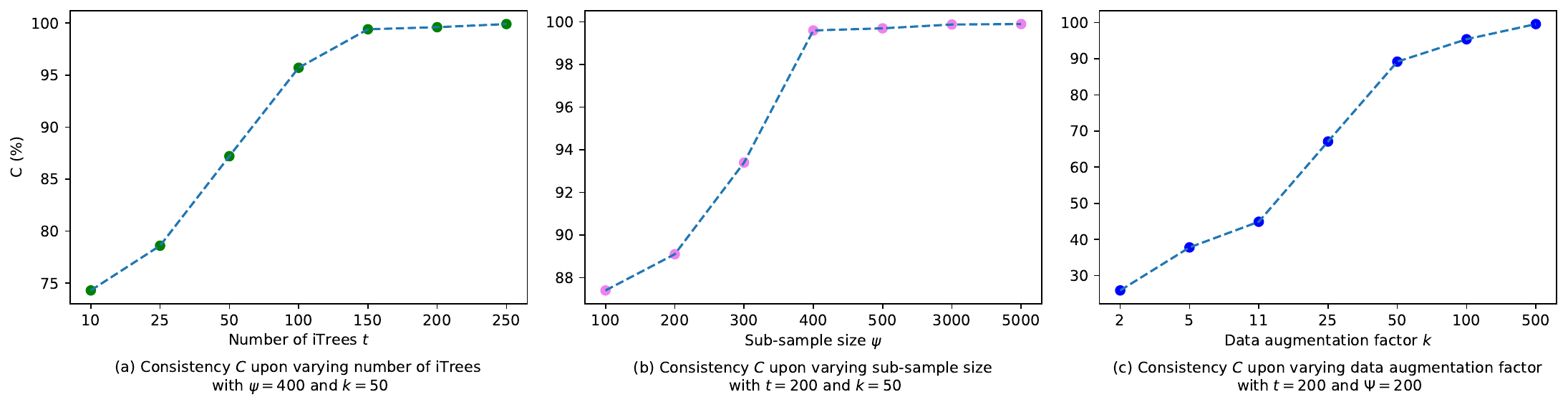}
    \caption{Effect of various hyperparameters on consistency of knowledge distillation algorithm.}
    \vspace{-0.5cm}
    \label{fig:consdis}
\end{figure}
We show the effect of various hyperparameters on the iForest knowledge distillation consistency in \fref{fig:consdis}. We see that beyond certain threshold of hyperparameters, our distilled iForest yields same performance as Magnifier (or autoencoder). 

\begin{figure}
  \begin{center}
    \includegraphics[width=0.48\textwidth]{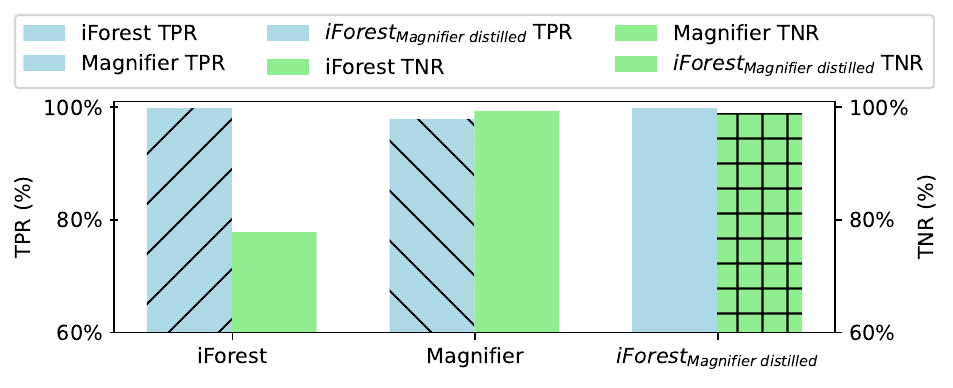}
  \end{center}
  \caption{TPR and TNR comparison of iForest, Magnifier \cite{290987} and Magnifier-distilled iForest. Distilled iForest retains high TPR of iForest as well as high TNR from Magnifier.}
  \vspace{-0.5cm}
  \label{fig:dif}
\end{figure}

\myparab{Results. }We compare trained knowledge distilled iForest to both iForest and autoencoder in terms of macro (averaged over all attacks mentioned in \secref{expimp}) TPR and TNR. For comparison, we take $t=200$ iTress and $\Psi = 5000$ sub-sample size. We take only a single autoencoder called Magnifier \cite{290987} and distill its knowledge into our iForest. We make two changes in the Magnifier: (i) add \textit{burst-flow mapping} (supplementary material) to figure out flow-level (FL) features from burst-level (BL) features (\secref{bfe}) for each of $5$ time windows as mentioned in \cite{290987}, and (ii) we do not apply model quantization \cite{wu2020integer} to the Magnifier unlike HorusEye \cite{290987}. As for the dataset, we divide benign dataset into $2:8$ for validation and training. We add one attack dataset at a time to the validation set. For validation set, attack traffic makes up $25\%$ of the traffic. BL (burst-level) and PL (packet-level) features used are mentioned in \secref{blplf}. We compare TPR and TNR of iForest, knowledge-distilled iForest using Magnifier (without model quantization), and Magnifier (also without model quantization) in \fref{fig:dif}. We see that distilled iForest is able to retain the best of both: high TPR from iForest and high TNR from Magnifier. 

Additionally, we also demonstrate this comparison for non-network-based datasets in the supplementary material. 

\subsubsection{iForest whitelist rules generation} \label{iwlrg}
To generate whitelist rules from trained distilled iForest, we first generate iForest hypercubes from trained and distilled iForest model. Then we label all the hypercubes and extract out the whitelist rules. The details follow below and visual illustration is done in steps 7-11 in \fref{fig:wl}.

\myparab{iForest hypercubes generation. }It has been observed that each iTree of an iForest essentially conducts a round of
hyper-dimension feature space dividing. That is, each iTree
divides the hyper-dimension feature space into hypercubes
(called iTree hypercubes) through binary tree branching \cite{290987}. Due to many iTrees, installing all the rules on the data plane is not feasible. Therefore, we adopt a variant of whitelist rules generation strategy from \cite{290987}. The strategy involves deriving hypercubes from each of the iTree called \textit{iTree hypercubes}. Each hypercube of iTree hypercubes defines a feature space boundary, and every sample belonging to that hypercube maps to the same label. We then merge all the iTree hypercubes into an \textit{iForest hypercubes} (similar to aggregating the space divisions, i.e.,
branches, of all iTrees). This merging maintains the consistency that \textit{all samples inside an iForest hypercube share the same label} (theorem 1 in supplementary material). 

The hypercubes of an iForest might be characterized by non-integer boundaries and we know that the data plane does not support floating-point
arithmetic. Therefore, in case BL and PL features involve only integral values, we can further “shift” these hypercubes slightly for integer boundaries by rounding down the
branches of each feature to their nearest integers. The "shifting" does not change which hypercube an integer data point falls into (theorem 2 in supplementary material).     

\myparab{Hypercubes labeling and rule generation. }Once we obtain (distilled) iForest hypercubes with feature boundaries, we need to map samples inside each hypercube with a common label. Without loss of generality, we take any random sample inside a hypercube (e.g., the right boundary of each feature) and feed it to the trained and distilled iForest to obtain the label (Eq (4)). We then assign the obtained label to the respective hypercube. We repeat the same for all hypercubes of iForest. Next, we merge adjacent hypercubes with the same assigned label. Finally, we return hypercubes having $label=0$ as whitelist rules.

For more details and clarity on our whitelist rules generation procedure, refer to the supplementary material. We also present theoretical results on our knowledge distillation algorithm and whitelist rules generation in the supplementary material.

\myparab{Consistency of whitelist rules. }To check the fidelity of the whitelist rules generated, we use consistency $C$ given by,
\begin{equation}
    C = \frac{1}{N}\sum_{i=1}^N \mathbbm{1}\{iForest_{distilled}(x_i) = R(x_i)\}
\end{equation}
where $R$ is the set of whitelist rules. On various parameters like number of iTrees $t=\{10,50,100,200\}$ at $\Psi=400$, and sub-sample size $\Psi=\{1000,2000,5000\}$ at $t=200$, we get $C>0.994$ demonstrating the effectiveness of whitelist rules.
\subsection{Burst feature extractor} \label{bfe}
To apply the whitelist rules derived above, it is essential to extract features from the burst of packets from the traffic. Such features are called burst-level (BL) features (used to train iForest). We define a burst as a long sequence of continuously sent packets in a flow, where the
inter-arrival time of a packet does not exceed a certain threshold $\delta_{idle}$. In other words, \{$p_1$, $p_2$, ..., $p_n$\} forms a burst iff (i) $p_i.time - p_{i-1}.time \leq \delta_{idle}$, where $i \geq 2$, and (ii) $p_i.5t = p_j.5t$ (meaning packet's 5-tuple), where $i \neq j$ and $1\leq i,j<n$. If the former condition is not met, idle timeout is said to occur. Next, we explain burst segmentation parameters, bi-hashing mechanism, collision handling mechanism, and BL+PL (packet-level) features usage during model training.
\subsubsection{Burst segmentation}
One issue is that bursts could be very long (meaning the number of packets $n$ could be very high or burst duration $p_n.time - p_1.time$ could be high). This leads to persistent storage \cite{trimananda2020packet,ma2020pinpointing} in switch memory which is not practical for Tbps switches. Therefore, to reduce long-term resource consumption of switches by keep-alive traffic, we set two thresholds on our bursts. First, we set a packet number segmentation threshold $N_{threshold}$ on the number of packets in a burst to achieve
low resource usage and high real-time malicious traffic detection. That is, for every burst, $n \leq N_{threshold}$. Second, we keep a threshold on burst duration called active timeout $\delta_{active}$. So for every burst, $p_n.time - p_1.time \leq \delta_{active}$. 
\subsubsection{Efficient hashing and collision handling} \label{bhdh}
To efficiently obtain bidirectional BL features and reduce storage index collisions we borrow the bi-hash algorithm and double hash table from \cite{290987}.

\myparab{Bi-hash. }To obtain bi-directional BL features efficiently, instead of conventional hashing which is $h(pkt.5t)$, we instead obtain bi-hash as,
$h_1(dstIP, dstPort, protocol) \oplus h_2(srcIP, srcPort, protocol)$, where $\oplus$ is XOR operator. It is observed that bi-hash accounts for the bi-directionality of bursts before obtaining BL features while also inducing the same number of collisions as the hashing of 5-tuple \cite{290987}.

\myparab{Double hash table. }To mitigate storage index hash collisions, we implement the double hash table algorithm. We divide one hash table into two hash tables\footnote{We can actually have two hash tables or have a single hash table with an imaginary line dividing two. Former involves two sets of storage registers while in latter, we need to resubmit/recirculate packet during during collision despite having only single set of storage.}. Thus, we have $3$ bi-hash functions, $H_1$ of $a_1$ bits for the first hash table, $H_2$ of $a_2$ bits for the second hash table, and $H$ of $a_1+a_2$ bits for the hash table if there was no division. We first perform bi-hash using $H(.)$. If the value conflicts (collision) with the first hash table, the hashing is done on the second hash table using bi-hash $H_2$.\eat{, we find whether the hash index lies in the first hash table or the second hash table. If the conflict is in the first hash table, then the storage values are reallocated to the second hash table using $H_2$ bi-hash. Otherwise, re-allocation is done using bi-hash $H_1$. Our method of utilizing a double hash table is different from \cite{290987} which uses two bi-hashes instead of three.} It has been observed that a double hash table mitigates hash collisions by up to 10 times compared to a single hash table \cite{290987}. Further, two equally sized hash tables ($a_1=a_2$) seem to perform well in most of the cases \cite{290987}.  
\subsubsection{BL and PL features} \label{blplf}
The BL features we consider are \textit{number of packets in a burst}, \textit{burst size}, \textit{average}, \textit{minimum}, \textit{maximum}, \textit{variance} and \textit{standard deviation of per-burst packet size and jitter} (inter-packet delay), and \textit{burst duration}. For PL features, we consider \textit{dstPort}, \textit{srcPort}, \textit{protocol}, \textit{packet's TTL}, and \textit{packet's length}. Note that  \textit{dstPort}, \textit{srcPort}, and \textit{protocol} can also be clubbed with BL features as they map to same 5-tuple. We next discuss the issues while extracting BL features.

\textbf{Preparing BL features can cause delay}. One of the issues that most of the existing works \cite{290987,Akem2023FlowrestPF} do not address is that early packets of a burst remain unaccounted or ignored \cite{akem2024jewel}. For instance, before BL features are reliable, early malicious packets of a burst may flood into the network and harm it. To address this issue, we train an iForest model (but without knowledge distillation) \textit{only on PL features of early packets}. Until BL features are prepared, we classify early packets with this separate iForest model trained only on PL features. On the other hand, the main iForest model (with knowledge distillation) that we have trained on BL features may also use PL features but only limited to \textit{dstPort} and \textit{protocol} (rest of PL features may vary from packet to packet). Alternatively, we can merge the whitelist rules from only PL-based iForest to our main (distilled) iForest that was trained on BL features. Then, we could use the same set of whitelist rules for both early packets of a burst and also when BL features are ready.

\textbf{Data plane does not support division. }One other issue is to extract BL features that involve division which is not natively supported in the data planes. In fact, for this we can borrow an approach from \cite{zhou2023efficient} which obtains average/variance/standard deviation at $2^i$ packets only. In such a case, we keep burst segmentation threshold $N_{threshold} = 2^i$. For instance if $N_{threshold} = 16$, then we obtain such BL features at $2$, $4$, $8$, and $16$ packets only. Example, for $16$ packets, average packet size is $(\textit{burst size)} >> 4$ using bitwise shift operations. There is a provision in data plane to obtain square and square root as well but only on first $4$ MSBs (limitation). An alternative to obtain such division based features (approximate value) at all packets involve careful use of log and exponent operations. For instance, $A/B \approx 2^{log_2A - log_2B}$. Both the approaches are covered in supplementary material.    
\subsection{Hardware implementation} \label{dpi}
\begin{figure*}[hbt!]
    \centering
    \includegraphics[width=18cm]{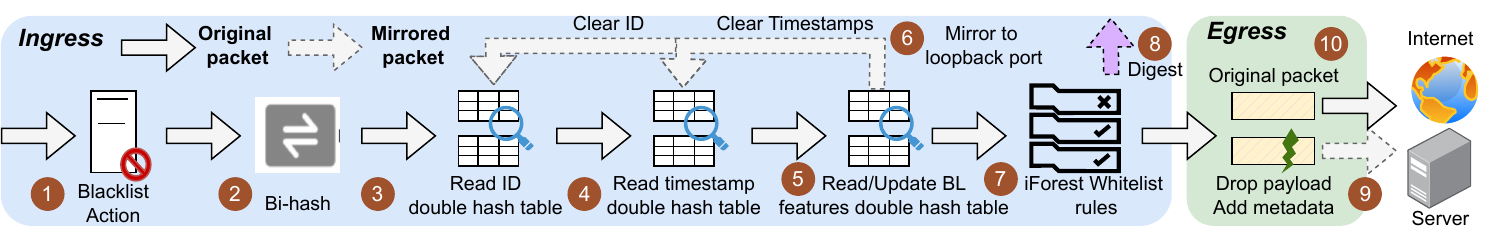}
    \caption{Data plane implementation of \arch at a glance}
    \label{fig:imp}
    \vspace{-0.5cm}
\end{figure*}
We implement \arch data plane module on programmable switches
using the P4 programming language. We \textit{intelligently} design the data plane by addressing one of the major pipeline constraints, that a storage register cannot be accessed twice. For instance, \textit{sequential calls} to first read and then update a register is not possible without resubmission. 

As a general example, suppose we wish to read a value from register $R_1$ and based on the read value, either update the values in registers $R_1$ and $R_2$, or reset the values in $R_1$ and $R_2$. One obvious approach (which is also followed by most recent work \cite{290987}) is to first read the value in $R_1$, perform packet resubmission after adding value of $R_1$ as metadata to that packet, and then use the resubmitted packet's metadata to update or reset values in $R_1$ and $R_2$. This is shown as P4 code snippet in listing \ref{lst:p4code1}. 
\begin{figure}[hbt!]
\begin{lstlisting}[caption={Approach followed by \cite{290987}}, label={lst:p4code1}]
action resubmit (bit<7> resub_port) {
    ig_tm_md.ucast_egress_port = resub_port;
}
apply{
    //If packet is resubmitted one
    if(ig_intr_md.ingress_port == RESUBMIT_PORT){
        //Based on metadata from R1
        if(meta.resub_value == x) { //Update R1, R2
            update_R1_action.execute(meta.index);
            update_R2_action.execute(meta.index);
        }
        else{ //Reset R1, R2
            reset_R1_action.execute(meta.index);
            reset_R2_action.execute(meta.index);
        }
    }
    else{ //normal packet: read R1, R2
    meta.val1 = read_R1_action.execute(meta.index);
    meta.val2 = read_R2_action.execute(meta.index);
    //Resubmit
    resubmit(RESUBMIT_PORT)
    }
}
\end{lstlisting}
\vspace{-0.4cm}
\end{figure}
Note that this approach involves resubmission for every incoming packet which can burden the switch pipeline and reduce processing throughput. 

As an alternative approach we leverage the fact that a register action can take at most two \textit{if} conditions. Therefore, in a single register action associated with $R_1$, we will perform read+update or read+reset atomically. In general, we intelligently structure the sequence of read/update/reset instructions atomically as shown in listing \ref{lst:p4code2} via P4 code. In the \texttt{read\_update\_reset\_R1\_action} $R_1$ action, we make update to $R_1$ value based on current value itself. This avoids packet resubmissions. 
\begin{figure}[hbt!]
\begin{lstlisting}[caption={Approach followed by \arch}, label={lst:p4code2}]
apply{
    //No packet resubmissions
    //Read value from R1 and atomically reset/update
    meta.v1 = read_update_reset_R1_action.execute(meta.index);
    if(meta.v1 == x) { //Based on read value from R1
            //Read and update R2
            meta.v2 = read_update_R2_action.execute(meta.index);
        }
    else { //Read and reset R2
            meta.v2 = read_reset_R2_action.execute(meta.index);
        }
}
\end{lstlisting}
\end{figure}
We follow this approach while implementing our data plane logic as shown below.

We show the overview of data plane implementation in \fref{fig:imp}. As shown, \circled{1} we first match the incoming packet's 5-tuple in the blacklist match action table. If there is a match, we can simply drop the packet or redirect it for further analysis. Otherwise, \circled{2} switch uses bi-hash algorithm to calculate the index
of the five-tuple and calculates the location of the data storage according to the double hash table algorithm\footnote{If instead of two separate hash tables, we consider only a single hash table logically divided into two, then we have to resubmit the packet during hash table conflict. We will assume that we actually have two separate hash tables to avoid packet resubmissions. } (\secref{bhdh}). \circled{3} Based on storage location, entry in flow ID storage is read and hash conflicts (collisions) are checked. \circled{4} Next, timestamp storage registers are read and timeout conditions are checked. Moreover, packet count storage register is also read and updated and a check is made whether packet count has reached a burst segmentation threshold. Based on the decision in previous step, \circled{5} BL features storage registers are either "updated", "read and reset" or "updated, read and reset" (operations not performed sequentially to avoid unnecessary packet resubmissions). If burst segmentation threshold is reached \circled{6} we also need to clear timestamp registers and flow ID register by simply mirroring the packet and sending it to loopback port (this avoids original packet delay). \circled{7} BL features obtained are matched with iForest whitelist rules. \circled{8} a message digest is sent to the control plane (informing whether the burst is normal or malicious) to alert the controller for malicious ID. Then, the controller can install appropriate blacklist rules. Finally, \circled{10} original packet can be forwarded to the internet while a \circled{9} mirrored packet containing BL features of a normal burst as metadata (match with whitelist rules) can be sent to the control plane for distilled iForest model update. 
\setlength{\textfloatsep}{0.5pt}%
\begin{algorithm}[t!]
\caption{Data plane logic of \arch} \label{algo:2}
\small
\textbf{Input: }A packet $pkt$ with 5-tuple $pkt.5t$
\begin{algorithmic}[1]
\State $pkt.ID \gets h(pkt.5t)$ \Comment{\dgc{Get original hash of 5-tuple}}
\State $H_1 \gets h_1(pkt.5t)$ \Comment{\dgc{Get bi-hash of first table}}
\State $S_1 \gets H_1[indexSize1:0]$ \Comment{\dgc{Get storage index of first table}}
\State $H_2 \gets h_2(pkt.5t)$ \Comment{\dgc{Get bi-hash of second table}}
\State $S_2 \gets H_2[indexSize2:0]$ \Comment{\dgc{Get storage index of second table}}
\State $S_{id} \gets S_1$ \Comment{\dgc{Assume storage index is of first table}}
\State \textit{TableSelected}$ \gets 0$ \Comment{\dgc{Initially no hash table selected}}
\If{not $pkt.isResubmit$} \Comment{\dgc{$pkt$ is not resubmitted}}
\If{$pkt.5t$ hit $T_{blacklist}$} \Comment{\dgc{Blacklist match}}
\State Drop $pkt$ or fwd to desired port  
\Else
\State \textit{ID}\textsubscript{\textit{stored}} $\gets$ ReadReg1\textsubscript{$ID$}($S_{1}$) \Comment{\dgc{Read ID from first table}}
\If{\textit{ID}\textsubscript{\textit{stored}} $\neq 0$ \textit{and} $pkt.ID$ $\neq$ \textit{ID}\textsubscript{\textit{stored}}} \Statex \Comment{\dgc{Collision in first table}}
\State \textit{ID}\textsubscript{\textit{stored}} $\gets$ ReadReg2\textsubscript{$ID$}($S_{2}$) 
\Statex \Comment{\dgc{Read ID from second table}}
\If{\textit{ID}\textsubscript{\textit{stored}} $= 0$ \textit{or} $pkt.ID$ $=$ \textit{ID}\textsubscript{\textit{stored}}} 
\Statex \Comment{\dgc{No collision in 2nd table}}
\State \textit{TableSelected} $\gets 2$
\State $S_{id} \gets S_2$
\EndIf
\Else \Comment{\dgc{No collision in first table}}
\State \textit{TableSelected} $\gets 1$
\EndIf
\If{\textit{TableSelected} $\neq 0$} \Comment{\dgc{Choose Reg1 / Reg2}}
\State \textit{FirstTime\textsubscript{stored}} $\gets$ ReadUpdateReg\textsubscript{\textit{firstTime}}($S_{id}$)
\State \textit{LastTime\textsubscript{stored}} $\gets$ ReadUpdateReg\textsubscript{\textit{lastTime}}($S_{id}$)
\State $\Delta t_1 \gets$ \textit{LastTime\textsubscript{stored}} - \textit{FirstTime\textsubscript{stored}}
\State $\Delta t_2 \gets pkt.time-$ \textit{LastTime\textsubscript{stored}}
\State \textit{timeout} $\gets \Delta t_1 > \delta_{active}$ or $\Delta t_2 > \delta_{idle}$
\If{\textit{timeout}} \Comment{\dgc{Update packet count storage}}
\State \textit{Count\textsubscript{stored}} $\gets$ ReadResetReg\textsubscript{\textit{count}}($S_{id}$, $1$)
\Else
\State \textit{Count\textsubscript{stored}} $\gets$ UpdateReadResetReg\textsubscript{\textit{count}}($S_{id}$)
\EndIf
\If{\textit{timeout}} \Comment{\dgc{Update BL features storage}}
\State \textit{BL\textsubscript{stored}}$\gets$ ReadResetReg\textsubscript{\textit{BL}}($S_{id}$, $pkt.features$) 
\ElsIf{\textit{Count\textsubscript{stored}} $<$ \textit{N\textsubscript{threshold}}}
\State UpdateReg\textsubscript{\textit{BL}}($S_{id}$)
\Else
\State \textit{BL\textsubscript{stored}}$\gets$ UpdateReadResetReg\textsubscript{\textit{BL}}($S_{id}$,$0$)
\Statex \Comment{\dgc{Mirror to loopback port to clear ID and timestamp registers}}
\State MirrorResubmit($S_{id}$, \textit{TableSelected}) 
\EndIf
\If{\textit{timeout} or \textit{Count}\textsubscript{\textit{stored}} $=$ \textit{N\textsubscript{threshold}}}
\If{\textit{BL}\textsubscript{\textit{stored}} not hit $BL_{whitelist}$}
\State SendDigest($pkt.5t$, \textit{malicious})
\Else
\State SendDigest($pkt.5t$, \textit{normal})
\Statex \Comment{\dgc{Mirror BL features of normal traffic for iForest update}}
\State MirrorToCPU($pkt.5t$, \textit{BL\textsubscript{\textit{stored}}}) 
\EndIf
\EndIf
\EndIf
\If{$pkt.features$ not hit $PL_{whitelist}$}
\State Drop $pkt$ or fwd to desired port
\EndIf
\EndIf
\Else \Comment{\dgc{Resub pkt. Choose Reg1 / Reg 2 as per \textit{TableSelected}}}
\If{\textit{TableSelected} = $1$}
\State InitReg1\textsubscript{$ID$}($S_{id}$, $0$)
\State InitReg1\textsubscript{$firstTime$}($S_{id}$, $0$)
\State InitReg1\textsubscript{$lastTime$}($S_{id}$, $0$)
\Else
\State InitReg2\textsubscript{$ID$}($S_{id}$, $0$)
\State InitReg2\textsubscript{$firstTime$}($S_{id}$, $0$)
\State InitReg2\textsubscript{$lastTime$}($S_{id}$, $0$)
\EndIf
\EndIf
\State User defined $pkt$ processing
\end{algorithmic}
\end{algorithm}

The details are shown in Alg. \ref{algo:2}. In lines 1-7, we obtain bi-hashes of first and second hash table. Line 8 checks if the packet has been mirrored to the loopback port and if not, lines 9-10 matches $pkt.5t$ to blacklist rules. If there is a match, the packet is malicious and we need not proceed further. Otherwise, we check for hash conflicts (lines 11-19). If the conflict lies in both the tables, we have no option but to use whitelist rules for iForest model trained on only PL features (lines 43-44). Otherwise, we use appropriate hash table where there is no hash conflict. In lines 21-25, we detect if there is a timeout. Appropriate register action is performed on per-burst packet count register (lines 26-29). Based on timeout condition, and whether the per-burst packet count has reached segmentation threshold, BL feature registers are updated/read/reset (lines 30-36). If burst segmentation threshold \textit{N\textsubscript{threshold}} is reached, we mirror the packet to the loopback port to clear ID and timestamp registers (line 36, lines 45-53). Once BL features are obtained, if there is timeout or burst segmentation threshold is reached, we match them (BL features) to whitelist rules obtained from distilled iForest trained on BL features. In any case, we send a digest to the control plane to alert the controller in case of high number of bursts of malicious IDs. We also mirror the normal traffic burst to the CPU for distilled iForest update (lines 37-42). Otherwise, we match early packets of a burst using whitelist rules from only PL features (lines 43-44). Lastly, we decide what to do with the original packet based on user-defined processing (line 54).   

For more details on data plane implementation like BL feature extraction, refer to supplementary material. 
\section{Control plane module} \label{cm}
We discuss two components of control plane module. 
\subsection{Profiler} \label{plr}
The profiler (\fref{fig:overview}) is an automated tool that derives configuration for \arch for maximum detection accuracy and minimum data plane memory footprint. The configuration is selected to maximize reward given by,
\begin{equation}
    reward = \frac{\alpha}{3}(TPR + TNR + PR_{AUC}) + (1-\alpha)(1-\rho)
\end{equation}
where $\rho$ is a measure of memory footprint of the system, expressed as a fraction of
the total available resources in the target switch. We put $\alpha = 0.5$ to balance out the two factors in our experiments. Detailed procedure similar to \cite{akem2024jewel}.
\subsubsection{Hyperparameter selection}
Hyperparameters for offline distilled iForest model preparation are (i) number of iTrees $t$, (ii) sub-sample size $\Psi$, (iii) burst segmentation threshold $N_{threshold}$, (iv) idle and active timeouts ($\delta_{idle}, \delta_{active}$), (v) thresholds $Ts$ of ensemble of autoencoders used during distillation and, (vi) data augmentation factor $k$ which is number of samples obtained using uniform random sampling while distilling iForest. Important PL+BL features selection is not possible in unsupervised learning and therefore, we need to obtain important ones using hit and trial. 
\subsubsection{Hardware optimization}
Factors affecting data plane memory utilization include (i) storage index sizes of two hash tables and, (ii) range limit and ternary limit that denotes limit of the keys which are used for range matching and ternary matching respectively.

Each configuration of hyperparameters and hardware parameters is denoted by [\textbf{hp}, \textbf{hw}]. An optimal one is found based on maximum reward as stated previously.
\subsection{Data plane management} \label{dpm}
\begin{figure}
    \centering
    \includegraphics[width=8.5cm]{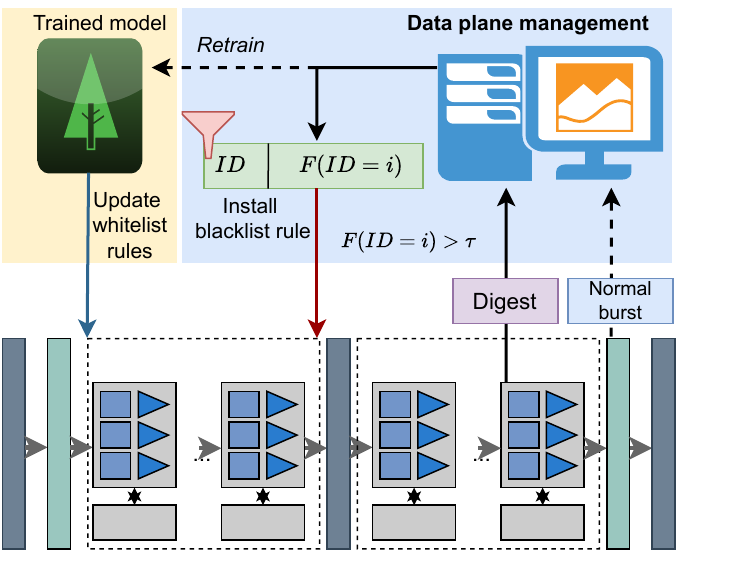}
    \caption{Data plane management}
    \label{fig:x}
\end{figure}
This component (\fref{fig:overview}) is responsible for following tasks (also shown in \fref{fig:x}).
\subsubsection{Installing and updating blacklist rules} \label{iubr}
Once we receive a message digest from the data plane upon applying whitelist match action rules, the flow’s abnormal frequency is calculated as,
\begin{equation}
    F(ID=i) = \frac{1}{N_i}\sum_{j=1}^{N_i} \mathbbm{1}\{y_j = malicious\}
\end{equation}
where $N_i$ is the number of bursts in a flow of ID $i$. We \textit{install a blacklist rule in data plane} corresponding to flow of ID $i$ if flow's abnormal frequency $F(ID=i) > \tau$ and $N_i > 1$, where $\tau$ is any flow's abnormal frequency threshold and $\tau \in [0,1]$.

Similarly, we can delete a blacklist rule (in data plane) if the blacklist rule entry in the data plane is old as per LRU or FIFO policy \cite{zhou2023efficient}. 

We experimentally analyze the effect of flow's abnormality threshold on \texttt{CyberSentinel}'s anomaly detection performance in supplementary material.
\subsubsection{Updating distilled iForest model online}
The BL features of normal traffic are mirrored to the control plane. Once we receive such payload-truncated packets with additional metadata containing BL features, we use that BL feature data to retrain knowledge distilled iForest model and update the whitelist rules in the data plane accordingly.
\section{Comparision with recent work}\label{rwc}
\begin{table}
    \centering
    \scalebox{1.05}{
    \begin{tabular}{|c||c|c|c|c|}
    \hline
       \textbf{System}  & \textbf{No Label} & \textbf{Data Plane} & \textbf{Line Speed} & \textbf{Accuracy} \\ \hline 
       \makecell{NetBeacon \cite{zhou2023efficient}, \\Jewel \cite{akem2024jewel}, Leo \cite{jafri2024leo}} & \rc{\textbf{\xmark}} & \dgc{\textbf{\cmark}} & \dgc{\textbf{\cmark}} & \tikz\draw[red!70!black,fill=red!70!black] (0,0) circle (.5ex); \\ \hline
       Kitsune \cite{mirsky2018kitsune} & \dgc{\textbf{\cmark}} & \rc{\textbf{\xmark}} & \rc{\textbf{\xmark}} & \tikz\draw[yellow,fill=yellow] (0,0) circle (.5ex); \\ \hline
       HorusEye \cite{290987} & \dgc{\textbf{\cmark}} & \dgc{\textbf{\cmark}} & \rc{\textbf{\xmark}} & \tikz\draw[green!45!black,fill=green!45!black] (0,0) circle (.5ex); \\ \hline
       \textbf{\arch} & \dgc{\textbf{\cmark}} & \dgc{\textbf{\cmark}} & \dgc{\textbf{\cmark}} & \tikz\draw[green!45!black,fill=green!45!black] (0,0) circle (.5ex); \\ \hline 
    \end{tabular}}
    \caption{Comparison of \arch with recent works. \protect\tikz\protect\draw[red!70!black,fill=red!70!black] (0,0) circle (.5ex);Low  \protect\tikz\protect\draw[yellow,fill=yellow] (0,0) circle (.5ex);Medium \protect\tikz\protect\draw[green!45!black,fill=green!45!black] (0,0) circle (.5ex);High}
    \vspace{-0.2cm}
    \label{tab:comp}
\end{table}
We compare \arch with the following recent works as mentioned in \tref{tab:comp}.

NetBeacon \cite{zhou2023efficient}, Jewel \cite{akem2024jewel} and Leo \cite{jafri2024leo} classify traffic using packet-level and flow-level features entirely in the switch data plane. Therefore, accurate classification happens at line speed, however, the decision-tree based models they deploy in the data planes are supervised and therefore labels (or knowledge) of attack datasets are used to train ML models. Therefore these works \cite{zhou2023efficient,akem2024jewel,jafri2024leo} struggle to detect unseen attacks accurately (\secref{eted}).

Kitsune \cite{mirsky2018kitsune} is malicious traffic detection system that uses state-of-the-art autoencoders to detect malicious traffic and hence, does not need labels during training. However, it is not deployed in switch data planes and therefore can certainly not detect anomalies at line speed.

HorusEye \cite{290987} deploys an unsupervised model iForest in the switch data plane\footnote{One can argue that since first stage of malicious traffic detection happens in the data plane, it is after all line speed. However, the data plane module alone is not sufficient for accurate anomaly detection and is a weaker part of two stage detection in HorusEye.} (Gulliver Tunnel) and a state-of-the-art autoencoder called Magnifier in the control plane. However, Gulliver Tunnel alone is not sufficient for both high TPR and high TNR (\cite{290987}, \secref{gtc}). In fact, Gulliver Tunnel has to take the support of Magnifier for high TNR (low FPR). Since Magnifier operates in control plane, malicious traffic detection is certainly not happening at line speed. 

\arch deploys a knowledge distilled iForest into switch data plane. The knowledge of ensemble of autoencoders gets transferred to iForest during model training. This ensures that \arch is able to accurately detect malicious traffic (high TPR and high TNR) at line speed. In fact, unseen attack detection of \arch is similar to HorusEye \cite{290987} which in turn is better than Kitsune \cite{mirsky2018kitsune}. 
\pgfplotsset{every tick label/.append style={font=\tiny}, label style={font=\tiny}}
\section{~~~Evaluation} \label{exp}
\subsection{Implementation} \label{expimp}
We implement \arch data plane using P$4_{16}$ language and deployed it on Edgecore 32X switch with Tofino 1 ASIC and forwarding rate of $6.4$ Tbps. Control plane is implemented in python3 on 40-core, 2 x Intel(R) Xeon(R) Silver 4316 CPU @ 2.30GHz, and 256GB DDR4 memory, and RTX 3060 is used to perform model training and knowledge distillation. We deploy the system on a 40 Gbps link and the same traffic rate is controlled through tcpreplay.

\myparab{Knowledge distillation. }We distill knowledge (\secref{kd}) from Magnifier \cite{290987} into iForest. We set the label for a test sample to be malicious only when both distilled iForest and original iForest predict a malicious traffic. This is done to retain high TPR of iForest while performing knowledge distillation. Further, we choose the data augmentation factor $k=50$ for embedding reconstruction errors to all leaves. As for Magnifier, (i) we adjust the number of nodes in the input and output layers of Magnifier since the number of features in Magnifier is limited by iForest (if we perform knowledge distillation) and, (ii) we add \textit{burst-flow mapping module} to figure out FL features for $5$ time windows from input BL features. The rest of the hyperparameters of Magnifier are similar to \cite{290987}. For details, refer to the supplementary material.

\myparab{Other hyperparameters. }We consider active timeout $\delta_{active}=15s$. We also keep idle timeout $\delta_{idle} = 1s$ \cite{zhang2018homonit}, and burst segmentation threshold $N_{threshold}=15$ based on pdf of burst segmentation lengths (number of packets per burst) \cite{290987}.

\myparab{Hardware parameters. }We keep storage indexes of both hash tables as $16$ bits. 

\myparab{PL and BL features. }For all the experiments, we consider \textit{Per-burst number of packets}, \textit{burst size}, \textit{burst duration}, average and standard deviations of packet sizes and inter-packet delays per burst. \textit{DstPort} is used as PL/BL feature. This makes up total of $8$ features. For hardware performance experiments and Gulliver Tunnel comparison (\secref{gtc} and \secref{hpe}) we use only \textit{Per-burst number of packets} and \textit{burst size} as BL features. \textit{DstPort} is used as PL/BL feature. 

\myparab{Model training. }Distilled iForest model is trained on BL features as mentioned, and its whitelist rules are merged with another iForest (without distillation) trained only on \textit{DstPort}. This allows us to use same set of whitelist rules for matching PL features of early packets as well as BL features of a burst.  

\myparab{Profiler setting. }Profiler is only used in \secref{eted}. For the rest of the experiments, configurations stated in this section are carried over. Profiler still has no control over feature selection.

\myparab{Normal dataset. }We use benign traffic PCAP traces provided in \cite{290987,sivanathan2018classifying}. For \cite{290987}, we use the first 4 days of IoT normal traffic as the training set and the following 2 days as the test set. For \cite{sivanathan2018classifying}, we randomly select five days of data as the training set and the day after each of them as
the test set. The training and test sets do not overlap.

\myparab{Attack dataset. }For test/validation set, we mix normal data with attack data. We use public attack datasets \cite{mirsky2018kitsune,bezerra2018providing,koroniotis2019towards,art6}. The collective attack datasets predominantly encompass four distinct categories of anomalous activities: botnet infections, which encompass notorious instances such as Aidra, Bashlite \cite{bezerra2018providing}, and Mirai. Then, data exfiltration methods are included, comprising activities such as keylogging and data theft \cite{koroniotis2019towards}. Next, scanning attacks are part of the dataset, encompassing both service and operating system scans. Lastly, the datasets include distributed denial of service (DDoS) attacks, covering variations such as HTTP, TCP, and UDP DDoS attacks. We also consider UNSW NB15 dataset\cite{moustafa2015unsw} (used only in \secref{dpod}).

\myparab{Metrics. }We consider $TPR = \frac{TP}{TP+FN}$, $TNR = \frac{TN}{TN+FP}$, and $PR_{AUC}$ which is the area under precision-recall curve. $PR_{AUC}$ is very useful for performance evaluation on imbalanced datasets. Note that $FPR$ and $FNR$  are simply complements of $TNR$ and $TPR$. We use packet-level metrics, meaning a malicious burst means all packets of the burst are malicious. Moreover, a packet under collision is deemed malicious or benign based on PL features. 
\subsection{Comparison with Gulliver Tunnel} \label{gtc}
\begin{figure*}[t!]
\captionsetup[sub]{font=tiny,labelfont={bf,sf},textfont=}
     \centering
     \begin{subfigure}[b]{0.3\textwidth}
         \centering
         \includegraphics[width=5.8cm]{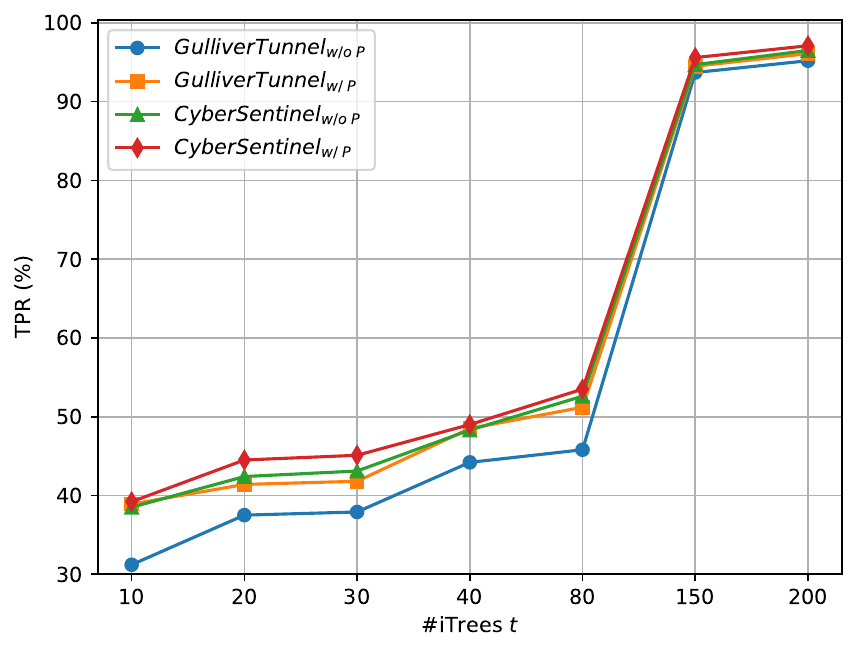}
         \caption{TPR (\%). We keep $\Psi=200$ and contamination ratio as $0.18$.}
         \label{fig:p1}
     \end{subfigure}
     \begin{subfigure}[b]{0.3\textwidth}
         \centering
         \includegraphics[width=5.8cm]{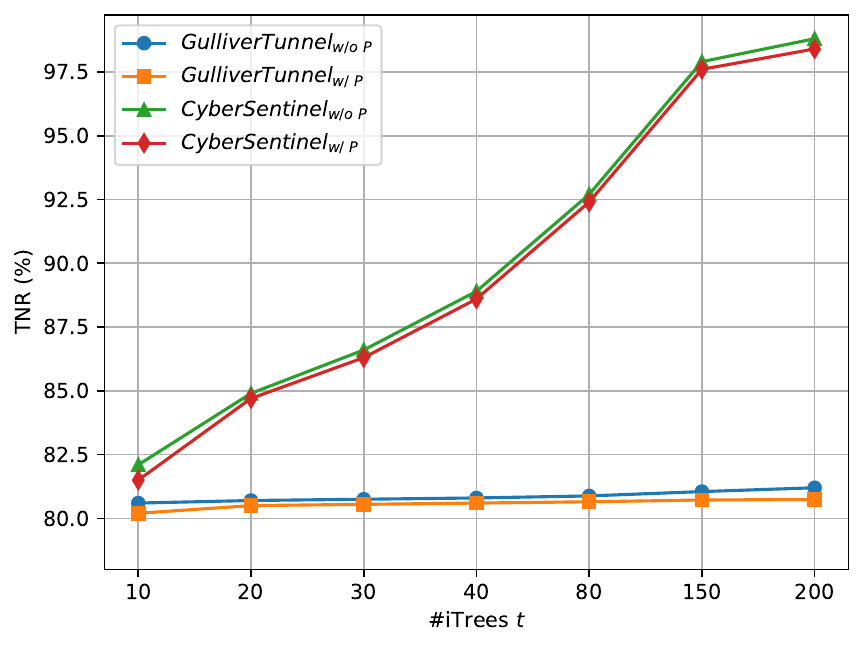}
         \caption{TNR (\%). We keep $\Psi=200$ and contamination ratio as $0.18$.}
         \label{fig:p2}
     \end{subfigure}
     \begin{subfigure}[b]{0.3\textwidth}
         \centering
         \includegraphics[width=5.8cm]{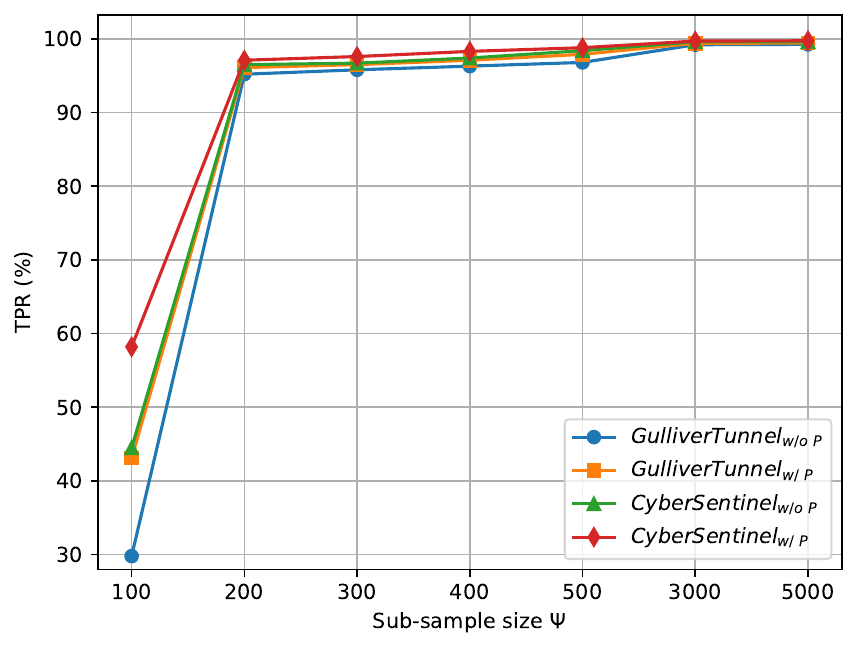}
         \caption{TPR (\%). We keep $t=200$ and contamination ratio as $0.18$.}
         \label{fig:p3}
     \end{subfigure}
     \begin{subfigure}[b]{0.3\textwidth}
         \centering
         \includegraphics[width=5.8cm]{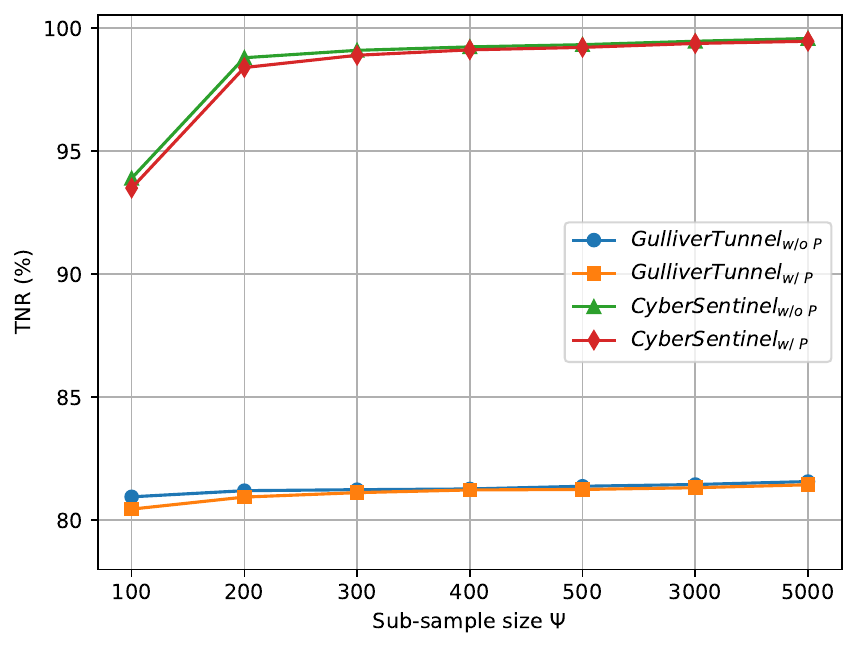}
         \caption{TNR (\%). We keep $t=200$ and contamination ratio as $0.18$.}
         \label{fig:p4}
     \end{subfigure}
     \begin{subfigure}[b]{0.3\textwidth}
         \centering
         \includegraphics[width=5.8cm]{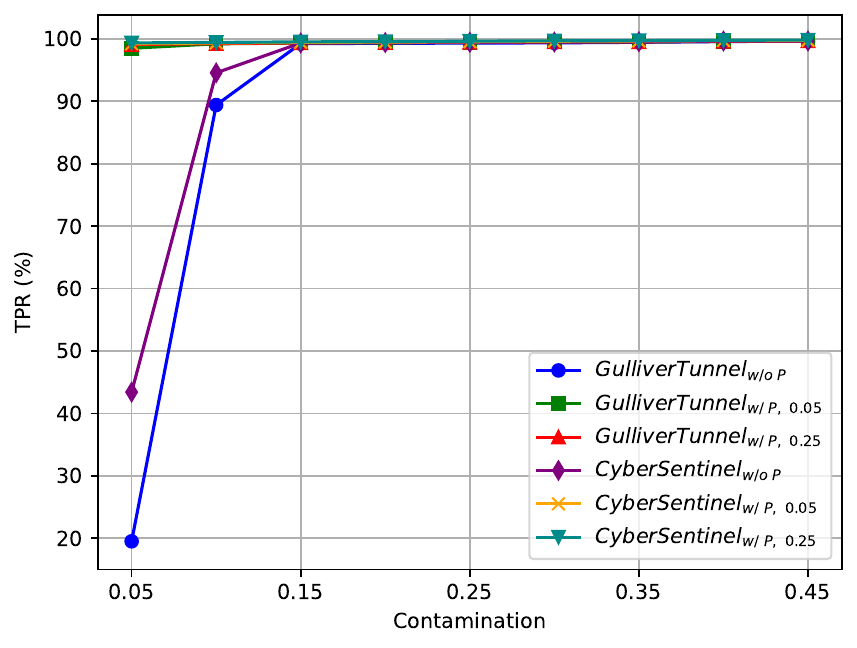}
         \caption{TPR (\%). We keep $t=200$ and $\Psi=5000$.}
         \label{fig:p5}
     \end{subfigure}
     \begin{subfigure}[b]{0.3\textwidth}
         \centering
         \includegraphics[width=5.8cm]{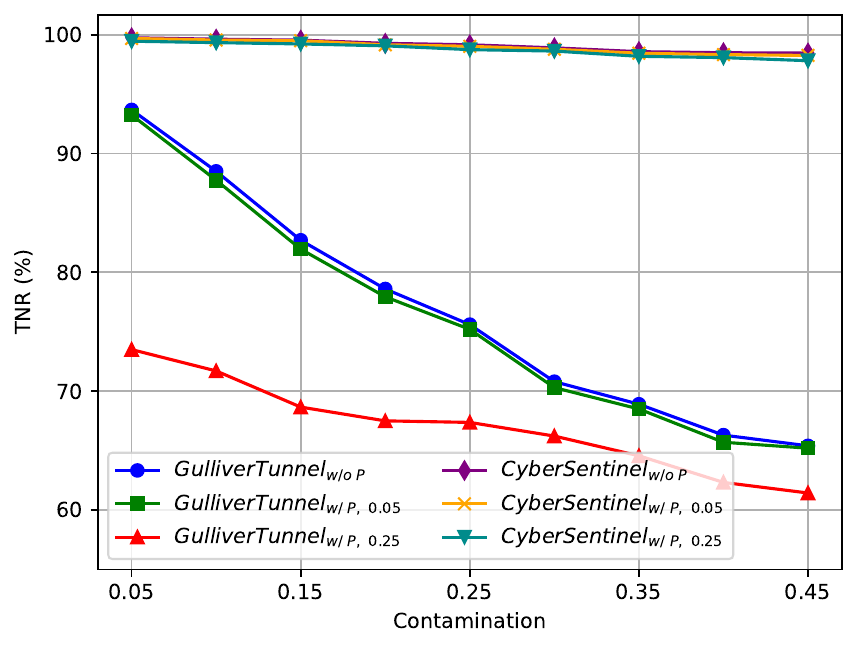}
         \caption{TNR (\%). We keep $t=200$ and $\Psi=5000$.}
         \label{fig:p6}
     \end{subfigure}
     \caption{Comparison of detection performance (TPR and TNR) of \arch with Gulliver Tunnel on various hyperparameters combinations.}
     \vspace{-0.5cm}
     \label{fig:gp}
 \end{figure*}
We compare detection performance (TPR and TNR averaged over all attacks) of \arch with only Gulliver Tunnel of HorusEye \cite{290987}, while also varying various hyperparameters like number of iTrees $t$, sub-sample size $\Psi$, and contamination ratio (ratio of validation samples being malicious). Note that w/o $P$ denotes we do not use \textit{dstPort} as a BL feature, while w/ $P$ means that we do. Here w/ $P$ means a separate iForest is trained only on \textit{dstPort} and its port-based whitelist rules are clubbed with distilled iForest whitelist rules trained on BL features. 

For this task, we divide benign training set into validation set and training set with a ratio of 2:8 and only use one attack dataset at a time which we add to validation/test set. 

\textbf{Number of iTrees. }We compare TPR and TNR of \arch with Gulliver Tunnel by varying number of iTrees $t$ as shown in \fref{fig:p1}-\ref{fig:p2} by keeping sub-sampling size $\Psi=200$ and contamination ratios of both with and without ports as $0.18$. As shown in \fref{fig:p1}, TPR of both Gulliver Tunnel and \arch is very low for $t \leq 80$. This means, to learn patterns of normal traffic properly, we need many iTrees. Fortunately, our whitelist rule generation technique can easily deploy rules of $200$ iTrees in the data plane. TPR of Gulliver Tunnel and \arch is very close because high TPR capability comes from iForest. Moreover, in \fref{fig:p2}, it is shown that TNR of \arch exceeds Gulliver Tunnel by $15\%$ because low FPR capability comes from autoencoder Magnifer which is distilled into our iForest.

\textbf{Sub-sampling size. }We next vary sub-sampling size $\Psi$ as shown in \fref{fig:p3}-\ref{fig:p4} by keeping number of iTrees $t=200$ and contamination ratios of both with and without ports as $0.18$.  In \fref{fig:p3}, TPR of both Gulliver Tunnel and \arch increases up to $\Psi=3000$ after which the increase in minimal. Moreover, TPR of both systems is almost same because high TPR is attributed to iForest. Moreover, in \fref{fig:p4}, we can see high TNR (by about $18\%$) of \arch compared to Gulliver Tunnel is due to Magnifier being distilled in iForest.   

\textbf{Contamination ratio. }We vary contamination ratio as shown in \fref{fig:p5}-\ref{fig:p6} by keeping number of iTrees $t=200$ and $\Psi=5000$. In \fref{fig:p5}, we notice that increasing contamination increases recalls (TPR) for both Gulliver Tunnel and \texttt{CyberSentinel}. With contamination of $0.05$, there are very less TPs and the whitelist rules (without \textit{dstPort} rules) produced will have high FNs, therefore low recalls. This issue is solved upon using \textit{dstPort} based rules. As usual, \arch has almost similar recall to Gulliver Tunnel. As shown in \fref{fig:p6}, increasing contamination reduces TNR drastically for Gulliver Tunnel. This is because many anomalies use same port as normal traffic and therefore, port-based rules are weak to correctly detect normal traffic. Moreover, increasing port contamination makes TNR worse because then port-based whitelist rules induce even more FPs. As we can see contamination does not affect \arch much because Magnifier that is distilled produces very robust BL features based whitelist rules that even the injected port-based rules do not have much effect on. As explained, \arch yields better TNR.
\begin{figure}[hbt!]
    \centering
    \includegraphics[width=8cm]{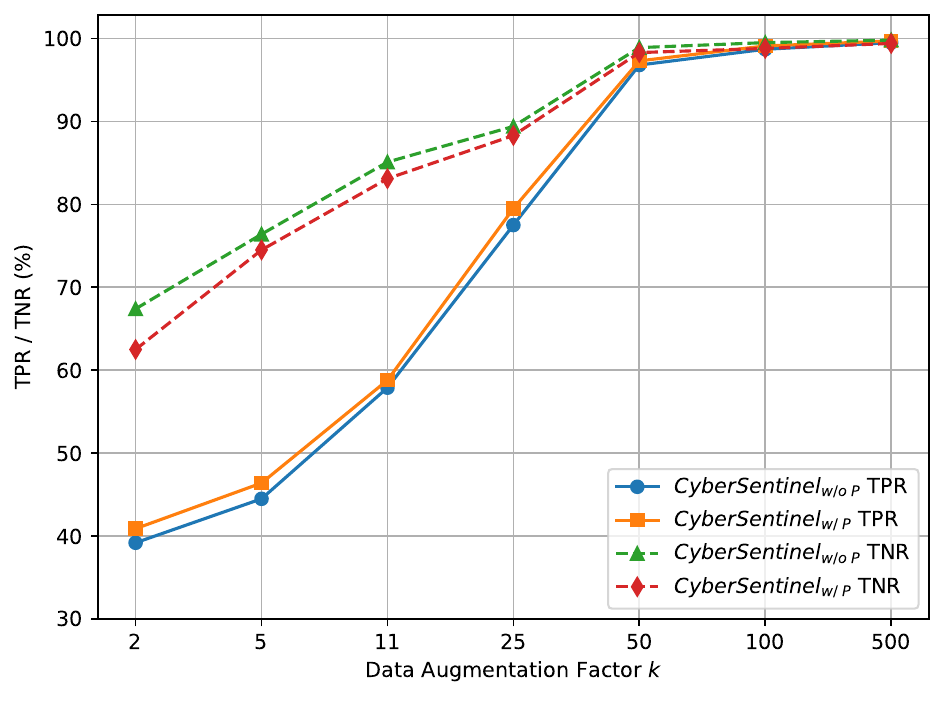}
    \caption{Effect of data augmentation factor $k$ on \texttt{CyberSentinel}'s TPR and TNR. Number of iTrees = $200$, subsample size = $400$, and contamination ratio of $0.18$ on both w / port and w/o port.}
    \vspace{-0.4cm}
    \label{fig:k}
\end{figure}

\textbf{Data augmentation factor. }We show the effect of data augmentation factor $k$ on TPR and TNR of \arch. The contamination factor is $0.18$ for both w/ port and w/o port variants of \arch. The results are shown in \fref{fig:k} at $t=200$ iTrees and $\Psi = 400$ subsample size. For both the \arch variants, TPR / TNR increases drastically up to $k=50$ (increase of $57\%$) after which the increase is not much (increase of $3.5\%$).  This is because as per the knowledge distillation procedure in \secref{kd}, augmenting more samples per leaf node to decide expected reconstruction error helps in better representing actual autoencoder performance in the trained iForest model.

\subsection{Hardware performance} \label{hpe}
\begin{table}
    \centering
    \scalebox{1.15}{
    \begin{tabular}{|c||c|c|c|}
    \hline
       \textbf{Packet size (B)}  & \textbf{Per-packet latency} & \textbf{\makecell{Packet processing\\ throughput}} \\ \hline 
       256 B & 1027 ns / 510 ns & 31.12 Gbps / 39.64 Gbps  \\ \hline
       512 B & 1033 ns / 515 ns & 31.07 Gbps / 39.63 Gbps  \\ \hline
       1024 B & 1052 ns / 525 ns & 31.04 Gbps / 39.62 Gbps  \\ \hline
       1500 B & 1074 ns / 538 ns & 30.98 Gbps / 39.59 Gbps  \\ \hline 
       IMIX & 1142 ns / 576 ns & 30.91 Gbps / 39.51 Gbps  \\ \hline
       \textbf{Average} & 1065.6 ns / 532.8 ns & 31.02 Gbps / 39.59 Gbps  \\ \hline
    \end{tabular}}
    \caption{Packet processing throughput and per-packet latency comparison of Gulliver Tunnel / \arch. }
    \label{tab:lat}
\end{table}
We evaluate the performance of data plane module of \arch in terms of (i) memory consumption on switch in terms of SRAM, TCAM, and stateful ALUs, (ii) throughput after loading P4 program, and (iii) per-packet latency meaning time taken to process a packet by P4 program. 

\textbf{Switch memory overheads. }\arch takes up $9$ stages of the pipeline with TCAM utilization of $2.85\%$, SRAM utilization of $9.86\%$, and sALU's utilization of $7.57\%$. This is similar to Gulliver Tunnel component of HorusEye that takes $9$ stages, $2.78\%$ TCAM, $9.90\%$ SRAM, and $7.62\%$ sALUs. Overall, resource utilization of \arch is low due to BL features extraction, double hash table, and bi-hash algorithm.  

\textbf{Throughput and latency. }Packet-processing throughput and per-packet latency of \arch is evaluated and compared to Gulliver Tunnel on a $40$ Gbps link. We show the comparison in \tref{tab:lat}. We use SPIRENT N11U traffic generator for high speed traffic simulations and experiment with various packet sizes (IMIX means internet mix). As we can see, \arch has about $50\%$ lower per-packet latency on average while yielding $27.6\%$ higher packet processing throughput than that of Gulliver Tunnel. This is because, Gulliver Tunnel performs packet resubmissions in data plane upon every incoming packet which adversely affects its latency and throughput. While we \textit{intelligently} design our data plane (\secref{dpi}) so that we eliminate packet resubmissions. And even when required (when the burst segmentation threshold is reached), we instead mirror the packet to the loopback port so that original packet is not delayed in the pipeline. This leads to high throughput. To better justify increased throughput of \arch, we also compare percentage of packet resubmissions and per-packet resubmission overhead to that of Gulliver Tunnel. \arch reduces percentage of packet resubmissions by $88.3\%$ and per-packet resubmission overhead by $99.7\%$. For details, refer to supplementary material.   
\subsection{End-to-end detection performance} \label{eted}
\begin{table*}[hbt!]
\scalebox{0.8}{
\begin{tabular}{|c|c|llc|llc|llc|llc|llc|}
\hline
\multirow{3}{*}{Dataset}                                            & \multirow{3}{*}{Attacks} & \multicolumn{3}{c|}{Kitsune \cite{mirsky2018kitsune}}                                                                   & \multicolumn{3}{c|}{Magnifier \cite{290987}}                                                                 & \multicolumn{3}{c|}{HorusEye \cite{290987}}                                                                  & \multicolumn{3}{c|}{\arch ($n=8$ features)}                                         & \multicolumn{3}{c|}{\arch ($n=21$ features)}                                    \\ \cline{3-17} 
                                                                    &                          & \multicolumn{2}{c|}{TPR}                                            & \multirow{2}{*}{PR\textsubscript{AUC}} & \multicolumn{2}{c|}{TPR}                                            & \multirow{2}{*}{PR\textsubscript{AUC}} & \multicolumn{2}{c|}{TPR}                                            & \multirow{2}{*}{PR\textsubscript{AUC}} & \multicolumn{2}{c|}{TPR}                                                & \multirow{2}{*}{PR\textsubscript{AUC}} & \multicolumn{2}{c|}{TPR}                                            & \multirow{2}{*}{PR\textsubscript{AUC}} \\ \cline{3-4} \cline{6-7} \cline{9-10} \cline{12-13} \cline{15-16}
                                                                    &                          & \multicolumn{1}{c|}{$\leq$ 5e-5} & \multicolumn{1}{c|}{$\leq$ 5e-4} &                          & \multicolumn{1}{c|}{$\leq$ 5e-5} & \multicolumn{1}{c|}{$\leq$ 5e-4} &                          & \multicolumn{1}{c|}{$\leq$ 5e-5} & \multicolumn{1}{c|}{$\leq$ 5e-4} &                          & \multicolumn{1}{c|}{$\leq$ 2e-3}   & \multicolumn{1}{c|}{$\leq$ 1e-2}   &                          & \multicolumn{1}{c|}{$\leq$ 5e-5} & \multicolumn{1}{c|}{$\leq$ 5e-4} &                          \\ \hline
\multirow{11}{*}{\begin{tabular}[c]{@{}l@{}}\cite{bezerra2018providing}\\ \cite{koroniotis2019towards}\\ \cite{mirsky2018kitsune}\end{tabular}} & Aidra                    & \multicolumn{1}{c|}{0.228}       & \multicolumn{1}{c|}{0.406}       & 0.718                    & \multicolumn{1}{c|}{0.370}       & \multicolumn{1}{c|}{0.451}       & 0.631                    & \multicolumn{1}{c|}{0.383}       & \multicolumn{1}{c|}{0.469}       & 0.657                    & \multicolumn{1}{c|}{0.381 / 0.392} & \multicolumn{1}{c|}{0.472 / 0.489} & 0.661 / 0.715            & \multicolumn{1}{c|}{0.395}       & \multicolumn{1}{c|}{0.498}       & 0.721                    \\ \cline{2-17} 
                                                                    & Bashlite                 & \multicolumn{1}{c|}{0.605}       & \multicolumn{1}{c|}{0.677}       & 0.818                    & \multicolumn{1}{c|}{0.698}       & \multicolumn{1}{c|}{0.730}       & 0.806                    & \multicolumn{1}{c|}{0.713}       & \multicolumn{1}{c|}{0.735}       & 0.817                    & \multicolumn{1}{c|}{0.711 / 0.727} & \multicolumn{1}{c|}{0.734 / 0.754} & 0.818 / 0.825            & \multicolumn{1}{c|}{0.745}       & \multicolumn{1}{c|}{0.757}       & 0.834                    \\ \cline{2-17} 
                                                                    & Mirai                    & \multicolumn{1}{c|}{0.105}       & \multicolumn{1}{c|}{0.183}       & 0.949                    & \multicolumn{1}{c|}{0.962}       & \multicolumn{1}{c|}{0.966}       & 0.976                    & \multicolumn{1}{c|}{0.964}       & \multicolumn{1}{c|}{0.966}       & 0.980                    & \multicolumn{1}{c|}{0.966 / 0.972} & \multicolumn{1}{c|}{0.967 / 0.971} & 0.979 / 0.985            & \multicolumn{1}{c|}{0.967}       & \multicolumn{1}{c|}{0.972}       & 0.989                    \\ \cline{2-17} 
                                                                    & Keylogging               & \multicolumn{1}{c|}{0.527}       & \multicolumn{1}{c|}{0.527}       & 0.602                    & \multicolumn{1}{c|}{0.527}       & \multicolumn{1}{c|}{0.528}       & 0.779                    & \multicolumn{1}{c|}{0.527}       & \multicolumn{1}{c|}{0.528}       & 0.806                    & \multicolumn{1}{c|}{0.527 / 0.532} & \multicolumn{1}{c|}{0.529 / 0.535} & 0.806 / 0.810            & \multicolumn{1}{c|}{0.533}       & \multicolumn{1}{c|}{0.537}       & 0.815                    \\ \cline{2-17} 
                                                                    & Data theft               & \multicolumn{1}{c|}{0.508}       & \multicolumn{1}{c|}{0.508}       & 0.587                    & \multicolumn{1}{c|}{0.508}       & \multicolumn{1}{c|}{0.508}       & 0.785                    & \multicolumn{1}{c|}{0.508}       & \multicolumn{1}{c|}{0.510}       & 0.810                    & \multicolumn{1}{c|}{0.506 / 0.512} & \multicolumn{1}{c|}{0.510 / 0.515} & 0.808 / 0.812            & \multicolumn{1}{c|}{0.511}       & \multicolumn{1}{c|}{0.517}       & 0.814                    \\ \cline{2-17} 
                                                                    & Service scan             & \multicolumn{1}{c|}{0.217}       & \multicolumn{1}{c|}{0.274}       & 0.833                    & \multicolumn{1}{c|}{0.318}       & \multicolumn{1}{c|}{0.358}       & 0.915                    & \multicolumn{1}{c|}{0.334}       & \multicolumn{1}{c|}{0.363}       & 0.934                    & \multicolumn{1}{c|}{0.337 / 0.342} & \multicolumn{1}{c|}{0.364 / 0.372} & 0.936 / 0.942            & \multicolumn{1}{c|}{0.348}       & \multicolumn{1}{c|}{0.377}       & 0.944                    \\ \cline{2-17} 
                                                                    & OS scan                  & \multicolumn{1}{c|}{0.367}       & \multicolumn{1}{c|}{0.507}       & 0.939                    & \multicolumn{1}{c|}{0.461}       & \multicolumn{1}{c|}{0.561}       & 0.933                    & \multicolumn{1}{c|}{0.498}       & \multicolumn{1}{c|}{0.577}       & 0.946                    & \multicolumn{1}{c|}{0.496 / 0.514} & \multicolumn{1}{c|}{0.575 / 0.583} & 0.944 / 0.951            & \multicolumn{1}{c|}{0.514}       & \multicolumn{1}{c|}{0.584}       & 0.953                    \\ \cline{2-17} 
                                                                    & HTTP DDoS                & \multicolumn{1}{c|}{0.055}       & \multicolumn{1}{c|}{0.211}       & 0.779                    & \multicolumn{1}{c|}{0.235}       & \multicolumn{1}{c|}{0.382}       & 0.927                    & \multicolumn{1}{c|}{0.285}       & \multicolumn{1}{c|}{0.408}       & 0.942                    & \multicolumn{1}{c|}{0.287 / 0.293} & \multicolumn{1}{c|}{0.410 / 0.415} & 0.943 / 0.947            & \multicolumn{1}{c|}{0.293}       & \multicolumn{1}{c|}{0.417}       & 0.952                    \\ \cline{2-17} 
                                                                    & TCP DDoS                 & \multicolumn{1}{c|}{0.903}       & \multicolumn{1}{c|}{0.936}       & 0.969                    & \multicolumn{1}{c|}{0.959}       & \multicolumn{1}{c|}{0.971}       & 0.989                    & \multicolumn{1}{c|}{0.903}       & \multicolumn{1}{c|}{0.912}       & 0.929                    & \multicolumn{1}{c|}{0.907 / 0.909} & \multicolumn{1}{c|}{0.910 / 0.916} & 0.931 / 0.937            & \multicolumn{1}{c|}{0.911}       & \multicolumn{1}{c|}{0.917}       & 0.941                    \\ \cline{2-17} 
                                                                    & UDP DDoS                 & \multicolumn{1}{c|}{0.904}       & \multicolumn{1}{c|}{0.936}       & 0.968                    & \multicolumn{1}{c|}{0.959}       & \multicolumn{1}{c|}{0.972}       & 0.989                    & \multicolumn{1}{c|}{0.965}       & \multicolumn{1}{c|}{0.973}       & 0.990                    & \multicolumn{1}{c|}{0.966 / 0.972} & \multicolumn{1}{c|}{0.972 / 0.982} & 0.989 / 0.995            & \multicolumn{1}{c|}{0.973}       & \multicolumn{1}{c|}{0.985}       & 0.996                    \\ \cline{2-17} 
                                                                    & \textbf{macro}           & \multicolumn{1}{c|}{0.442}       & \multicolumn{1}{c|}{0.516}       & 0.816                    & \multicolumn{1}{c|}{0.600}       & \multicolumn{1}{c|}{0.643}       & 0.873                    & \multicolumn{1}{c|}{0.608}       & \multicolumn{1}{c|}{0.644}       & 0.881                    & \multicolumn{1}{c|}{0.608 / 0.617} & \multicolumn{1}{c|}{0.644 / 0.653} & 0.881 / 0.892            & \multicolumn{1}{c|}{0.619}       & \multicolumn{1}{c|}{0.656}       & 0.896                    \\ \hline
\multirow{6}{*}{\cite{290987}}                                               & Mirai                    & \multicolumn{1}{c|}{0.000}       & \multicolumn{1}{c|}{0.012}       & 0.636                    & \multicolumn{1}{c|}{0.196}       & \multicolumn{1}{c|}{0.412}       & 0.842                    & \multicolumn{1}{c|}{0.303}       & \multicolumn{1}{c|}{0.424}       & 0.868                    & \multicolumn{1}{c|}{0.304 / 0.311} & \multicolumn{1}{c|}{0.427 / 0.431} & 0.871 / 0.875            & \multicolumn{1}{c|}{0.353}       & \multicolumn{1}{c|}{0.437}       & 0.877                    \\ \cline{2-17} 
                                                                    & Service scan             & \multicolumn{1}{c|}{0.918}       & \multicolumn{1}{c|}{0.956}       & 0.998                    & \multicolumn{1}{c|}{0.989}       & \multicolumn{1}{c|}{0.995}       & 0.999                    & \multicolumn{1}{c|}{0.991}       & \multicolumn{1}{c|}{0.996}       & 1.000                    & \multicolumn{1}{c|}{0.990 / 0.995} & \multicolumn{1}{c|}{0.996 / 0.996} & 0.999 / 1.000            & \multicolumn{1}{c|}{0.994}       & \multicolumn{1}{c|}{0.999}       & 1.000                    \\ \cline{2-17} 
                                                                    & OS scan                  & \multicolumn{1}{c|}{0.617}       & \multicolumn{1}{c|}{0.810}       & 0.994                    & \multicolumn{1}{c|}{0.943}       & \multicolumn{1}{c|}{0.983}       & 0.999                    & \multicolumn{1}{c|}{0.968}       & \multicolumn{1}{c|}{0.985}       & 0.999                    & \multicolumn{1}{c|}{0.971 / 0.977} & \multicolumn{1}{c|}{0.984 / 0.988} & 1.000 / 1.000            & \multicolumn{1}{c|}{0.976}       & \multicolumn{1}{c|}{0.991}       & 1.000                    \\ \cline{2-17} 
                                                                    & TCP DDoS                 & \multicolumn{1}{c|}{0.994}       & \multicolumn{1}{c|}{0.996}       & 1.000                    & \multicolumn{1}{c|}{0.997}       & \multicolumn{1}{c|}{0.998}       & 1.000                    & \multicolumn{1}{c|}{0.997}       & \multicolumn{1}{c|}{0.998}       & 1.000                    & \multicolumn{1}{c|}{0.997 / 0.998} & \multicolumn{1}{c|}{0.998 / 0.998} & 1.000 / 1.000            & \multicolumn{1}{c|}{0.998}       & \multicolumn{1}{c|}{0.999}       & 1.000                    \\ \cline{2-17} 
                                                                    & UDP DDoS                 & \multicolumn{1}{c|}{0.995}       & \multicolumn{1}{c|}{0.997}       & 1.000                    & \multicolumn{1}{c|}{0.997}       & \multicolumn{1}{c|}{0.998}       & 1.000                    & \multicolumn{1}{c|}{0.998}       & \multicolumn{1}{c|}{0.998}       & 1.000                    & \multicolumn{1}{c|}{0.997 / 0.999} & \multicolumn{1}{c|}{0.999 / 1.000} & 1.000 / 1.000            & \multicolumn{1}{c|}{0.999}       & \multicolumn{1}{c|}{1.000}       & 1.000                    \\ \cline{2-17} 
                                                                    & \textbf{macro}           & \multicolumn{1}{c|}{0.705}       & \multicolumn{1}{c|}{0.754}       & 0.925                    & \multicolumn{1}{c|}{0.825}       & \multicolumn{1}{c|}{0.877}       & 0.968                    & \multicolumn{1}{c|}{0.852}       & \multicolumn{1}{c|}{0.880}       & 0.973                    & \multicolumn{1}{c|}{0.852 / 0.856} & \multicolumn{1}{c|}{0.881 / 0.883} & 0.974 / 0.975            & \multicolumn{1}{c|}{0.864}       & \multicolumn{1}{c|}{0.885}       & 0.975                    \\ \hline
\end{tabular}}
\caption{Performance of \arch compared to state-of-the-arts. Note that $\leq$5e-4 means FPR $\leq$ 5e-4. Moreover, we implement \arch with $8$ features in the hardware and for this version, TPR and PR\textsubscript{AUC} is shown for both without profiler / with profiler. \arch with $21$ features is simulated in python3 due to switch constraints. We use normal traffic dataset \cite{290987}. Normal traffic dataset from \cite{sivanathan2018classifying} yields similar trend.}
\vspace{-0.5cm}
\label{tab:etep}
\end{table*}
We compare the attack detection performance of \arch with Kitsune \cite{mirsky2018kitsune}, Magnifier \cite{290987} and HorusEye \cite{290987} (Magnifier + Gulliver Tunnel) in \tref{tab:etep}. All systems are trained on normal traffic from \cite{290987}. For normal traffic \cite{sivanathan2018classifying}, we defer the results to the supplementary material (the trend is similar). TPR is evaluated under various FPR schemes (obtained by setting the RMSE threshold of autoencoders).  Magnifier and Kitsune are implemented on $21$ FL features (based on standard deviations of packet sizes and packet jitters \cite{290987}). Due to the knowledge distillation process, the number of features used by autoencoders (which we distill into iForest) is limited to those used by iForest. We implement $3$ variants of \arch, \arch with $8$ features stated \secref{expimp} without and with profiler (in hardware) and, \arch with $21$ features (same as Magnifier) simulated in python3 (due to switch memory limit). We note that with $8$ features it was not possible for \arch to obtain very low FPR (Magnifier and Kitsune used 21 features). With FPR of $\leq$1e-2 and $\leq$2e-3, \arch with $3$ features can give similar performance to HorusEye under FPR $\leq$5e-4 and $\leq$5e-5. Moreover, it is self-explanatory that \arch with profiler yields better TPR and PR\textsubscript{AUC} on all the attacks.

Moreover, \arch with $21$ features yields better or the same TPR (and PR\textsubscript{AUC}) as \arch with $8$ features and HorusEye under very low FPR. It is natural to observe that since HorusEye yields better TPR on most of the attacks compared to Kitsune, so does \arch. Since Magnifier's knowledge is transferred in \arch, we can argue superior attack detection performance in \arch similar to that of HorusEye \cite{290987}. In fact our knowledge distillation schemes help us leverage best of both Magnifier and iForest.
\begin{table*}
    \centering
    \scalebox{0.93}{
    \begin{tabular}{|c|c|c|c|c|c|c|c|c|c|c|}
    \hline
         & Mousika \cite{10158739,xie2022mousika} & Flowrest \cite{Akem2023FlowrestPF} & NetBeacon \cite{zhou2023efficient} & Jewel \cite{akem2024jewel} & Leo \cite{jafri2024leo} & Kitsune \cite{mirsky2018kitsune} & Magnifier \cite{290987} & HorusEye \cite{290987} & \makecell{\arch\\ (8 features)} & \makecell{\arch\\ (21 features)} \\ \hline 
       TPR & 0.211 & 0.305 & 0.315 & 0.313 & 0.318 & 0.515 & 0.686 & 0.688 & 0.687 & 0.692 \\ \hline
       FPR  & 0.578 & 0.544 & 0.527 & 0.521 & 0.522 & 3.45e-4 & 2.73e-4 & 2.75e-4 &  8.52e-3 & 2.71e-4 \\ \hline
       PR\textsubscript{AUC} & 0.646 & 0.657 & 0.661 & 0.659 & 0.660 & 0.723 & 0.776 & 0.781 & 0.779 & 0.788 \\ \hline 
    \end{tabular}}
    \caption{Superior attack detection performance of \arch compared with supervised ML-based solutions. }
    \vspace{-0.5cm}
    \label{tab:scom}
\end{table*}

Further, we also compare \arch's macro TPR, FPR and PR\textsubscript{AUC} to state-of-the-art supervised methods (implemented in data planes) like Mousika \cite{xie2022mousika,10158739}, Flowrest \cite{Akem2023FlowrestPF}, NetBeacon \cite{zhou2023efficient}, Jewel \cite{akem2024jewel} and, Leo \cite{jafri2024leo}. For Mousika we use RF (random forest) distilled Binary decision tree (BDT). Rest of the methods use RFs. Mousika uses \textit{dstPort} and \textit{packet's length} as PL features. Flowrest uses $7$ FL features while the rest of the methods use $8$ PL+FL features stated in \secref{expimp}. We train
and test supervised models using the same normal data \cite{290987} and three categories of botnet infection (i.e., Mirai, Aidra and Bashlite from \cite{290987}). For supervised learning, we use two of the attacks for training and the other one as the unknown attack for testing. For
unsupervised learning, these attacks are only used for testing, i.e., they are not in the training set. This comparison is given in \tref{tab:scom} and therefore, we demonstrate the superiority of unsupervised schemes when detecting new and unseen attacks.
\subsection{Detection performance on other datasets} \label{dpod}
\begin{figure}
    \centering
    \includegraphics[width=9cm]{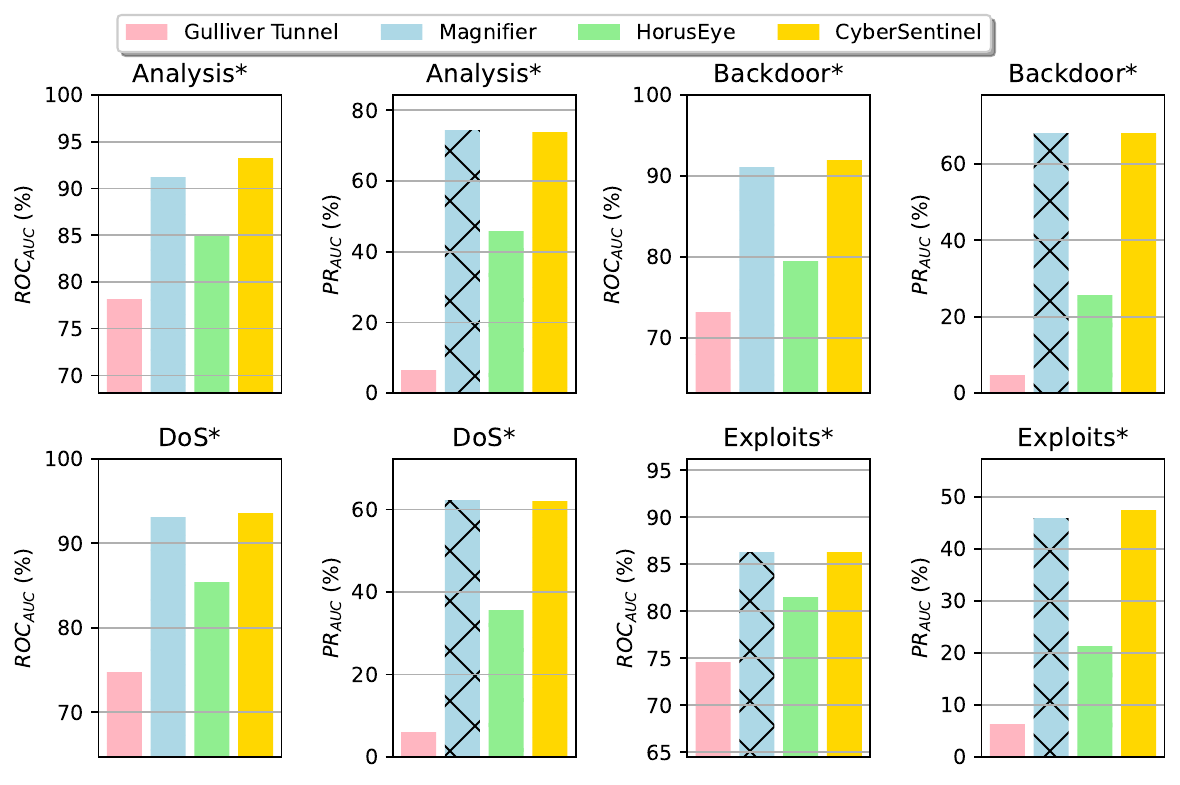}
    \caption{Comparison of \arch with Gulliver Tunnel, Magnifier, and HorusEye in terms of ROC\textsubscript{AUC} and PR\textsubscript{AUC} on UNSW NB-15 dataset.}
    \label{fig:fincomp}
\end{figure}
\textbf{Setting.} We use the popular intrusion detection dataset called UNSW NB-15 \cite{moustafa2015unsw} from where we pick up four popular attacks: \textit{Analysis, Backdoor, DoS,} and \textit{Exploit}. We use two metrics: PR\textsubscript{AUC} and ROC\textsubscript{AUC}. The ROC curve indicates the true positives against
false positives, while the PR curve summarises precision and
recall of the anomaly class only. Training and validation sets are divided into a $4:1$ ratio. Moreover, contamination ratios for each of the attacks are the same as the anomaly ratios provided in the datasets. We use number of iTrees $t=300$, sub-sample size $\Psi = 256$ and, data augmentation factor $k=150$. We choose $N_{threshold} = 4$ packets based on pdf of burst segmentation lengths (or number of packets per burst). 

We show the results in \fref{fig:fincomp}. Here, we only distill Magnifier into iForest and do not take into account the original iForest's anomaly score to predict a sample as benign/anomaly. This is because iForest itself does not give high TPR for these datasets, and therefore, Gulliver Tunnel is not able to give a very high TPR. Thus, the performance of Magnifier (better area under ROC and PR curve) exceeds that of Gulliver Tunnel and HorusEye for most of the attacks. For almost all the attack datasets, \arch yields almost the same area under the ROC and PR curve as that of Magnifier. This demonstrates the effectiveness of our knowledge distillation and whitelist rules generation procedure. 
\subsection{Robustness of adversarial attack detection}
\begin{figure}
    \centering
    \includegraphics[width=9cm]{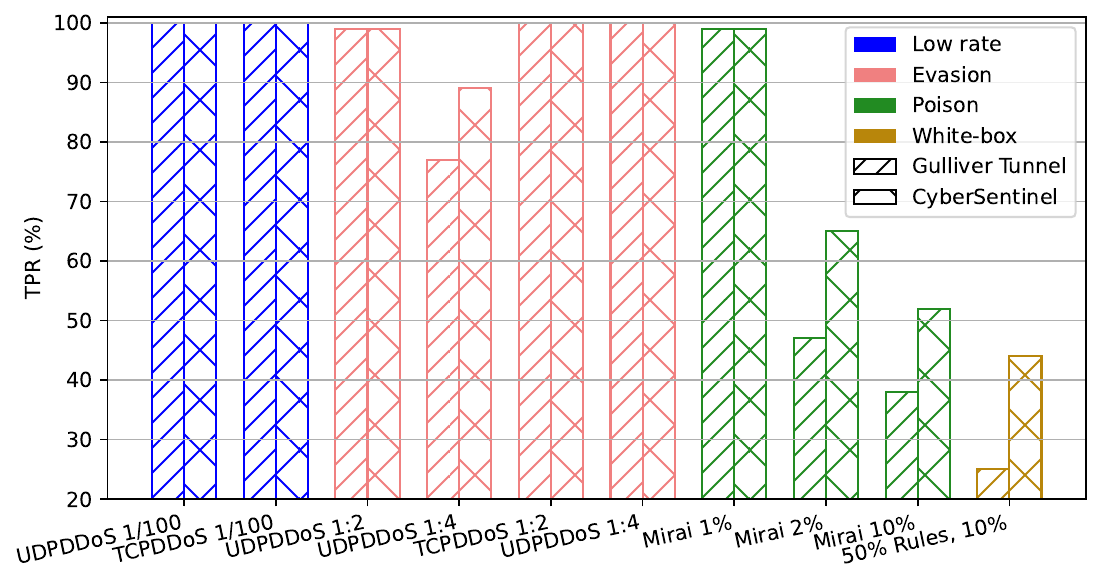}
    \caption{Robustness of \arch against different
adversarial attacks and its comparison with Gulliver Tunnel. Note 1/100 or 1:2 denotes fraction of malicious traffic to that of benign traffic.}
    \label{fig:robust}
\end{figure}
We discuss robustness of \arch on three black-box adversarial attacks and one white-box attack, and compare its resilience to Gulliver Tunnel (see \fref{fig:robust}).

\myparab{Low rate attacks. }We respectively reduce the rates of TCP and UDP DDoS by 100 times (to 900 packets/second). Both \arch and Gulliver Tunnel are resilient to such attacks and retain TPR of nearly 100\%. This is because reducing rate of traffic does not have any effect on segmented BL features.

\myparab{Evasion attacks. }To disguise TCP and UDP DDoS traffic as benign, we inject benign QUIC and TLS traffic into it. Both Gulliver Tunnel and \arch can resist most of such attacks "one-way" because of the use of bi-hash algorithm. \arch gives better TPR because of distilled Magnifier's better representation capability and dilated convolution's large receptive field.

\myparab{Poison attacks. }We inject Mirai attack traffic to benign training set. Both Gulliver Tunnel and \arch achieve TPR of 99\% when pollution is mild $1\%$. However drop in TPR of \arch is not as much as Gulliver Tunnel for poison $\geq 2\%$. This is because distilled Magnifier's deep encoding layers can capture better representations of normal traffic from polluted traffic.

\myparab{Whitebox attacks. }We consider an attack where attacker can perturb whitelist rules of iForest. The TPR falls greatly in case of Gulliver  upon $10\%$ perturbation of feature boundaries on $50\%$ rules while the reduction is not as high in \texttt{CyberSentinel}. This is because our whitelist rules are not as explainable as original iForest because knowledge of Magnifier is used to obtain them. 
\subsection{Throughput and detection capability} \label{tdc}
\begin{figure}
    \centering
    \includegraphics[width=9cm]{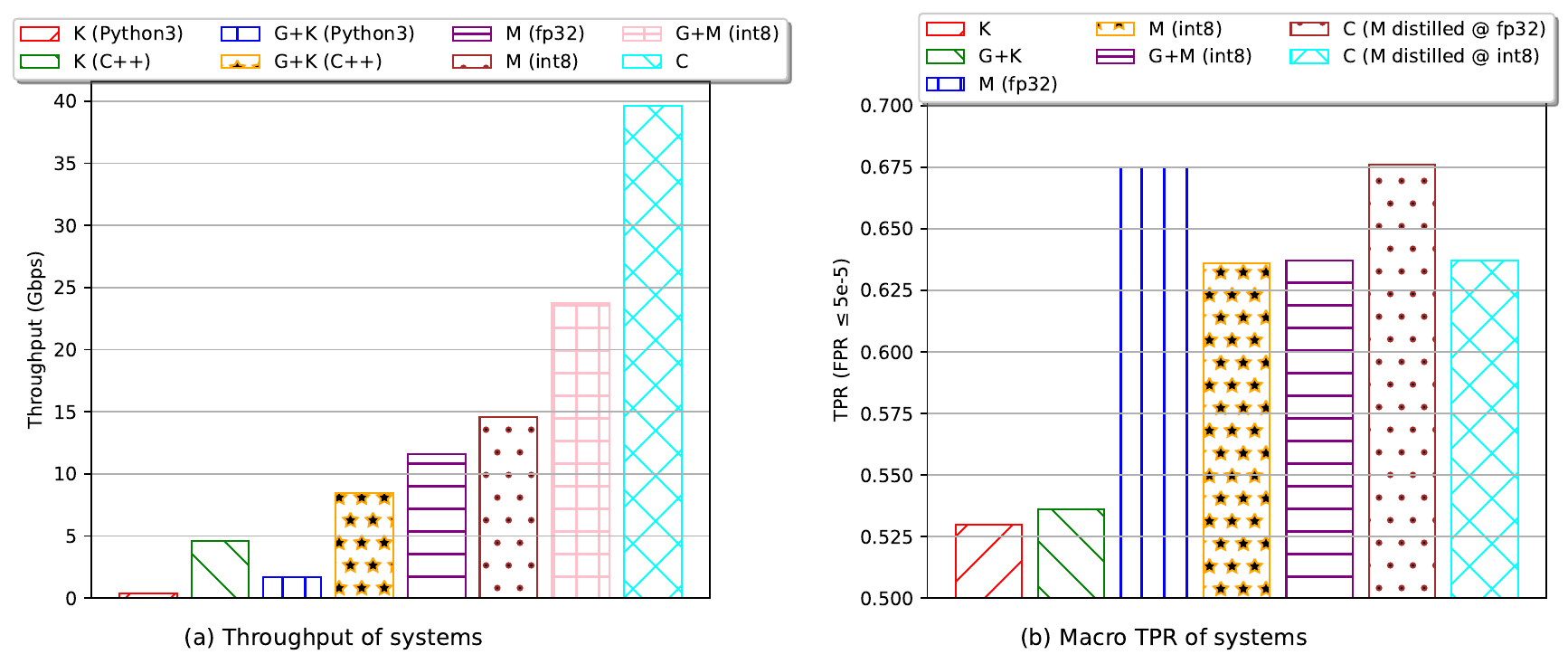}
    \caption{Throughput and detection capability. G = Gulliver Tunnel, K = Kitsune, M = Magnifier and C = \arch (it uses FPR $\leq$ 2e-3). }
    \label{fig:thpt}
\end{figure}
The comparison of \arch in terms of packet processing throughput on $40$ Gbps link and macro TPR with multiple systems under several constraints is given in \fref{fig:thpt}a-\ref{fig:thpt}b. Only \arch using $8$ BL+PL features is shown in the plots.

On a $40$ Gbps link, HorusEye (Gulliver Tunnel + Magnifier with 8-bit int quantization) yields $23.79$ Gbps packet processing throughput. In contrast, we saw that \arch yields a solid throughput of $39.6$ Gbps (+ $66.47\%$). This is because \arch operates entirely in data plane while HorusEye takes support of Magnifier operating in control plane. 

Macro TPR of HorusEye (Gulliver Tunnel + Magnifier with 8-bit int quantization) under FPR $\leq$ 5e-5 is $0.637$ while Magnifier with 32-bit floating point quantization under same FPR yields TPR as $0.675$. \arch (with $8$ features) obtained upon distilling Magnifier with 32-bit floating point quantization yields TPR of $0.676$ under FPR$\leq$2e-3. On the other hand, with $21$ features (as in Magnifier from HorusEye), we get TPR of $0.678$ under FPR $\leq$ 5e-5. This is because knowledge of Magnifier is already contained in \arch due to knowledge distillation. 

\section{~~~~Related Work} \label{rw}
We discuss the following categories of related work (\textit{i.e.,} anomaly detection systems).

\myparab{Control plane-based. }Works such as \cite{fu2021realtime,mirsky2018kitsune,celik2019iotguard,tang2020zerowall,zhang2018homonit} cannot scale to multi-Tbps because they perform
detection in the control plane. Moreover \cite{fu2021realtime,mirsky2018kitsune} can leverage \arch to perform pre-filtering of suspicious traffic but this will cause us to lose line speed. Overall, control plane-based works cannot achieve high throughput detection. This makes them unable to scale to
increased interactions between IoT devices. Therefore, they cannot be deployed in smart cities or for security purposes in software defined networking. 

\myparab{Programmable switch-based. }There are works that only partially leverage data planes for attack detection \cite{xing2020netwarden,barradas2021flowlens,zhou2023cerberus,seufert2024marina} and are unable to produce high throughput detection. Then there are works \cite{zhou2023efficient,xie2022mousika,akem2024jewel,Akem2023FlowrestPF,zheng2021planter,zheng2022automating,zheng2022iisy,lee2020switchtree,zhang2021pheavy,friday2022inc,friday2022learning,akem2022henna,akem2024encrypted,jafri2024leo,10158739} use supervised decision trees in data planes, thus they require labeled data for training. Only works that discuss deploying unsupervised models in switch data planes are \cite{290987,zheng2022automating} but \cite{290987} takes the support of control plane for stage 2 of detection while \cite{zheng2022automating} approach towards deploying iForest does not scale (because it installs all the rules of iForest, and do not provide actual implementation). \arch falls in this line work can achieve high throughput and accurate anomaly detection through unsupervised methods.
\section{Conclusion} \label{conc}
In this paper, we propose \arch which is the first work to perform knowledge distillation from an ensemble of autoencoders into an iForest. The iForest is then deployed in the switch data plane in the form of whitelist rules. As we saw, \arch can perform malicious traffic detection entirely in the data plane (at line rate) and yields almost similar TPR compared to state-of-the-art while reducing per-packet latency by $50\%$ and improving packet processing throughput on $40$ Gbps link by $66.47\%$ (compared to HorusEye).

\bibliographystyle{IEEEtranS}
\bibliography{bibliography}
\newpage
\clearpage
\section*{Supplementary Material}
\section{Distilled iForest model and rules generation} \label{difrg}
We present a detailed procedure for distilled iForest generation in Alg. \ref{algo:3} and whitelist rules generation in Alg. \ref{algo:4}. As for knowledge distillation, we explain in Alg. \ref{algo:3} how knowledge distillation of an ensemble of $r$ autoencoders is performed, and their knowledge is transferred into trained iForest.  In Alg. \ref{algo:4} we present how whitelist rules are generated from the trained and distilled iForest that is generated in Alg. \ref{algo:3}.

One major advantage (as we can see from Alg. \ref{algo:3}) is that our knowledge distillation scheme does not add extra time complexity overhead from iForest model training \cite{liu2008isolation,290987}. 
\setlength{\textfloatsep}{0.5pt}%
\begin{algorithm}[hbt!]
\caption{Knowledge distillation in iForest}\label{algo:3}
\small
\textbf{Input: }A trained iForest, trained ensemble of $r$ autoencoders\\
\textbf{Notations: }Training set of benign samples $X_{train}$, data augmentation factor $k$, $leaf_{ij}$ denoting $j^{th}$ leaf of $i^{th}$ iTree, embedding function associated with a leaf $\mathcal{E}:\textbf{leaves} \rightarrow \mathcal{R}^{r}$, features range associated with a leaf (or a branch) $features\_range(leaf_{ij})$
\begin{algorithmic}[1]
\State For each training sample $x \in X_{train}$, traverse all the iTrees and map the sample to a leaf of each iTree. Repeat for all the samples.
\Statex If $X_{ij} \subseteq X_{train}$ mapped to $leaf_{ij}$, then for each iTree,
\[
X_{train} = \bigcup_{j=1}^{L_i} X_{ij}
\]
where $L_i$ is the number of leaves in $i^{th}$ iTree.
\State For each leaf $leaf_{ij}$, sample $k$ points (data augmentation) without replacement and repeat for all leaves as follows
\[
A_{ij} \sim features\_range(leaf_{ij})
\]
where $|A_{ij}| = k$.
\State For each of $r$ trained autoencoders, obtain the expected reconstruction error as follows,
\[
\overline{RE_{u_{ij}}} = \mathbbm{E}_{x \sim X_{ij} \cup A_{ij}}[RE_u(x)]
\]
where $\overline{RE_{u_{ij}}}$ is expected reconstruction error ($RE$) for $u^{th}$ autoencoder on samples $X_{ij} \cup A_{ij}$. Whereas $RE_u(x)$ is reconstruction error of $u^{th}$ autoencoder for sample $x$. This is repeated for each leaf of iForest.
\State For each leaf, obtain embedding vector $\mathcal{E}$ and repeat for all leaves as follows,
\[
\mathcal{E}(leaf_{ij}) = \{\overline{RE_{1_{ij}}}, \overline{RE_{2_{ij}}}, ..., \overline{RE_{r_{ij}}}\}
\]
We consider a vector of $\mathcal{E}(leaf_{ij})$ for every $(i,j)$ pair (which constitutes a leaf) as \textbf{E}.
\State Use an appropriate labeling function $\mathcal{F}$ to decide the label corresponding to the embedding vector of a leaf node.
\[
\mathcal{F}:\textbf{E} \rightarrow \{0,1\}
\]
where $\mathcal{F}$ is user defined. We repeat for each leaf node. For example, we took $\mathcal{F}$ as a straightforward mapping in \secref{kd}. 
\Statex We took $w_u$, $u \geq 1$ to be importance weights of each of $u^{th}$ autoencoder. Then, we map embedding vector \textbf{E} to labels \{0, 1\} as follows.
\[
agg\_score_{ij} =  \sum_{u=1}^r w_u \times \mathbbm{1}\{\overline{RE_{u_{ij}}} > T_u\}
\]
where $\sum_{u=1}^r w_u = 1$. Then we decide the label as follows.
\[
label_{ij} = 
\begin{cases}
1 & \text{if } agg\_score_{ij} > 0.5 \\
0 & \text{otherwise}
\end{cases}
\]
where $label_{ij}$ denotes label (0 or 1) associated with $leaf_{ij}$.
\State \textbf{Testing phase.} For a test sample $x_{test}$, traverse each iTree of iForest and retrieve labels for each iTree. Decide the label using \textit{majority voting} over all iTrees. Let that be $label_{distilled}(x_{test})$. Also, obtain $score(x_{test})$ which corresponds to the anomaly score of the sample $x_{test}$ based on the expected path length in the original iForest. Then for the final label for any test sample decide another labeling function $\mathcal{G}$ such that,
\[
\mathcal{G}: \{0, 1\} \times \mathcal{R} \rightarrow \{0, 1\}
\]
So $final\_label(x) = \mathcal{G}(label_{distilled}(x_{test}),\; score(x_{test}))$. For our experiments we chose $final\_label(x_{test}) = label_{distilled}(x_{test}) \times \mathbbm{1}\{ score(x_{test})<0.5 \}$.
\end{algorithmic}
\end{algorithm}

\setlength{\textfloatsep}{0.5pt}%
\begin{algorithm}[hbt!]
\caption{Distilled iForest whitelist rules generation}\label{algo:4}
\small
\textbf{Input: }A trained and knowledge-distilled iForest.\\
\textbf{Notations: }$leaf_{ij}$ denoting $j^{th}$ leaf of $i^{th}$ iTree, features range associated with a leaf (or a branch) $features\_range(leaf_{ij})$. A hypercube $Q_{ij}$ corresponding to feature range $features\_range(leaf_{ij})$ of $leaf_{ij}$. Consider a dataset sample to have $m$ features $\{f_1, f_2, ..., f_m\}$.
\begin{algorithmic}[1]
\State For each leaf $leaf_{ij}$, define a hypercube $Q_{ij}$ as follows.
\[
Q_{ij} = \{(f_1, f_2, ..., f_m)\;|\;f_p \in [f_{p_{ij, min}}, f_{p_{ij, max}}], 1 \leq p \leq m\}
\]
where $features\_range(leaf_{ij}) = \bigcup_{p=1}^m [f_{p_{ij, min}}, f_{p_{ij, max}}]$. 
\State We combine hypercubes corresponding to each leaf node and form an iTree hypercubes as follows.
\[
Q_i = \bigcup_{j=1}^{L_i} Q_{ij}
\]
where $Q_i$ is the iTree hypercubes of $i^{th}$ iTree.
\State Merge the individual iTree hypercubes and form the iForest hypercubes by following \cite{290987} as shown.
\[
\textbf{Q} = \bigoplus_{i=1}^t Q_i
\]
where $\bigoplus$ denotes merge operation and there are $t$ iTrees. Note that merging means taking the branch set of each feature $f_p$ from \textit{features\_branch} and forming iForest hypercubes using cartesian product operation \cite{290987}.
\State If all features take only integral values, then we can shift the iForest hypercubes boundaries towards their nearest integral values \cite{290987}. For example, if $b_i$ is a boundary of a feature then we make it $\check{b_i}$ = \textit{floor($b_i$)}. This is done keeping in mind that floating point operations are not supported in the P4 language. In case, the features may take non-integral values, we may omit this step.
\State Label each $hypercube$ in iForest hypercubes \textbf{Q} as 0 or 1 as follows. First randomly sample a point $x$ from an iForest hypercube.
\[
x \sim hypercube
\]
Then label that hypercube as per step 6 of Alg. \ref{algo:3} by considering sample point $x$. Traverse $x$ in all the $t$ iTrees and obtain a label for each iTree. Then choose majority voting among $t$ iTrees for the final assigned label. This is repeated for all the hypercubes of an iForest hypercubes.
\[
label_{distilled}(x) = majority\_vote(iTrees.predict(x))
\]
\[
hypercube.label = \mathcal{G}((label_{distilled}(x), score(x)))
\]

\State Merge the adjacent hypercubes of iForest hypercubes \textbf{Q} whose labels are same \cite{290987}. This is done to make the model simpler to deploy in the switch.
\State Now generate whitelist rules from hypercubes whose label is $0$ as follows.
\[
whitelist\_rules = rule\_generation(\textbf{Q}[label == 0])
\]
where rules are taken for that $hypercube$ in \textbf{Q} where $hypercube.label = 0$.
\end{algorithmic}
\end{algorithm}

\section{Theoretical results} \label{tr}
Before reading theorems and corollaries, please refer to Algorithms \ref{algo:3} and \ref{algo:4} for more clarity.

\myparab{Theorem 1. }\textit{All samples inside an iForest hypercube share the same label assigned by our trained and distilled iForest in Alg. \ref{algo:3}. (\textbf{Label Consistency}).}\\
\textit{Proof. }It is trivial to see that for each iTree, labels of all samples in an iTree hypercube are consistent. Following the merging of iTree hypercubes into iForest hypercubes, \textit{all sample points inside
an iForest hypercube must belong to the same iTree hypercube over all iTrees. Thus all points will have the same anomaly score of the original iForest}. In other words,
\[
\therefore \forall x \in \textit{iForest hypercube},\; \textit{score(x) is consistent.}
\]
Since all points in an iForest hypercube belong to the same iTree hypercube over all iTrees, \textit{they all land up on the same leaf node in a given iTree. Therefore, they will have the same label from the distilled iForest by majority voting.} In other words,
\[
\therefore \forall x \in \textit{iForest hypercube},\; \textit{label\textsubscript{distilled}(x) is consistent}
\]
This implies that the final label assigned to all samples in the iForest hypercube will also be the same. Thus, $\forall x \in $ \textit{iForest hypercube},
\[
\therefore \mathcal{G}(label_{distilled}(x),\; score(x))
\]
is also consistent. This completes the proof. \qedsymbol{}

\myparab{Theorem 2. }We further “shift” iForest hypercubes slightly for integer boundaries by rounding down the
branches of each feature to their nearest integers as follows,
\[
\check{b_i} \gets \lfloor b_i \rfloor,\;b_i \in \textbf{b} 
\]
where \textbf{b} is the set of branching thresholds of the distilled iForest. \textit{This shifting does not change
which iForest hypercube an integer data point falls into, i.e., for any
integer value $\alpha$, $\alpha \in (b_{i-1}, b_i] \implies \alpha \in (\check{b_{i-1}}, \check{b_i}]$. }\\
\textit{Proof. }If $b_i$ is an integer, then $(b_{i-1}, b_i] \subseteq (\check{b_{i-1}}, \check{b_i}]$. Therefore, $\alpha \in (b_{i-1}, b_i] \implies \alpha \in (\check{b_{i-1}}, \check{b_i}]$.

If $b_i$ is not an integer, then we use proof by contradiction. Let us assume that for any integer $\alpha$, $\alpha \in (b_{i-1}, b_i] \nRightarrow \alpha \in (\check{b_{i-1}}, \check{b_i}]$. This means $\alpha \in (b_{i-1}, b_i]$ and $\alpha \notin (\check{b_{i-1}}, \check{b_i}]$. Therefore $\alpha \in (\check{b_i}, b_i]$. But there exists no integer in interval $(\check{b_i}, b_i]$ and hence $\alpha$ cannot be an integer which is a contradiction. Therefore, for any
integer value $\alpha$, $\alpha \in (b_{i-1}, b_i] \implies \alpha \in (\check{b_{i-1}}, \check{b_i}]$. This completes the proof. \qedsymbol{}

\myparab{Corollary 1. }\textit{Theorem 2 holds for integer data points. Therefore, shifting the branching thresholds to the nearest integer values only works for integral BL and PL features. This is because if $\alpha$ is a non-integer data point then $\alpha \in (\check{b_i}, b_i]$ may hold}. 

\myparab{Theorem 3. }\textit{Let $h_{\theta}$ denote autoencoder function. Assume that autoencoder reconstructs a sample $x = I + f$, that is, $||x - h_{\theta}(x)||_2 < T$, where $I$ and $f$ are integer and fraction part of $x$ respectively. Also assume that autoencoder satisfies the Lipschitz property \cite{cobzacs2019lipschitz}, that is, $||h_{\theta}(x) - h_{\theta}(y)||_2 \leq L||x-y||_2$. Then for small $||f||_2$ the autoencoder is able to reconstruct the integer part of sample $x$ within the bound of $T(1 + 2L)$, where $L$ is the Lipschitz constant. In other words,}
\[
||I - h_{\theta}(I)||_2 < T(1 + 2L)
\]\\
\textit{Proof. }Let $x$ be a sample and $x'$ be a reconstructed sample such that $x' = h_{\theta}(x)$. Given that $||x - x'||_2 < T$, where $||\cdot||_2$ denotes the L2 norm, and $x = I + f$, where $I$ is the integer part of $x$. We want to show that $||I - h_{\theta}(I)||_2 < T(1+2L)$. 
We are given,
\[
||I+f - h_{\theta}(I+f)||_2 < T
\]
From there it is easy to see,
\[
||I - h_{\theta}(I+f)||_2 \leq ||I+f - h_{\theta}(I+f)||_2 < T
\]
To proceed, we can expand the expression $||I - h_{\theta}(I)||_2$ as follows:
\[
||I - h_{\theta}(I)||_2 = ||I - h_{\theta}(I+f) + h_{\theta}(I+f) - h_{\theta}(I)||_2
\]
Using the triangle inequality, we have,
\begin{equation}
||I - h_{\theta}(I)||_2 \leq ||I - h_{\theta}(I+f)||_2 + ||h_{\theta}(I+f) - h_{\theta}(I)||_2
\end{equation}
Using this Lipschitz property, we can bound the second term of Eq (1) as follows:
\[
||h_{\theta}(I+f) - h_{\theta}(I)||_2 \leq L ||f||_2
\]
Plugging this in Eq (1),
\begin{equation}
||I - h_{\theta}(I)||_2 < T + L||f||_2
\end{equation}
Now, if \( ||f||_2 \) is small enough, then we can even bound $||f||_2$ term in Eq (2) as follows by using triangle inequality,
\[
||f||_2 - ||I - h_{\theta}(I+f)||_2 \leq ||I + f - h_{\theta}(I+f)||_2 < T
\]
Or,
\[
||f||_2 - ||I - h_{\theta}(I+f)||_2 < T
\]
Using $||I - h_{\theta}(I+f)||_2 < T$, we get,
\[
||f||_2 < 2T
\]
Plugging the above inequality in Eq (2), we get,
\begin{equation}
||I - h_{\theta}(I)||_2 < T(1+2L)
\end{equation}
This completes the proof. \qedsymbol{}

\myparab{Corollary 2. }\textit{Since $T(1+2L) > T$, and if $T(1+2L)$ is also small enough, then we can set a new RMSE threshold of a Lipschitz autoencoder as $T(1+2L)$ under which it can also reconstruct integer part of samples with small fractional parts.} 

\myparab{Corollary 3. }\textit{Let new RMSE threshold of Lipschitz autoencoder is $T(1+2L)$ where $T(1+2L)$ is very small. Then if the autoencoder can reconstruct $x=I+f$, it can also reconstruct $I$. Also, assume that the final label of a distilled iForest for a sample point is influenced by the ensemble of autoencoders rather than the original iForest. That is, 
\[
final\_label(x) = label_{distilled}(x)
\]
Then we can safely shift those iForest hypercubes whose $hypercube.label = 0$ slightly for integer boundaries by rounding down the
branches of each feature to their nearest integers. It follows from theorem 3 and corollary 2.} 

\myparab{Practical significance. }Lipschitz autoencoders are practically used in anomaly detection \cite{kim2020lipschitz,tong2019lipschitz,tong2022fixing}. Therefore, theorem 3 and corollaries 2 and 3 have practical significance as well. In fact, we can apply Corollary 3 while performing knowledge distillation of a Lipschitz autoencoder into an iForest and deploying whitelist rules on a target switch.

\section{Hardware implementation} \label{blfi}
We provide P4 implementation for checking timeouts and extracting BL features. 

\subsection{Checking timeouts}
We first obtain the first time stamp of a burst and the latest (last) timestamp of the burst. Based on the timestamps, we check if there is an idle timeout or an active timeout (or both). This is illustrated in listing \ref{lst:p4_code_tout}.
\begin{figure}
    \centering
    \begin{lstlisting}[caption={P4 Code for checking timeouts}, label={lst:p4_code_tout}]
//First timestamp register
Register<bit<32>, bit<16>>(REGISTER_LENGTH, 0) flow_start_time_stamp;
RegisterAction<bit<32>, bit<16>, bit<32>>(flow_start_time_stamp) flow_start_time_stamp_register_action = {
    void apply(inout bit<32> fst, out bit<32> flts) {
        if(fst == 0) {
                fst = hdr.timestamp.ts;
        }
        flts = fst;	
    }
};
//Latest timestamp register
Register<bit<32>, bit<16>>(REGISTER_LENGTH, 0) last_time_stamp;
RegisterAction<bit<32>, bit<16>, bit<32>>(last_time_stamp) last_time_stamp_register_action = {
    void apply(inout bit<32> lts, out bit<32> ltd) {
        bit<32> temp;
        if(lts == 0) {
                lts = hdr.timestamp.ts;
                temp = lts;
                ltd = temp;
        }
        else{
        temp = lts;
        ltd = temp;
        lts = hdr.timestamp.ts;
        }
    }
};

action compute_diff1(){
    meta.diff1 = IDLE_TIMEOUT - meta.inactive_duration;
}

action compute_diff2(){
    meta.diff2 = ACTIVE_TIMEOUT - meta.active_duration;
}
                     
apply{
//First time
meta.start_time = flow_start_time_stamp_register_action.execute(meta.index);
//Last time
meta.last_time = last_time_stamp_register_action.execute(meta.index);
meta.inactive_duration = hdr.timestamp.ts - meta.last_time;
meta.active_duration = meta.last_time - meta.first_time;
compute_diff1();
compute_diff2();
if(meta.diff1[31:31] == 1 || meta.diff2[31:31] == 1){ //timeout
meta.timeout = true;
} 
}
\end{lstlisting}
\end{figure}

Another limitation of a Tofino switch is its inability to compare constants greater than 12 bits in width (this limitation is relaxed inside register action). In our case, idle timeout occurs when \texttt{inactive\_duration} $> \delta_{idle}$ and active timeout occurs if \texttt{active\_duration} $> \delta_{active}$, which is not possible in Tofino switch as both $\delta_{idle}$ and $\delta_{active}$ exceed 12 bits width at nanoseconds level.

To overcome this issue, we perform the subtractions as $\delta_{idle} - 
 $\texttt{inactive\_duration} and $\delta_{active} - $\texttt{active\_duration} and do a check if msbs are 1. This is shown in lines 29-49 of listing \ref{lst:p4_code_tout}.
\subsection{Extracting burst-level features}
We will be providing P4 code snippets for extracting simple BL features: number of packets per burst, and burst size. That is then followed by complex BL features such as: average packet size, variance of packet size, and standard deviation of packet size. 

\myparab{Simple features. }We show the P4 code implementation of the number of packets per burst (upper bounded by $N_{threshold}$) and burst size in listing \ref{lst:p4_code_countpc} and listing \ref{lst:p4_burs}. As shown in listing \ref{lst:p4_code_countpc} we reset the packet count to $1$ if there is a timeout. Otherwise, we increment the packet count. The updated value is returned and the count is reset to $0$ if there is a burst segmentation threshold. In listing \ref{lst:p4_burs}, where we obtain burst size. It is reset to the packet's length when there is a timeout. Otherwise, it is updated. The value is returned and reset to $0$ if there is a burst segmentation threshold.
\begin{figure}
    \centering
    \begin{lstlisting}[caption={P4 Code for extracting packet counts}, label={lst:p4_code_countpc}]
Register<bit<16>, bit<16>>(REGISTER_LENGTH, 0) pkt_count;
RegisterAction<bit<16>, bit<16>, bit<16>>(pkt_count) pkt_count_action = {
    void apply(inout bit<16> n_pkts, out bit<16> curr_pkts) {
        bit<16> temp = n_pkts; 
        //At most 2 if condition comparison inside register action
        if(meta.timeout){
            n_pkts = 1;
        }
        else{
            n_pkts = n_pkts + 1;
            temp = n_pkts;
            if(temp == SEG_THRES){
            //Reached burst segmentation threshold
                n_pkts = 0;
            }
        }
        curr_pkts = temp;
    }
};            

apply {
meta.pktCount = pkt_count_action.execute(meta.index);
}
\end{lstlisting}
\end{figure}
\begin{figure}
    \centering
    \begin{lstlisting}[caption={P4 Code for extracting burst size}, label={lst:p4_burs}]
Register<bit<16>, bit<16>>(REGISTER_LENGTH, 0) burst_size;
RegisterAction<bit<16>, bit<16>, bit<16>>(burst_size) burst_size_action = {
    void apply(inout bit<16> b_size, out bit<16> curr_b) {
        bit<16> temp = b_size; 
        //At most 2 if condition comparison inside register action
        if(meta.timeout){
            b_size = meta.ip_len;
        }
        else{
            b_size = b_size + meta.ip_len;
            temp = b_size;
            if(meta.pktCount == SEG_THRES){
            //Reached burst segmentation threshold
                b_size = 0;
            }
        }
        b_size = temp;
    }
};            

apply {
meta.burstSize = burst_size_action.execute(meta.index);
}
\end{lstlisting}
\end{figure}

\myparab{Complex division-based features. }We consider calculating a BL feature given by $\frac{A}{B}$ (\textit{e.g.}, average packet size, given by $\frac{\text{flow size}}{\text{number of packets}}$). One of the approaches is to use bitwise right shift operations for division \cite{zhou2023efficient}. Since the burst segmentation threshold is generally not very high (e.g., when $N_{threshold} = 16$, we can use $4$ if conditions to obtain results such as average packet size), it is very much doable in the data plane (we show NetBeacon's approach) as shown in listing \ref{lst:p4_codenb}. At every packet count at $2^i$ (\textit{e.g.}, 2, 4, 8, 16), we also store such features so that at timeout the last value could be retrieved. 
\begin{figure}
    \centering
    \begin{lstlisting}[caption={NetBeacon's \cite{zhou2023efficient} approach: average/variance/standard deviation of packet size}, label={lst:p4_codenb}]
MathUnit<bit<16>>(MathOp_t.SQR, 1) square;
MathUnit<bit<16>>(MathOp_t.SQRT, 1) square_root;
apply{
    //1. Average Packet Size
    meta.pkts = read_pkt_count.execute(meta.index);
    meta.burst_size  = read_burst_size.execute(meta.index);
    meta.burst_size_square  = read_burst_size_square.execute(meta.index);
    if(meta.pkts == 2){
    //Average packet size
    meta.avg_pkt_size = meta.burst_size >> 1;
    meta.avg_pkt_size_square = meta.burst_size_square >> 1;
    }
    else if(meta.pkts == 4){
    meta.avg_pkt_size = meta.burst_size >> 2;
    meta.avg_pkt_size_square = meta.burst_size_square >> 2;
    }
    else if(meta.pkts == 8){
    meta.avg_pkt_size = meta.burst_size >> 3;
    meta.avg_pkt_size_square = meta.burst_size_square >> 3;
    }
    else if(meta.pkts == 16){
    meta.avg_pkt_size = meta.burst_size >> 4;
    meta.avg_pkt_size_square = meta.burst_size_square >> 4;
    }
    //Variance of packet size
    meta.var_pkt_size = meta.avg_pkt_size_square - square.execute(meta.avg_pkt_size);
    //Standard deviation of packet size
    meta.std_pkt_size = square_root.execute(meta.var_pkt_size);
}
\end{lstlisting}
\end{figure}

We adopt a different but novel strategy as shown in \fref{lst:p4_codex}. We assume $A \approx 2^i$ and $B \approx 2^j$. Then, $\frac{A}{B} \approx 2^{i-j}$. To implement in data plane, we use two $log_2$ tables and one exponent ($2^x$) table.  Each entry of a log table is a range match ($[2^x, 2^{x+1})$ is matched to $x$) while each entry of the exponent table is an exact match ($x$ is matched to $2^x$). In other words, for the log table, value $A$ is matched to $i=\lfloor log_2A\rfloor$, and $B$ is matched to $j=\lfloor log_2B\rfloor$. Then $i-j$ in the exponent table is matched to $2^{i-j} \approx \frac{A}{B}$ (in case of multiplication, we match $i+j$ in the exponent table). We can introduce a scaling factor $s$ for better division (or multiplication) results (but more memory consumption). The only change is, for the log table, value $A$ is matched to $i=\lfloor s * log_2A\rfloor$ and $B$ is matched to $j=\lfloor s * log_2B\rfloor$. Then $i-j$ in exponent table is matched to $2^{\lfloor \frac{i-j}{s} \rfloor} \approx \frac{A}{B}$. Note that scaling factor $s$ can be optimized using profiler.

\begin{figure*}[htbp]
      \begin{subfigure}{0.5\textwidth}
        \begin{lstlisting}[caption={}]
MathUnit<bit<16>>(MathOp_t.SQR, 1) square;
MathUnit<bit<16>>(MathOp_t.SQRT, 1) square_root;

action log1(bit<8> logval){
    meta.i = logval;
}

table logtable1{
    key = {
        meta.pkts : range;
    }
    actions = {
        log1; NoAction;
    }
    size = 17;
    default_action = NoAction();
}

action log2(bit<8> logval){
    meta.j = logval;
}

table logtable2{
    key = {
        meta.burst_size : range;
    }
    actions = {
        log2; NoAction;
    }
    size = 28;
    default_action = NoAction();
}

action log3(bit<16> logval){
    meta.j = logval;
}

table logtable3{
    key = {
        meta.burst_size_square : range;
    }
    actions = {
        log3; NoAction;
    }
    size = 28;
    default_action = NoAction();
}

action exp1(bit<16> expval){
    meta.avg_pkt_size = expval;
}

        \end{lstlisting}
    \end{subfigure}%
    \begin{subfigure}{0.5\textwidth}
        \begin{lstlisting}[caption={}, firstnumber=52]
table exptable1{
    key = {
        meta.exp_idx : exact;
    }
    actions = {
        exp1; NoAction;
    }
    size = 64;
    default_action = NoAction();
}

action exp2(bit<16> expval){
    meta.avg_pkt_size_square = expval;
}

table exptable2{
    key = {
        meta.exp_idx : exact;
    }
    actions = {
        exp2; NoAction;
    }
    size = 64;
    default_action = NoAction();
}

apply{
    //1. Average Packet Size
    meta.pkts = read_pkt_count.execute(meta.index);
    meta.burst_size  = read_burst_size.execute(meta.index);
    logtable1.apply();
    logtable2.apply();
    meta.exp_idx = meta.j - meta.i;
    exptable1.apply(); //avg pkt size obtained

    //2. Variance of packet size
    meta.burst_size_square  = read_burst_size_square.execute(meta.index);
    logtable3.apply();
    meta.exp_idx = meta.j - meta.i;
    exptable2.apply(); //avg pkt size square obtained
    meta.var_pkt_size = meta.avg_pkt_size_square - square.execute(meta.avg_pkt_size);

    //3. Standard deviation of packet size
    meta.std_pkt_size = square_root.execute(meta.var_pkt_size);
}
        \end{lstlisting}
    \end{subfigure}
    \caption{P4 Code for average/variance/standard deviation of packet size}
    \vspace{-0.5cm}
    \label{lst:p4_codex}
\end{figure*}



\begin{figure}[hbt!]
    \centering
    \includegraphics[width=9cm]{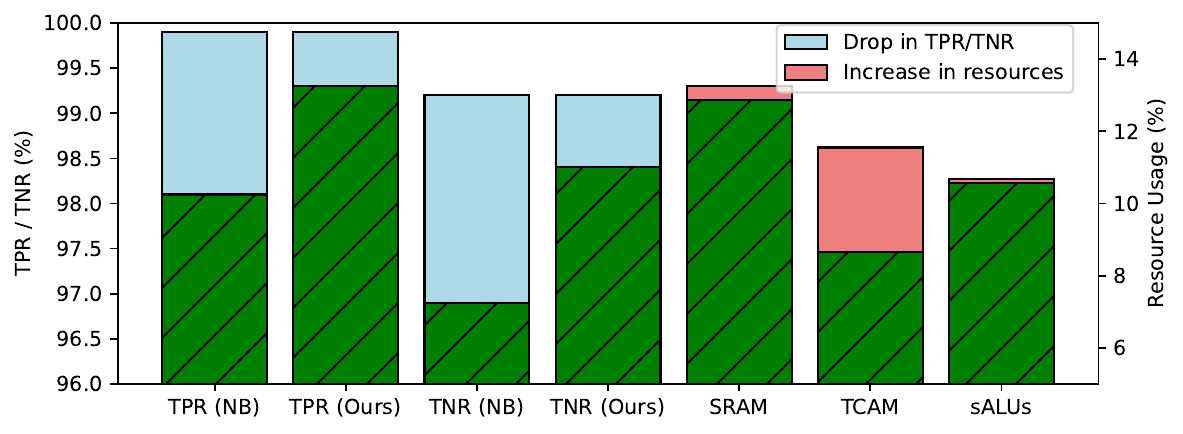}
    \caption{Drop in TPR/TNR and increase in TCAM, SRAM, and ALUs resource consumption when we approximate division-based BL features $A/B \approx 2^i$, where $i$ is an integer.}
    \label{fig:featimp}
\end{figure}

\myparab{Results. }Although above approach to extract division-based BL features might cause some information loss (floating point error), we show that it does not affect our results. We use the same benign training and validation set that we mentioned in the experiments of our paper. For attacks, we add $25\%$ HTTP DDoS attack samples to our validation set. As for implementation, we simulate a prototype in python3 using $21$ features (based on standard deviations of burst sizes and inter-packet delays/packet jitter) that were used by Magnifier as mentioned in \cite{290987} to judge the drop in TPR / TNR. As for resource consumption increase compared to NetBeacon's method (listing \ref{lst:p4_codenb}), we deploy a prototype in switch using average, variance, and standard deviation of packet sizes along with number of packets per burst. We demonstrate this in \fref{fig:featimp}. We see a drop of 0.6\% and 0.8\% for TPR and TNR from the ideal case, but also an increase in $1.2\%$ and $1.5\%$ for TPR and TNR from NetBeacon's method respectively. SRAM, TCAM, and sALU resource consumption increased by 0.4\%, 2.9\%, and 0.1\%  compared to NetBeacon's method respectively.

\section{Experiments}
\subsection{Consistency of our knowledge distillation scheme} \label{ckd}
We demonstrate the consistency and fidelity of our novel knowledge distillation algorithm using consistency metric $C$ defined as follows,
\begin{equation}
    C = \frac{1}{N} \sum_{i=1}^N \mathbbm{1}\{iForest_{distilled}(x_i) = Autoencoder(x_i)\}
\end{equation}
where $N$ is the number of samples in the validation/test set. 

\myparab{Knowledge distillation. }We distill the knowledge of Magnifier \cite{290987} into the iForest. For this experiment, we do not consider anomaly score from the original iForest. In other words, the label of distilled iForest can be obtained for this experiment by simply traversing all $t$ iTrees for a sample $x$ and taking a majority vote of the labels obtained on all iTrees. The BL features used to train iForest and Magnifier are the same and are given in our paper.  

\myparab{Dataset. }We divide benign dataset into $2:8$ for validation and training. We add one attack dataset (HTTP DDoS) to the validation set. For the validation set, attack traffic makes up $25\%$ of the traffic. In other words, the contamination ratio is 0.25.
\begin{figure*}[hbt!]
    \centering
    \includegraphics[width=18.5cm]{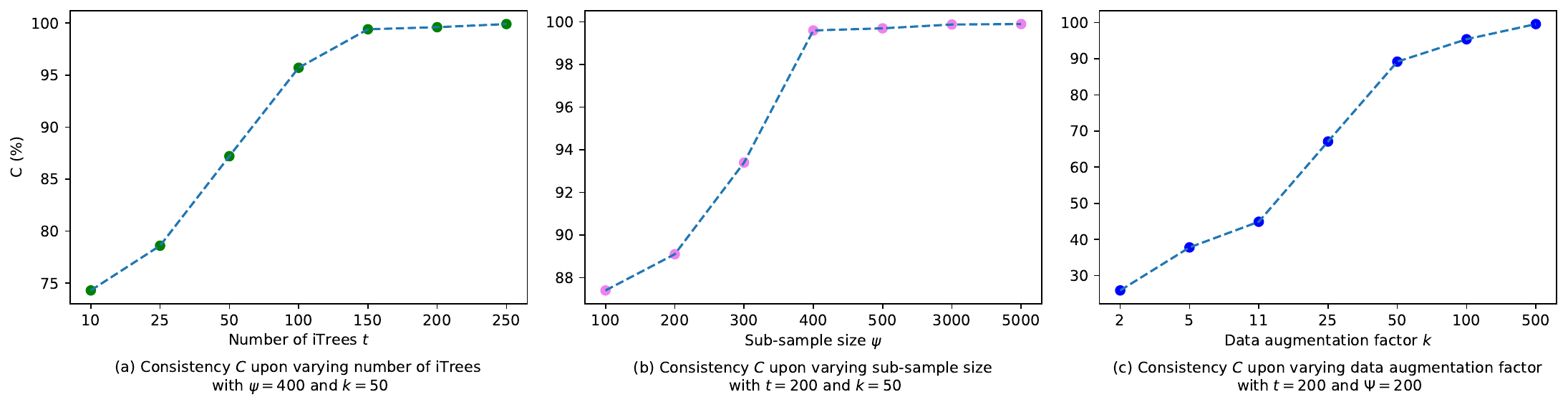}
    \caption{Effect of various hyperparameters on consistency of knowledge distillation algorithm.}
    \label{fig:consdis}
\end{figure*}

\myparab{Results. }We show the effect of various hyperparameters on the iForest knowledge distillation consistency in \fref{fig:consdis}.

\textbf{Number of iTrees (\fref{fig:consdis}a). } The more the number of iTrees in an iForest, the more the number of feature branches. More hyper-dimension feature space divisions ensure better knowledge transfer from Magnifier to the iForest by embedding reconstruction errors in the leaf nodes. Therefore, consistency increases with an increase in the number of iTrees up to a specific limit. As we can see, we need $t \approx 100$ iTrees for $95\%$ consistency. For $t > 150$ iTrees, the increase in consistency is very less. 

\textbf{Sub-sample size (\fref{fig:consdis}b). }iForest \cite{liu2008isolation} shows that the maximum depth of an iTree is on an average $\lfloor log_2 \Psi \rfloor$, where $\Psi$ is the sub-sample size. Therefore, subsample size controls the "narrowness" of a feature boundary in a branch, which in turn also affects \textit{how well} the knowledge of an autoencoder (here Magnifier) gets contained in the iForest. Intuitively, an increase in sub-sample size should increase the consistency, which is what happens until $\Psi = 400$ (after which the increase in consistency is minimal).

\textbf{Data augmentation factor $k$ (\fref{fig:consdis}c). }This factor denotes the number of points sampled in the features range at every leaf node of an iTree and the average reconstruction errors of those points are used to decide the label at the leaf node. Obviously, more data augmentation factor will give us a better representative performance of an autoencoder in an iForest. As shown, we need at least $k=50$ for good consistency of knowledge distillation procedure. 
\subsection{Effect of flow abnormality frequency threshold} \label{efat}
\begin{figure}[hbt!]
    \centering
    \includegraphics[width=8cm]{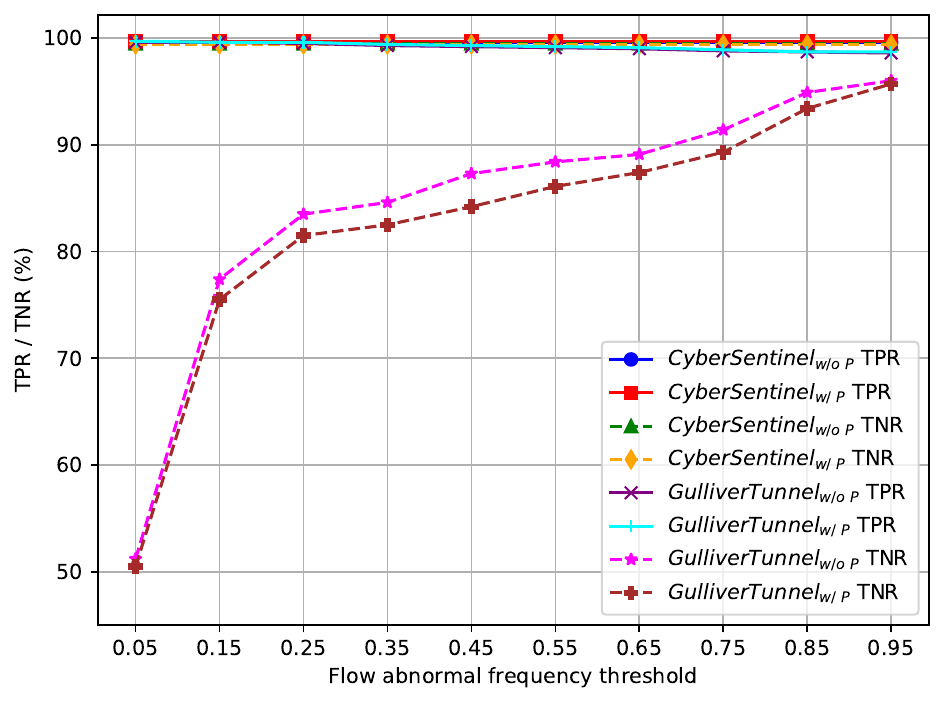}
    \caption{Effect of flow abnormality threshold on TPR and TNR of \arch and Gulliver Tunnel. Number of iTrees $=200$, sub-sample size $=5000$, contamination w/o \textit{P} = 0.15, and with \textit{P}= 0.05. Also data augmentation factor = $150$.}
    \label{fig:abt}
\end{figure}
We show the effect of flow abnormality frequency threshold (presented in our paper) on TPR and TNR of \arch. Moreover we also compare the performance with Gulliver Tunnel. Dataset is as the experiment in \secref{ckd}. We show the results in \fref{fig:abt}. As we can see, the flow abnormality frequency threshold has no effect on \texttt{CyberSentinel}'s TPR and TNR. This is because malicious traffic detection happens entirely in the data plane. On the other hand, Gulliver Tunnel's TPR is consistently high for all frequency thresholds because iForest model inherently has high TPR (most of the anomalies have an extremely large warning frequency). Gulliver Tunnel's TNR increases because increasing abnormality frequency thresholds increase the involvement of Magnifier in the control plane \cite{290987}. This experiment also justifies and concludes that \arch can achieve similar performance as Gulliver Tunnel + Magnifier but entirely in the data plane.

\myparab{Rationale for using flow abnormal frequency threshold in \arch while installing blacklist rules. }This has practical significance. For instance, if we set low abnormal frequency threshold for a flow, then upon receiving few malicious bursts for that flow, \arch data plane management module will install a blacklist rule. Subsequently, if normal packets from that same flow ID is incident on the switch, those packets get treated as malicious and are not lead into the network which may cause operational security issues \cite{277250}. On the other hand, if the frequency is set high, then the blacklist rule is installed only after receiving sufficient malicious bursts. While this helps us take better decisions in avoiding malicious traffic, some early malicious packets might be led into the network before the blacklist rule could be installed. However, this problem is alleviated upon deploying another iForest model with just PL features. 
\subsection{Per-packet resubmission overhead} \label{ppro}
\begin{table}
    \centering
    \scalebox{1.15}{
    \begin{tabular}{|c||c|c|c|}
    \hline
       \textbf{Packet size (B)}  & \textbf{\makecell{Resubmission overhead\\ (Gulliver Tunnel)}} & \textbf{\makecell{Resubmission overhead\\ (\arch)}} \\ \hline 
       256 B & 1060 bits &  11 bits \\ \hline
       512 B & 2119 bits & 11 bits  \\ \hline
       1024 B & 4237 bits & 11 bits  \\ \hline
       1500 B & 6207 bits & 11 bits  \\ \hline 
       IMIX & 5311 bits & 11 bits  \\ \hline
       \textbf{Average} & 3787 bits &  11 bits \\ \hline
    \end{tabular}}
    \caption{Per-packet resubmission overhead comparison of \arch with Gulliver Tunnel. }
    \label{tab:prc}
\end{table}
We first compare the percentage of resubmitted packets in \arch to that of Gulliver Tunnel \cite{290987}. This comparison will help justify the improvement in packet processing throughput and reduction in per-packet latency compared to Gulliver Tunnel (and consequently HorusEye). More resubmitted packets cause additional packet processing overhead (increasing per-packet latency). Thus, more time is taken to process a given number of packets which reduces packet-processing throughput.  

Similar to hardware experiments in our paper, we simulate high-speed traffic of varying packet sizes. We add attack traffic to the validation set (HTTP DDoS and TCP DDoS) with a contamination ratio of $0.2$. For IMIX traffic, we get $6.046\%$ of resubmitted packets in \arch. In comparison, Gulliver Tunnel yields $51.72\%$ resubmitted packets. 

As we can see, \arch has about 88.3\% less packet resubmissions than Gulliver Tunnel. This is because Gulliver Tunnel resubmits a packet in the pipeline for every incoming packet as per its data plane processing logic \cite{290987}. In contrast, \texttt{CyberSentinel}'s intelligent data plane is constructed such that packets are mirrored to the loopback port only when the burst segmentation threshold is reached (\secref{hpe}). 

Second, we compare the per-packet resubmission overhead of \arch with Gulliver Tunnel and show the results in \tref{tab:prc}. \arch yields $99.7\%$ lower per-packet resubmission overhead compared to Gulliver Tunnel. This is because \arch \textit{loops back payload truncated mirrored packet} (with additional metadata such as burst ID and hash table selector) every time the burst segmentation threshold is reached. Therefore, the payload of a packet has no effect on \arch resubmission overhead. In contrast, Gulliver Tunnel \textit{resubmits every incoming packet} with additional metadata (burst ID and register read/reset choice).  
\subsection{End-to-end detection performance} \label{eedpa}
\begin{table*}[ht!]
\scalebox{0.8}{
\begin{tabular}{|c|c|ccc|ccc|ccc|ccc|ccc|}
\hline
\multirow{3}{*}{Dataset}                                            & \multirow{3}{*}{Attacks} & \multicolumn{3}{c|}{Kitsune \cite{mirsky2018kitsune}}                                                                   & \multicolumn{3}{c|}{Magnifier \cite{290987}}                                                                 & \multicolumn{3}{c|}{HorusEye \cite{290987}}                                                                  & \multicolumn{3}{c|}{\arch ($n=8$ features)}                                         & \multicolumn{3}{c|}{\arch ($n=21$ features)}                                    \\ \cline{3-17} 
                                                                    &                          & \multicolumn{2}{c|}{TPR}                                            & \multirow{2}{*}{PR\_AUC} & \multicolumn{2}{c|}{TPR}                                            & \multirow{2}{*}{PR\_AUC} & \multicolumn{2}{c|}{TPR}                                            & \multirow{2}{*}{PR\_AUC} & \multicolumn{2}{c|}{TPR}                                                & \multirow{2}{*}{PR\_AUC} & \multicolumn{2}{c|}{TPR}                                            & \multirow{2}{*}{PR\_AUC} \\ \cline{3-4} \cline{6-7} \cline{9-10} \cline{12-13} \cline{15-16}
                                                                    &                          & \multicolumn{1}{c|}{$\leq$ 5e-5} & \multicolumn{1}{c|}{$\leq$ 5e-4} &                          & \multicolumn{1}{c|}{$\leq$ 5e-5} & \multicolumn{1}{c|}{$\leq$ 5e-4} &                          & \multicolumn{1}{c|}{$\leq$ 5e-5} & \multicolumn{1}{c|}{$\leq$ 5e-4} &                          & \multicolumn{1}{c|}{$\leq$ 2e-3}   & \multicolumn{1}{c|}{$\leq$ 1e-2}   &                          & \multicolumn{1}{c|}{$\leq$ 5e-5} & \multicolumn{1}{c|}{$\leq$ 5e-4} &                          \\ \hline
\multirow{11}{*}{\begin{tabular}[c]{@{}c@{}}\cite{bezerra2018providing}\\ \cite{koroniotis2019towards}\\ \cite{mirsky2018kitsune}\end{tabular}} & Aidra                    & \multicolumn{1}{c|}{0.595}       & \multicolumn{1}{c|}{0.611}       & 0.854                    & \multicolumn{1}{c|}{0.620}       & \multicolumn{1}{c|}{0.657}       & 0.823                    & \multicolumn{1}{c|}{0.623}       & \multicolumn{1}{c|}{0.662}       & 0.880                    & \multicolumn{1}{c|}{0.621 / 0.632} & \multicolumn{1}{c|}{0.662 / 0.689} & 0.881 / 0.885            & \multicolumn{1}{c|}{0.624}       & \multicolumn{1}{c|}{0.665}       & 0.885                    \\ \cline{2-17} 
                                                                    & Bashlite                 & \multicolumn{1}{c|}{0.784}       & \multicolumn{1}{c|}{0.795}       & 0.890                    & \multicolumn{1}{c|}{0.809}       & \multicolumn{1}{c|}{0.843}       & 0.909                    & \multicolumn{1}{c|}{0.814}       & \multicolumn{1}{c|}{0.845}       & 0.935                    & \multicolumn{1}{c|}{0.811 / 0.820} & \multicolumn{1}{c|}{0.843 / 0.855} & 0.933 / 0.942            & \multicolumn{1}{c|}{0.816}       & \multicolumn{1}{c|}{0.847}       & 0.938                    \\ \cline{2-17} 
                                                                    & Mirai                    & \multicolumn{1}{c|}{0.965}       & \multicolumn{1}{c|}{0.966}       & 0.993                    & \multicolumn{1}{c|}{0.967}       & \multicolumn{1}{c|}{0.967}       & 0.993                    & \multicolumn{1}{c|}{0.967}       & \multicolumn{1}{c|}{0.967}       & 0.996                    & \multicolumn{1}{c|}{0.965 / 0.968} & \multicolumn{1}{c|}{0.967 / 0.969} & 0.995 / 0.997            & \multicolumn{1}{c|}{0.968}       & \multicolumn{1}{c|}{0.969}       & 0.997                    \\ \cline{2-17} 
                                                                    & Keylogging               & \multicolumn{1}{c|}{0.580}       & \multicolumn{1}{c|}{0.607}       & 0.869                    & \multicolumn{1}{c|}{0.592}       & \multicolumn{1}{c|}{0.660}       & 0.923                    & \multicolumn{1}{c|}{0.600}       & \multicolumn{1}{c|}{0.677}       & 0.941                    & \multicolumn{1}{c|}{0.601 / 0.608} & \multicolumn{1}{c|}{0.676 / 0.681} & 0.940 / 0.949            & \multicolumn{1}{c|}{0.605}       & \multicolumn{1}{c|}{0.682}       & 0.945                    \\ \cline{2-17} 
                                                                    & Data theft               & \multicolumn{1}{c|}{0.591}       & \multicolumn{1}{c|}{0.617}       & 0.868                    & \multicolumn{1}{c|}{0.598}       & \multicolumn{1}{c|}{0.667}       & 0.919                    & \multicolumn{1}{c|}{0.604}       & \multicolumn{1}{c|}{0.682}       & 0.938                    & \multicolumn{1}{c|}{0.605 / 0.610} & \multicolumn{1}{c|}{0.680 / 0.685} & 0.936 / 0.944            & \multicolumn{1}{c|}{0.607}       & \multicolumn{1}{c|}{0.685}       & 0.939                    \\ \cline{2-17} 
                                                                    & Service scan             & \multicolumn{1}{c|}{0.891}       & \multicolumn{1}{c|}{0.896}       & 0.982                    & \multicolumn{1}{c|}{0.889}       & \multicolumn{1}{c|}{0.905}       & 0.984                    & \multicolumn{1}{c|}{0.887}       & \multicolumn{1}{c|}{0.902}       & 0.990                    & \multicolumn{1}{c|}{0.888 / 0.901} & \multicolumn{1}{c|}{0.902 / 0.906} & 0.988 / 0.992            & \multicolumn{1}{c|}{0.889}       & \multicolumn{1}{c|}{0.905}       & 0.993                    \\ \cline{2-17} 
                                                                    & OS scan                  & \multicolumn{1}{c|}{0.602}       & \multicolumn{1}{c|}{0.667}       & 0.989                    & \multicolumn{1}{c|}{0.630}       & \multicolumn{1}{c|}{0.873}       & 0.995                    & \multicolumn{1}{c|}{0.659}       & \multicolumn{1}{c|}{0.880}       & 0.987                    & \multicolumn{1}{c|}{0.657 / 0.662} & \multicolumn{1}{c|}{0.881 / 0.887} & 0.985 / 0.988            & \multicolumn{1}{c|}{0.660}       & \multicolumn{1}{c|}{0.884}       & 0.989                    \\ \cline{2-17} 
                                                                    & HTTP DDoS                & \multicolumn{1}{c|}{0.674}       & \multicolumn{1}{c|}{0.744}       & 0.984                    & \multicolumn{1}{c|}{0.728}       & \multicolumn{1}{c|}{0.851}       & 0.994                    & \multicolumn{1}{c|}{0.731}       & \multicolumn{1}{c|}{0.853}       & 0.984                    & \multicolumn{1}{c|}{0.733 / 0.737} & \multicolumn{1}{c|}{0.855 / 0.858} & 0.981 / 0.985            & \multicolumn{1}{c|}{0.735}       & \multicolumn{1}{c|}{0.855}       & 0.986                    \\ \cline{2-17} 
                                                                    & TCP DDoS                 & \multicolumn{1}{c|}{0.961}       & \multicolumn{1}{c|}{0.965}       & 0.989                    & \multicolumn{1}{c|}{0.969}       & \multicolumn{1}{c|}{0.980}       & 0.993                    & \multicolumn{1}{c|}{0.970}       & \multicolumn{1}{c|}{0.981}       & 0.994                    & \multicolumn{1}{c|}{0.968 / 0.975} & \multicolumn{1}{c|}{0.981 / 0.982} & 0.994 / 0.994            & \multicolumn{1}{c|}{0.972}       & \multicolumn{1}{c|}{0.985}       & 0.994                    \\ \cline{2-17} 
                                                                    & UDP DDoS                 & \multicolumn{1}{c|}{0.961}       & \multicolumn{1}{c|}{0.966}       & 0.989                    & \multicolumn{1}{c|}{0.970}       & \multicolumn{1}{c|}{0.980}       & 0.993                    & \multicolumn{1}{c|}{0.969}       & \multicolumn{1}{c|}{0.980}       & 0.993                    & \multicolumn{1}{c|}{0.969 / 0.971} & \multicolumn{1}{c|}{0.980 / 0.982} & 0.994 / 0.995            & \multicolumn{1}{c|}{0.971}       & \multicolumn{1}{c|}{0.983}       & 0.995                    \\ \cline{2-17} 
                                                                    & \textbf{macro}           & \multicolumn{1}{c|}{0.760}       & \multicolumn{1}{c|}{0.783}       & 0.941                    & \multicolumn{1}{c|}{0.777}       & \multicolumn{1}{c|}{0.838}       & 0.952                    & \multicolumn{1}{c|}{0.782}       & \multicolumn{1}{c|}{0.843}       & 0.964                    & \multicolumn{1}{c|}{0.782 / 0.788} & \multicolumn{1}{c|}{0.843 / 0.849} & 0.963 / 0.967            & \multicolumn{1}{c|}{0.785}       & \multicolumn{1}{c|}{0.846}       & 0.967                    \\ \hline
\multirow{6}{*}{\cite{290987}}                                               & Mirai                    & \multicolumn{1}{c|}{0.661}       & \multicolumn{1}{c|}{0.761}       & 0.986                    & \multicolumn{1}{c|}{0.812}       & \multicolumn{1}{c|}{0.871}       & 0.994                    & \multicolumn{1}{c|}{0.817}       & \multicolumn{1}{c|}{0.888}       & 0.995                    & \multicolumn{1}{c|}{0.818 / 0.821} & \multicolumn{1}{c|}{0.888 / 0.889} & 0.995 / 0.995            & \multicolumn{1}{c|}{0.820}       & \multicolumn{1}{c|}{0.891}       & 0.995                    \\ \cline{2-17} 
                                                                    & Service scan             & \multicolumn{1}{c|}{0.993}       & \multicolumn{1}{c|}{0.994}       & 1.000                    & \multicolumn{1}{c|}{0.991}       & \multicolumn{1}{c|}{0.997}       & 1.000                    & \multicolumn{1}{c|}{0.991}       & \multicolumn{1}{c|}{0.997}       & 1.000                    & \multicolumn{1}{c|}{0.991 / 0.992} & \multicolumn{1}{c|}{0.997 / 0.997} & 0.999 / 1.000            & \multicolumn{1}{c|}{0.993}       & \multicolumn{1}{c|}{0.998}       & 1.000                    \\ \cline{2-17} 
                                                                    & OS scan                  & \multicolumn{1}{c|}{0.987}       & \multicolumn{1}{c|}{0.990}       & 1.000                    & \multicolumn{1}{c|}{0.986}       & \multicolumn{1}{c|}{0.994}       & 1.000                    & \multicolumn{1}{c|}{0.987}       & \multicolumn{1}{c|}{0.994}       & 1.000                    & \multicolumn{1}{c|}{0.988 / 0.989} & \multicolumn{1}{c|}{0.993 / 0.995} & 1.000 / 1.000            & \multicolumn{1}{c|}{0.988}       & \multicolumn{1}{c|}{0.994}       & 1.000                    \\ \cline{2-17} 
                                                                    & TCP DDoS                 & \multicolumn{1}{c|}{0.997}       & \multicolumn{1}{c|}{0.998}       & 1.000                    & \multicolumn{1}{c|}{0.998}       & \multicolumn{1}{c|}{0.998}       & 1.000                    & \multicolumn{1}{c|}{0.998}       & \multicolumn{1}{c|}{0.999}       & 1.000                    & \multicolumn{1}{c|}{0.999 / 0.999} & \multicolumn{1}{c|}{0.999 / 1.000} & 1.000 / 1.000            & \multicolumn{1}{c|}{0.999}       & \multicolumn{1}{c|}{0.999}       & 1.000                    \\ \cline{2-17} 
                                                                    & UDP DDoS                 & \multicolumn{1}{c|}{0.998}       & \multicolumn{1}{c|}{0.999}       & 1.000                    & \multicolumn{1}{c|}{0.999}       & \multicolumn{1}{c|}{0.999}       & 1.000                    & \multicolumn{1}{c|}{0.999}       & \multicolumn{1}{c|}{0.999}       & 1.000                    & \multicolumn{1}{c|}{0.999 / 1.000} & \multicolumn{1}{c|}{1.000 / 1.000} & 1.000 / 1.000            & \multicolumn{1}{c|}{0.999}       & \multicolumn{1}{c|}{1.000}       & 1.000                    \\ \cline{2-17} 
                                                                    & \textbf{macro}           & \multicolumn{1}{c|}{0.927}       & \multicolumn{1}{c|}{0.948}       & 0.997                    & \multicolumn{1}{c|}{0.957}       & \multicolumn{1}{c|}{0.972}       & 0.999                    & \multicolumn{1}{c|}{0.958}       & \multicolumn{1}{c|}{0.975}       & 0.999                    & \multicolumn{1}{c|}{0.959 / 0.960} & \multicolumn{1}{c|}{0.975 / 0.976} & 0.999 / 0.999            & \multicolumn{1}{c|}{0.960}       & \multicolumn{1}{c|}{0.976}       & 0.999                    \\ \hline
\end{tabular}}
\caption{Performance of \arch with state-of-the-arts. Note that $\leq$5e-4 means FPR $\leq$ 5e-4. Moreover, we implement \arch with $8$ features in the hardware and for this version, TPR and PR\textsubscript{AUC} are shown for both without profiler / with profiler. \arch with $21$ features is simulated in python3 due to switch constraints. We use normal traffic dataset \cite{sivanathan2018classifying}.}
\vspace{-0.5cm}
\label{tab:etep2}
\end{table*}
\tref{tab:etep2} presents an overview of the detection efficacy of models trained using normal traffic \cite{sivanathan2018classifying}. Each scheme demonstrates strong performance on this dataset, largely due to its limited number of devices and absence of complex behavioral interactions observed in normal traffic \cite{290987}.
\subsection{Detection performance of distilled iForest on other datasets} \label{dpod}
\begin{figure}
    \centering
    \includegraphics[width=9cm]{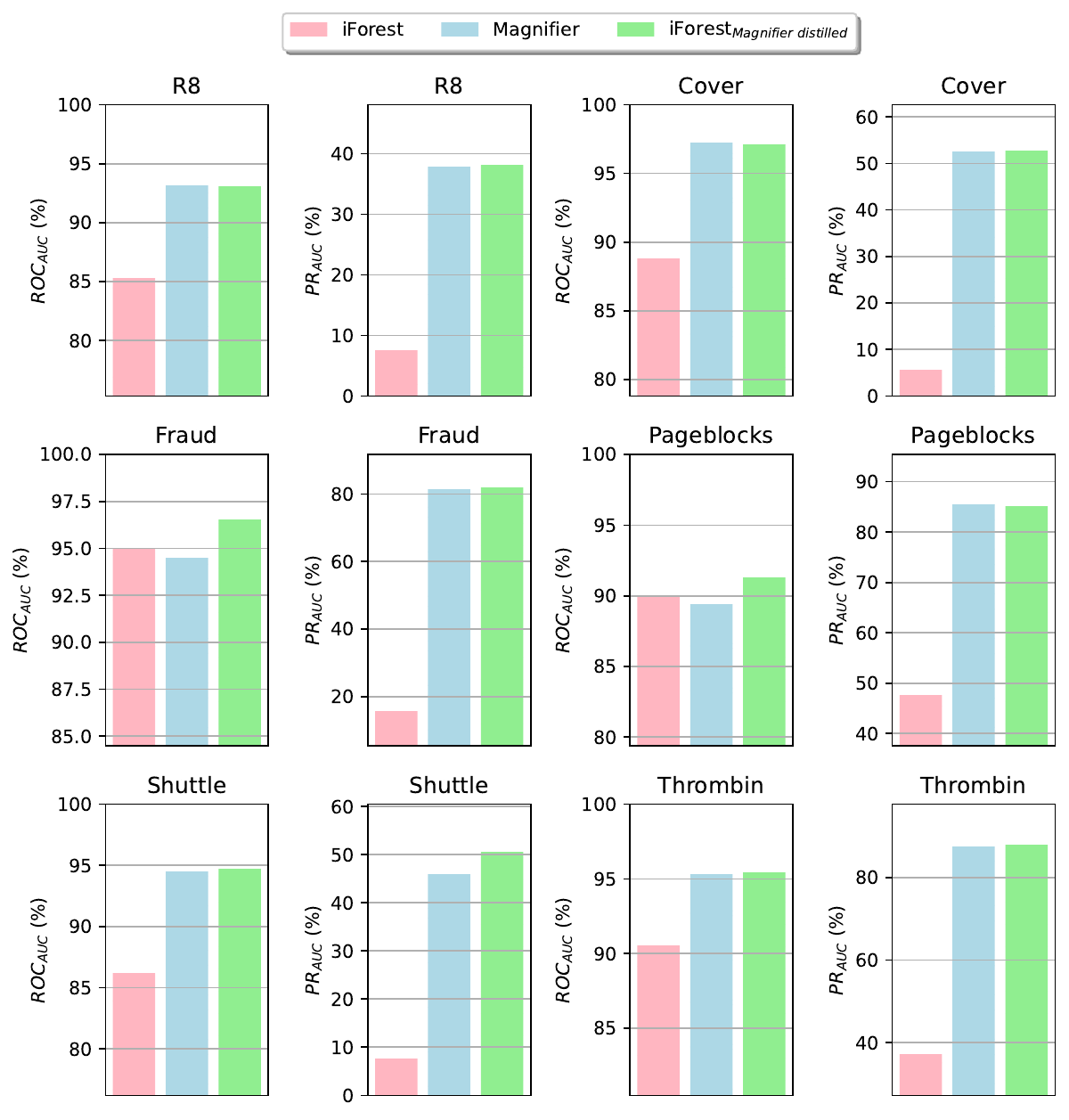}
    \caption{Comparison of Magnifier-distilled iForest with iForest and Magnifier in terms of ROC\textsubscript{AUC} and PR\textsubscript{AUC} on various non-network attacks.}
    \label{fig:fincomp2}
\end{figure}
We compare the performance of our knowledge distilled iForest with iForest and Magnifier on other datasets as well. 

\myparab{Dataset. }We will consider attacks such as (i) \textit{R8} \cite{pang2018learning,rayana2016less} which is a highly-imbalanced text classification dataset,
where the rare class are treated as anomalies; (ii) \textit{Cover} is from the ecology domain. (iii) \textit{Fraud} is for
fraudulent credit card transaction detection. (iv) \textit{Pageblocks} and
\textit{Shuttle} are provided by an anomaly benchmark study \cite{campos2016evaluation}. (v) \textit{Thrombin} \cite{pang2018learning} is to detect unusual molecular bio-activity for
drug design, which is an ultrahigh-dimensional anomaly
detection dataset.

\myparab{Experiment setting. }For all datasets, training and validation are divided into a $4:1$ ratio. Moreover, contamination ratios for each of the attacks are the same as the anomaly ratios provided in the datasets. 

\myparab{Hyperparameter setting. }We use number of iTrees $t=300$, sub-sample size $\Psi = 256$ and, data augmentation factor $k=150$. The setting for the Magnifier is the same as in \cite{290987}. 

\myparab{Metrics. }We use two metrics: PR\textsubscript{AUC} and ROC\textsubscript{AUC}. The ROC curve indicates the true positives against
false positives, while the PR curve summarises precision and
recall of the anomaly class only. 

\myparab{Implementation. }We only consider our knowledge distillation scheme, that is, Magnifier distilled iForest. We compare distilled iForest to iForest and Magnifier. Implementation is done in python3. 

\myparab{Results.} We show the results in \fref{fig:fincomp2}. Our knowledge distilled iForest yields a similar area under the PR curve and ROC curve compared to Magnifier. This demonstrates that our novel knowledge distillation algorithm is effective. Moreover, the detection performance exceeds that of iForest.

\myparab{Conclusion. }We have shown that for many attacks (including UNSW dataset), HorusEye may not work well if the iForest model does not work well (does not yield very high TPR). In these cases, it is better to leverage powerful autoencoders such as Magnifier. Fortunately, our novel knowledge distillation scheme helps us envision the powerful models' performance in a relatively lightweight iForest. Moreover, we also showed that our knowledge distillation algorithm works well for non-network-based traffic as well. 
\section{Burst to flow mapping module} \label{dmti}
\begin{figure*}[t!]
    \hspace{-1.6cm}
    \includegraphics[width=21cm]{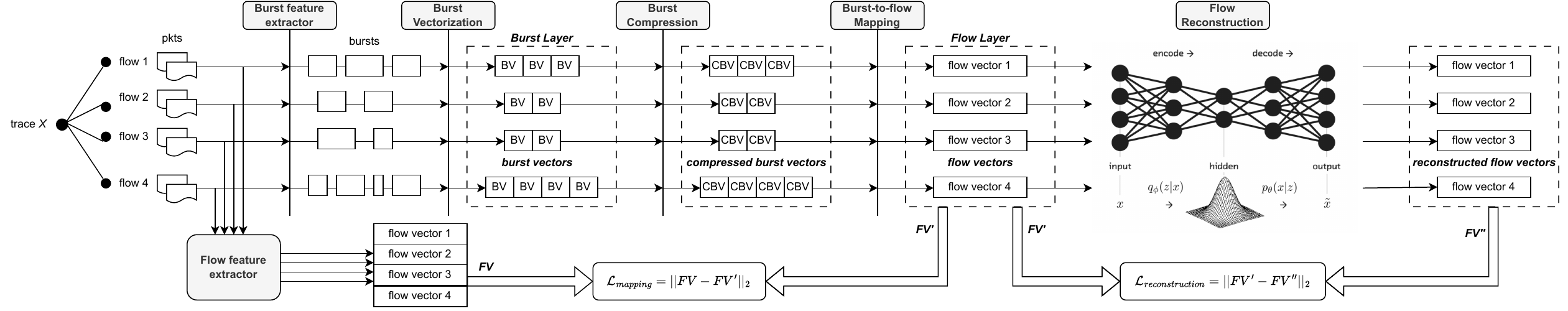}
    \caption{Burst-to-flow mapping module ($\mathcal{M}$) design}
    \vspace{-0.5cm}
    \label{fig:md}
\end{figure*}
We design a novel burst-flow mapping module by following a variant of the packet-flow mapping module implemented in \cite{qu2023input}.

\myparab{Rationale. }Many popular autoencoders such as Magnifier \cite{290987} or Kitsune \cite{mirsky2018kitsune} extract time window based FL features. For example, Magnifier obtains $21$ FL features based on standard deviations of packet sizes and inter-packet delays for each of the $5$ time windows. As such, Magnifier inputs a sample of $21@5$ and reconstructs it. On the other hand, we train the iForest on BL (and PL) features (\secref{bfe}). If we are to perform knowledge distillation of autoencoders (\textit{e.g.}, Magnifier) into an iForest, then we need to have a module that can map \textit{bursts} into a flow. In other words, for a given flow ID (5-tuple), the module takes into input a set of BL features and converts them into a FL feature. We show the design of our module below. 

\myparab{Module design. }We present the module design in \fref{fig:md}. We also explain the details in the following subsections.

\subsection{Burst feature extractor and flow feature extractor}
As shown in \fref{fig:md}, we first extract bursts from a sequence of packets of a given flow using a \textit{burst feature extractor} (\secref{bfe}). We re-iterate that a burst is a sequence of packets where inter-packet delay does not exceed a timeout threshold, and the number of packets does not exceed the burst segmentation threshold. Two flows may have different numbers of bursts. 

In parallel, we also extract flow-level features for a given flow using \textit{flow feature extractor}. This yields a fixed length vector for a given flow called \textit{flow vector}. A flow vector comprises a collection of flow-level features for a given flow (flow ID). For example, in Magnifier \cite{290987}, a flow vector for any given flow ID is $21@5$ dimensions, where it contains $21$ FL features for each of $5$ time windows. 
\subsection{Burst vectorization and compression}
\textit{Burst vectorization}. Using a sequence of bursts for a given flow, we create a sequence of \textit{burst vectors}. Referring to \fref{fig:md}, flow 1 has $3$ bursts, and therefore $3$ burst vectors (BVs) are concatenated and fed into \textit{burst compression layer}. For a given burst $B_i$ having $n_i$ packets, the corresponding burst vector $BV_i$ is,
\begin{equation}
    BV_i = \langle n_i, bf_1, bf_2, ..., bf_d \rangle
\end{equation}
where $bf_j$ is the $j^{th}$ burst-level (BL) feature and $d$ is number of BL features. A sequence of burst vectors (for a given flow) is fed into the burst compression layer. 

\textit{Burst compression}. It is computationally expensive for the neural networks responsible for burst-to-flow mapping (\fref{fig:md}). To handle long
sequences of burst vectors, we use Convolutional Neural
Network (CNN) to compress a long sequence of burst vectors into a short sequence of long vectors. We term such long
vectors as \textit{Compressed Burst Vectors (CBVs)}. This intuition is also leveraged in \cite{qu2023input}. 
\subsection{Burst-flow mapping}
To transform a sequence of burst vectors into a flow vector (see \fref{fig:md}),
we consider using four popular neural network structures, the chain-structured recurrent
neural network \cite{liu2019fs}, the tree-structured recurrent neural network
using a balanced binary tree \cite{tai2015improved}, the scaled dot-product attention-structured neural network \cite{vaswani2017attention}, and the hybrid neural network \cite{wu2020short,fu2020nsa}. Irrespective of which network we use, we feed it (burst-flow mapping) a sequence of variable length CBVs and get output as a fixed length flow vector (FV'). 

\myparab{Mapping loss. }Purpose of \textit{burst-flow mapping} is to correctly map the generated BVs of a flow to a flow vector. Suppose it mapped to FV' as shown in \fref{fig:md}. On the other hand, we get flow vectors FVs by directly feeding packets to a \textit{flow feature extractor}. Then the purpose of \textit{burst-flow mapping} is to minimize the below loss,
\begin{equation}
    \mathcal{L}_{mapping} = ||FV - FV'||_2
\end{equation}

\subsection{Flow reconstruction}
The flow vectors from the \textit{burst-flow mapping} (FV') are fed into an autoencoder (\textit{e.g.}, Magnifier \cite{290987}) which then tries to reconstruct the flow vectors (FV') by minimizing the below reconstruction loss.
\begin{equation}
    \mathcal{L}_{reconstruction} = ||FV' - FV''||_2
\end{equation}
where $FV''$ is the reconstructed flow vectors from $FV'$. 

\subsection{Final loss function}
We train the module (CNN in burst compression, neural networks in burst-flow mapping, and autoencoder in flow reconstruction) by minimizing the following loss function,
\begin{equation}
    \mathcal{L}_{module} = \mathcal{L}_{mapping} + \lambda \mathcal{L}_{reconstruction}
\end{equation}
where $\lambda$ is the relative weight given to two losses. 
\subsection{Real-time deployment}
Once the module based on minimizing $\mathcal{L}_{module}$ is trained, we can infer reconstruction loss for a flow feature sample $x$ as follows. Let $\textbf{xb} = \langle xb_1, xb_2, ..., xb_m \rangle$ be the corresponding sequence of burst feature samples for a flow having $m$ bursts. If $\mathcal{M}(.)$ is a module function, then $\mathcal{M}(\textbf{xb})$ yields reconstructed version of sample $x$. Then we can decide whether a flow sample is benign or an anomaly based on,
\begin{equation}
    label(x) = 
\begin{cases}
1 &  \sqrt{\frac{1}{d}\sum_{i=1}^d(\mathcal{M}(\textbf{xb})_i - x_i)^2} > T\\
0 & \text{otherwise}
\end{cases}
\end{equation}
where $T$ is the RMSE threshold of the autoencoder and $d$ is the number of features in flow and burst samples.

\myparab{Note: }Burst and flow samples should be consistent in the number and the type of features they use across the module.

\subsection{Challenges in implementing knowledge distillation}
The inclusion of the module in \fref{fig:md} makes the current knowledge distillation algorithm challenging to implement. All the steps in Alg. \ref{algo:3} can still be implemented except step 3. Module function $\mathcal{M}(.)$ takes a sequence of BVs for a given flow ID and generates, then reconstructs a fixed length flow vector. However, for each leaf node $leaf_{ij}$ burst samples $X_{ij} \cup A_{ij}$ cannot be fed directly to the autoencoder in module $\mathcal{M}$ (\fref{fig:md}). We need to first \textit{sequence} the burst samples as per the given flow ID. For this, we divide $X_{ij}$ into subsets (instead of individual samples) follows,
\begin{equation}
    X_{ij} = \{ X_{ij_{1}}, X_{ij_{2}}, X_{ij_{3}}, ..., X_{ij_{p}} \}
\end{equation}
where $X_{ij_{p}} \subseteq X_{ij}$ and $X_{ij_{p}}$ is sequence of burst samples mapped to flow $p$. Similarly, augmented burst samples can also partitioned by arbitrarily mapping them to flows as,
\begin{equation}
    A_{ij} = \{ A_{ij_{1}}, A_{ij_{2}}, A_{ij_{3}}, ..., A_{ij_{p}} \}
\end{equation}
where $|A_{ij_{p}}| = |A_{ij_{q}}| = \lfloor \frac{k}{p} \rfloor$, $p \neq q$.
Then autoencoder in module $\mathcal{M}$ can take input burst samples at the subset level and map them to flow samples. We show the modified algorithm for knowledge distillation in Alg. \ref{algo:5}.
\setlength{\textfloatsep}{0.5pt}%
\begin{algorithm}[t!]
\caption{Modified Knowledge distillation in iForest (inclusion of \fref{fig:md})}\label{algo:5}
\small
\begin{algorithmic}
\State $\bullet$ Steps 1-2 of Algo. \ref{algo:3}.
\Statex \Comment{We modify step 3 of Algo. \ref{algo:3}}
\State $\bullet$ We divide $X_{ij}$ into subsets (instead of individual samples) as follows,
\[
    X_{ij} = \{ X_{ij_{1}}, X_{ij_{2}}, X_{ij_{3}}, ..., X_{ij_{p}} \}
\]
where $X_{ij_{p}} \subseteq X_{ij}$ and $X_{ij_{p}}$ is sequence of burst samples mapped to flow $p$. Similarly, augmented burst samples can also partitioned by arbitrarily mapping them to flows as,
\[
    A_{ij} = \{ A_{ij_{1}}, A_{ij_{2}}, A_{ij_{3}}, ..., A_{ij_{p}} \}
\]
where $|A_{ij_{p}}| = |A_{ij_{q}}| = \lfloor \frac{k}{p} \rfloor$, $p \neq q$.
For each of $r$ trained autoencoders, obtain the expected reconstruction error as follows,
\[
\overline{RE_{u_{ij}}} = \mathbbm{E}_{x \sim X_{ij} \cup A_{ij}}[RE_u(x)]
\]
where $\overline{RE_{u_{ij}}}$ is expected reconstruction error ($RE$) for $u^{th}$ autoencoder on subset samples $X_{ij} \cup A_{ij}$. Whereas $RE_u(x)$ is reconstruction error of $u^{th}$ autoencoder for sample $x$. This is repeated for each leaf of iForest.
\State $\bullet$ Steps 4 - 6 of Algo. \ref{algo:3}
\end{algorithmic}
\end{algorithm}
\subsection{Experimental setting}
For our experiments in paper, we consider mapping BL features to FL features that are consistent with Magnifier \cite{290987}. This is because we perform knowledge distillation from Magnifier (trained on FL features) into an iForest model trained on BL features. We found that setting $\lambda=1$ in Eq(8) worked well for our experiments. Further, we used similar setting of \cite{qu2023input} for CNN in \textit{burst compression}. Moreover, in \textit{burst-flow mapping}, we found that using hybrid neural networks \cite{wu2020short,fu2020nsa} with the following setting of \cite{qu2023input} gave us the best results (that is, minimum $\mathcal{L}_{module}$ for validation set). As for Magnifier, we only adjusted the number of nodes in the input and output layer (as per the number of features) and kept the rest of the setting similar to \cite{290987}. Since the details of the burst-flow module are not the main part of our novelty or experimental analysis, we keep it short.

\myparab{Q.} If \textit{burst-flow mapping} itself maps bursts to a flow, then what is the need for further reconstructing the mapped flows? 

The whole purpose of having \textit{burst-flow mapping} is to make our novel knowledge distillation algorithm (\secref{kd}, Alg. \ref{algo:3}) compliant with the popular autoencoders \cite{mirsky2018kitsune,290987} that are trained on FL features. Our aim is to distill their knowledge into an iForest model that is trained on BL features; the mapping is there to map BL to FL features. However, if we remove the flow reconstruction (i.e., autoencoders) from the module in \fref{fig:md}, then it defeats the whole purpose of knowledge distillation.
\end{document}